
\magnification=\magstephalf

\newbox\SlashedBox
\def\slashed#1{\setbox\SlashedBox=\hbox{#1}
\hbox to 0pt{\hbox to 1\wd\SlashedBox{\hfil/\hfil}\hss}#1}
\def\hboxtosizeof#1#2{\setbox\SlashedBox=\hbox{#1}
\hbox to 1\wd\SlashedBox{#2}}

\def\mathslashed#1{\setbox\SlashedBox=\hbox{$#1$}
\hbox to 0pt{\hbox to 1\wd\SlashedBox{\hfil/\hfil}\hss}#1}

\def\ifsmall{\iffalse}  
\def\titlepagefont{}  
 
\def\DefineTeXgraphics{%
\special{ps::[global] /TeXgraphics { } def}}  
 
\def\today{\ifcase\month\or January\or February\or March\or April\or May
\or June\or July\or August\or September\or October\or November\or
December\fi\space\number\day, \number\year}
\def\eatPrefix19{}
\def\Year{\expandafter\eatPrefix\the\year}
\newcount\hours \newcount\minutes
\def\monthname{\ifcase\month\or
January\or February\or March\or April\or May\or June\or July\or
August\or September\or October\or November\or December\fi}
\def\shortmonthname{\ifcase\month\or
Jan\or Feb\or Mar\or Apr\or May\or Jun\or Jul\or
Aug\or Sep\or Oct\or Nov\or Dec\fi}

\def\TimeStamp{\hours\the\time\divide\hours by60%
\minutes -\the\time\divide\minutes by60\multiply\minutes by60%
\advance\minutes by\the\time%
${\rm \shortmonthname}\cdot\if\day<10{}0\fi\the\day\cdot\the\year%
\qquad\the\hours:\if\minutes<10{}0\fi\the\minutes$}




\def\Title#1{%
\vskip 1in{\titlefont\centerline{#1}}\vskip .5in}
 
\def\Date#1{\leftline{#1}\tenrm\supereject%
\global\hsize=\hsbody\global\hoffset=\hbodyoffset%
\footline={\hss\tenrm\folio\hss}}

\newif\ifdraftmode
\newif\ifleftlabels  

\def\nolabels{\def\wrlabeL##1{}\def\eqlabeL##1{}\def\reflabeL##1{}}
\def\writelabels{\def\wrlabeL##1{\leavevmode\vadjust{\rlap{\smash%
{\line{{\escapechar=` \hfill\rlap{\sevenrm\hskip.03in\string##1}}}}}}}%
\def\eqlabeL##1{{\escapechar-1\rlap{\sevenrm\hskip.05in\string##1}}}%
\def\reflabeL##1{\noexpand\rlap{\noexpand\sevenrm[\string##1]}}}
\def\writeleftlabels{\def\wrlabeL##1{\leavevmode\vadjust{\rlap{\smash%
{\line{{\escapechar=` \hfill\rlap{\sevenrm\hskip.03in\string##1}}}}}}}%
\def\eqlabeL##1{{\escapechar-1%
\rlap{\sixrm\hskip.05in\string##1}%
\llap{\sevenrm\string##1\hskip.03in\hbox to \hsize{}}}}%
\def\reflabeL##1{\noexpand\rlap{\noexpand\sevenrm[\string##1]}}}
\nolabels

\newdimen\fullhsize
\newdimen\hstitle
\hstitle=\hsize 
\newdimen\hsbody
\hsbody=\hsize 
\newdimen\hbodyoffset
\hbodyoffset=\hoffset 
\newbox\leftpage
\def\abstract#1{#1}
\def\rotated{\special{ps: landscape}
\magnification=1000  
\baselineskip=14pt
\global\hstitle=9truein\global\hsbody=4.75truein
\global\vsize=7truein\global\voffset=-.31truein
\global\hoffset=-0.54in\global\hbodyoffset=-.54truein
\global\fullhsize=10truein
\def\DefineTeXgraphics{%
\special{ps::[global]
/TeXgraphics {currentpoint translate 0.7 0.7 scale
              -80 0.72 mul -1000 0.72 mul translate} def}}
\let\lr=L
\def\ifsmall{\iftrue}
\def\titlepagefont{\twelvepoint}
\trueseventeenpoint
\def\almostshipout##1{\if L\lr \count1=1
      \global\setbox\leftpage=##1 \global\let\lr=R
   \else \count1=2
      \shipout\vbox{\hbox to\fullhsize{\box\leftpage\hfil##1}}
      \global\let\lr=L\fi}
 
\output={\ifnum\count0=1 
 \shipout\vbox{\hbox to \fullhsize{\hfill\pagebody\hfill}}\advancepageno
 \else
 \almostshipout{\leftline{\vbox{\pagebody\makefootline}}}\advancepageno
 \fi}
 
\def\abstract##1{{\leftskip=1.5in\rightskip=1.5in ##1\par}} }
 
\def\linemessage#1{\immediate\write16{#1}}
 
\global\newcount\secno \global\secno=0
\global\newcount\appno \global\appno=0
\global\newcount\meqno \global\meqno=1
\global\newcount\subsecno \global\subsecno=0
\global\newcount\figno \global\figno=0
 
\newif\ifAnyCounterChanged
\let\terminator=\relax
\def\normalize#1{\ifx#1\terminator\let\next=\relax\else%
\if#1i\aftergroup i\else\if#1v\aftergroup v\else\if#1x\aftergroup x%
\else\if#1l\aftergroup l\else\if#1c\aftergroup c\else%
\if#1m\aftergroup m\else%
\if#1I\aftergroup I\else\if#1V\aftergroup V\else\if#1X\aftergroup X%
\else\if#1L\aftergroup L\else\if#1C\aftergroup C\else%
\if#1M\aftergroup M\else\aftergroup#1\fi\fi\fi\fi\fi\fi\fi\fi\fi\fi\fi\fi%
\let\next=\normalize\fi%
\next}
\def\makeNormal#1#2{\def\doNormalDef{\edef#1}\begingroup%
\aftergroup\doNormalDef\aftergroup{\normalize#2\terminator\aftergroup}%
\endgroup}

\def\warnIfChanged#1#2{%
\ifundef#1
\else\begingroup%
\edef\oldDefinitionOfCounter{#1}\edef\newDefinitionOfCounter{#2}%
\ifx\oldDefinitionOfCounter\newDefinitionOfCounter%
\else%
\linemessage{Warning: definition of \noexpand#1 has changed.}%
\global\AnyCounterChangedtrue\fi\endgroup\fi}
 
\def\Section#1{\global\advance\secno by1\relax\global\meqno=1%
\global\subsecno=0%
\bigbreak\bigskip
\centerline{\twelvepoint \bf %
\the\secno. #1}%
\par\nobreak\medskip\nobreak}
\def\tagsection#1{%
\warnIfChanged#1{\the\secno}%
\xdef#1{\the\secno}%
\ifWritingAuxFile\immediate\write\auxfile{\noexpand\xdef\noexpand#1{#1}}\fi%
}
\def\section{\Section}
\def\Subsection#1{\global\advance\subsecno by1\relax\medskip %
\leftline{\bf\the\secno.\the\subsecno\ #1}%
\par\nobreak\smallskip\nobreak}
\def\tagsubsection#1{%
\warnIfChanged#1{\the\secno.\the\subsecno}%
\xdef#1{\the\secno.\the\subsecno}%
\ifWritingAuxFile\immediate\write\auxfile{\noexpand\xdef\noexpand#1{#1}}\fi%
}
 
\def\subsection{\Subsection}

\def\romappno{\uppercase\expandafter{\romannumeral\appno}}
\def\makeNormalizedRomappno{%
\expandafter\makeNormal\expandafter\normalizedromappno%
\expandafter{\romannumeral\appno}%
\edef\normalizedromappno{\uppercase{\normalizedromappno}}}
\def\Appendix#1{\global\advance\appno by1\relax\global\meqno=1\global\secno=0
\bigbreak\bigskip
\centerline{\twelvepoint \bf Appendix %
\romappno. #1}%
\par\nobreak\medskip\nobreak}
\def\tagappendix#1{\makeNormalizedRomappno%
\warnIfChanged#1{\normalizedromappno}%
\xdef#1{\normalizedromappno}%
\ifWritingAuxFile\immediate\write\auxfile{\noexpand\xdef\noexpand#1{#1}}\fi%
}
\def\appendix{\Appendix}
 
\def\eqn#1{\makeNormalizedRomappno%
\ifnum\secno>0%
  \warnIfChanged#1{\the\secno.\the\meqno}%
  \eqno(\the\secno.\the\meqno)\xdef#1{\the\secno.\the\meqno}%
     \global\advance\meqno by1
\else\ifnum\appno>0%
  \warnIfChanged#1{\normalizedromappno.\the\meqno}%
  \eqno({\rm\romappno}.\the\meqno)%
      \xdef#1{\normalizedromappno.\the\meqno}%
     \global\advance\meqno by1
\else%
  \warnIfChanged#1{\the\meqno}%
  \eqno(\the\meqno)\xdef#1{\the\meqno}%
     \global\advance\meqno by1
\fi\fi%
\eqlabeL#1%
\ifWritingAuxFile\immediate\write\auxfile{\noexpand\xdef\noexpand#1{#1}}\fi%
}
\def\defeqn#1{\makeNormalizedRomappno%
\ifnum\secno>0%
  \warnIfChanged#1{\the\secno.\the\meqno}%
  \xdef#1{\the\secno.\the\meqno}%
     \global\advance\meqno by1
\else\ifnum\appno>0%
  \warnIfChanged#1{\normalizedromappno.\the\meqno}%
  \xdef#1{\normalizedromappno.\the\meqno}%
     \global\advance\meqno by1
\else%
  \warnIfChanged#1{\the\meqno}%
  \xdef#1{\the\meqno}%
     \global\advance\meqno by1
\fi\fi%
\eqlabeL#1%
\ifWritingAuxFile\immediate\write\auxfile{\noexpand\xdef\noexpand#1{#1}}\fi%
}
\def\anoneqn{\makeNormalizedRomappno%
\ifnum\secno>0
  \eqno(\the\secno.\the\meqno)%
     \global\advance\meqno by1
\else\ifnum\appno>0
  \eqno({\rm\normalizedromappno}.\the\meqno)%
     \global\advance\meqno by1
\else
  \eqno(\the\meqno)%
     \global\advance\meqno by1
\fi\fi%
}
\def\mfig#1#2{\global\advance\figno by1%
\relax#1\the\figno%
\warnIfChanged#2{\the\figno}%
\edef#2{\the\figno}%
\reflabeL#2%
\ifWritingAuxFile\immediate\write\auxfile{\noexpand\xdef\noexpand#2{#2}}\fi%
}

\def\fig#1{\mfig{fig.~}#1}
 
\catcode`@=11 

\font\ninerm=cmr9
\font\eightrm=cmr8
\font\sixrm=cmr6

\def\loadtrueseventeenpoint{
 \font\seventeenrm=cmr10 at 17.28truept
 \font\seventeeni=cmmi10 at 17.28truept
 \font\seventeenbf=cmbx10 at 17.28truept
 \font\seventeenit=cmti10 at 17.28truept
 \font\seventeensl=cmsl10 at 17.28truept
 \font\seventeensy=cmsy10 at 17.28truept
}
\def\loadfourteenpoint{
\font\fourteenrm=cmr10 at 14.4pt
\font\fourteeni=cmmi10 at 14.4pt
\font\fourteenit=cmti10 at 14.4pt
\font\fourteensl=cmsl10 at 14.4pt
\font\fourteensy=cmsy10 at 14.4pt
\font\fourteenbf=cmbx10 at 14.4pt
}
\def\loadtruetwelvepoint{
\font\twelverm=cmr10 at 12truept
\font\twelvei=cmmi10 at 12truept
\font\twelveit=cmti10 at 12truept
\font\twelvesl=cmsl10 at 12truept
\font\twelvesy=cmsy10 at 12truept
\font\twelvebf=cmbx10 at 12truept
}

\font\ninei=cmmi9
\font\eighti=cmmi8
\font\sixi=cmmi6
\skewchar\ninei='177 \skewchar\eighti='177 \skewchar\sixi='177
 
\font\ninesy=cmsy9
\font\eightsy=cmsy8
\font\sixsy=cmsy6
\skewchar\ninesy='60 \skewchar\eightsy='60 \skewchar\sixsy='60

\font\ninebf=cmbx9
\font\eightbf=cmbx8
\font\sixbf=cmbx6
 
\font\ninett=cmtt9
\font\eighttt=cmtt8
 
\hyphenchar\tentt=-1 
\hyphenchar\ninett=-1
\hyphenchar\eighttt=-1
 
\font\ninesl=cmsl9
\font\eightsl=cmsl8
 
\font\nineit=cmti9
\font\eightit=cmti8

 
\newskip\ttglue
\def\tenpoint{\def\rm{\fam0\tenrm}%
  \textfont0=\tenrm \scriptfont0=\sevenrm \scriptscriptfont0=\fiverm
  \textfont1=\teni \scriptfont1=\seveni \scriptscriptfont1=\fivei
  \textfont2=\tensy \scriptfont2=\sevensy \scriptscriptfont2=\fivesy
  \textfont3=\tenex \scriptfont3=\tenex \scriptscriptfont3=\tenex
  \def\it{\fam\itfam\tenit}\textfont\itfam=\tenit
  \def\sl{\fam\slfam\tensl}\textfont\slfam=\tensl
  \def\bf{\fam\bffam\tenbf}\textfont\bffam=\tenbf \scriptfont\bffam=\sevenbf
  \scriptscriptfont\bffam=\fivebf
  \normalbaselineskip=12pt
  \let\sc=\eightrm
  \let\big=\tenbig
  \setbox\strutbox=\hbox{\vrule height8.5pt depth3.5pt width\z@}%
  \normalbaselines\rm}
 
\def\twelvepoint{\def\rm{\fam0\twelverm}%
  \textfont0=\twelverm \scriptfont0=\ninerm \scriptscriptfont0=\sevenrm
  \textfont1=\twelvei \scriptfont1=\ninei \scriptscriptfont1=\seveni
  \textfont2=\twelvesy \scriptfont2=\ninesy \scriptscriptfont2=\sevensy
  \textfont3=\tenex \scriptfont3=\tenex \scriptscriptfont3=\tenex
  \def\it{\fam\itfam\twelveit}\textfont\itfam=\twelveit
  \def\sl{\fam\slfam\twelvesl}\textfont\slfam=\twelvesl
  \def\bf{\fam\bffam\twelvebf}\textfont\bffam=\twelvebf
  \scriptfont\bffam=\ninebf
  \scriptscriptfont\bffam=\sevenbf
  \normalbaselineskip=12pt
  \let\sc=\eightrm
  \let\big=\tenbig
  \setbox\strutbox=\hbox{\vrule height8.5pt depth3.5pt width\z@}%
  \normalbaselines\rm}
 
\def\fourteenpoint{\def\rm{\fam0\fourteenrm}%
  \textfont0=\fourteenrm \scriptfont0=\tenrm \scriptscriptfont0=\sevenrm
  \textfont1=\fourteeni \scriptfont1=\teni \scriptscriptfont1=\seveni
  \textfont2=\fourteensy \scriptfont2=\tensy \scriptscriptfont2=\sevensy
  \textfont3=\tenex \scriptfont3=\tenex \scriptscriptfont3=\tenex
  \def\it{\fam\itfam\fourteenit}\textfont\itfam=\fourteenit
  \def\sl{\fam\slfam\fourteensl}\textfont\slfam=\fourteensl
  \def\bf{\fam\bffam\fourteenbf}\textfont\bffam=\fourteenbf%
  \scriptfont\bffam=\tenbf
  \scriptscriptfont\bffam=\sevenbf
  \normalbaselineskip=17pt
  \let\sc=\elevenrm
  \let\big=\tenbig
  \setbox\strutbox=\hbox{\vrule height8.5pt depth3.5pt width\z@}%
  \normalbaselines\rm}
 
\def\seventeenpoint{\def\rm{\fam0\seventeenrm}%
  \textfont0=\seventeenrm \scriptfont0=\fourteenrm \scriptscriptfont0=\tenrm
  \textfont1=\seventeeni \scriptfont1=\fourteeni \scriptscriptfont1=\teni
  \textfont2=\seventeensy \scriptfont2=\fourteensy \scriptscriptfont2=\tensy
  \textfont3=\tenex \scriptfont3=\tenex \scriptscriptfont3=\tenex
  \def\it{\fam\itfam\seventeenit}\textfont\itfam=\seventeenit
  \def\sl{\fam\slfam\seventeensl}\textfont\slfam=\seventeensl
  \def\bf{\fam\bffam\seventeenbf}\textfont\bffam=\seventeenbf%
  \scriptfont\bffam=\fourteenbf
  \scriptscriptfont\bffam=\twelvebf
  \normalbaselineskip=21pt
  \let\sc=\fourteenrm
  \let\big=\tenbig
  \setbox\strutbox=\hbox{\vrule height 12pt depth 6pt width\z@}%
  \normalbaselines\rm}
 
\def\ninepoint{\def\rm{\fam0\ninerm}%
  \textfont0=\ninerm \scriptfont0=\sixrm \scriptscriptfont0=\fiverm
  \textfont1=\ninei \scriptfont1=\sixi \scriptscriptfont1=\fivei
  \textfont2=\ninesy \scriptfont2=\sixsy \scriptscriptfont2=\fivesy
  \textfont3=\tenex \scriptfont3=\tenex \scriptscriptfont3=\tenex
  \def\it{\fam\itfam\nineit}\textfont\itfam=\nineit
  \def\sl{\fam\slfam\ninesl}\textfont\slfam=\ninesl
  \def\bf{\fam\bffam\ninebf}\textfont\bffam=\ninebf \scriptfont\bffam=\sixbf
  \scriptscriptfont\bffam=\fivebf
  \normalbaselineskip=11pt
  \let\sc=\sevenrm
  \let\big=\ninebig
  \setbox\strutbox=\hbox{\vrule height8pt depth3pt width\z@}%
  \normalbaselines\rm}
 
\def\eightpoint{\def\rm{\fam0\eightrm}%
  \textfont0=\eightrm \scriptfont0=\sixrm \scriptscriptfont0=\fiverm%
  \textfont1=\eighti \scriptfont1=\sixi \scriptscriptfont1=\fivei%
  \textfont2=\eightsy \scriptfont2=\sixsy \scriptscriptfont2=\fivesy%
  \textfont3=\tenex \scriptfont3=\tenex \scriptscriptfont3=\tenex%
  \def\it{\fam\itfam\eightit}\textfont\itfam=\eightit%
  \def\sl{\fam\slfam\eightsl}\textfont\slfam=\eightsl%
  \def\bf{\fam\bffam\eightbf}\textfont\bffam=\eightbf \scriptfont\bffam=\sixbf%
  \scriptscriptfont\bffam=\fivebf%
  \normalbaselineskip=9pt%
  \let\sc=\sixrm%
  \let\big=\eightbig%
  \setbox\strutbox=\hbox{\vrule height7pt depth2pt width\z@}%
  \normalbaselines\rm}
 
\def\tenbig#1{{\hbox{$\left#1\vbox to8.5pt{}\right.\n@space$}}}
\def\ninebig#1{{\hbox{$\textfont0=\tenrm\textfont2=\tensy
  \left#1\vbox to7.25pt{}\right.\n@space$}}}
\def\eightbig#1{{\hbox{$\textfont0=\ninerm\textfont2=\ninesy
  \left#1\vbox to6.5pt{}\right.\n@space$}}}
 
\def\footnote#1{\edef\@sf{\spacefactor\the\spacefactor}#1\@sf
      \insert\footins\bgroup\eightpoint
      \interlinepenalty100 \let\par=\endgraf
        \leftskip=\z@skip \rightskip=\z@skip
        \splittopskip=10pt plus 1pt minus 1pt \floatingpenalty=20000
        \smallskip\item{#1}\bgroup\strut\aftergroup\@foot\let\next}
\skip\footins=12pt plus 2pt minus 4pt 
\dimen\footins=30pc 
 
\newinsert\margin
\dimen\margin=\maxdimen
\def\titlefont{\seventeenpoint}
\loadtruetwelvepoint 
\loadtrueseventeenpoint
\catcode`\@=\active
\catcode`@=12  
\catcode`\"=\active
 
\def\eatOne#1{}
\def\ifundef#1{\expandafter\ifx%
\csname\expandafter\eatOne\string#1\endcsname\relax}
\def\notTrue{\iffalse}\def\isTrue{\iftrue}
\def\ifdef#1{{\ifundef#1%
\aftergroup\notTrue\else\aftergroup\isTrue\fi}}
\def\use#1{\ifundef#1\linemessage{Warning: \string#1 is undefined.}%
{\tt \string#1}\else#1\fi}

 
\global\newcount\refno \global\refno=1
\newwrite\rfile
\newlinechar=`\^^J
\def\ref#1#2{\the\refno\nref#1{#2}}
\def\nref#1#2{\xdef#1{\the\refno}%
\ifnum\refno=1\immediate\openout\rfile=refs.tmp\fi%
\immediate\write\rfile{\noexpand\item{[\noexpand#1]\ }#2.}%
\global\advance\refno by1}
\def\lref#1#2{\the\refno\xdef#1{\the\refno}%
\ifnum\refno=1\immediate\openout\rfile=refs.tmp\fi%
\immediate\write\rfile{\noexpand\item{[\noexpand#1]\ }#2\semi}%
\global\advance\refno by1}
\def\cref#1{\immediate\write\rfile{#1\semi}}

\def\semi{;\hfil\noexpand\break}

\def\inputAuxIfPresent#1{\immediate\openin1=#1
\ifeof1\message{No file \auxfileName; I'll create one.
}\else\closein1\relax\input\auxfileName\fi%
}
\def\NPB{Nucl.\ Phys.\ B}

\def\PL{Phys.\ Lett.\ }
\def\ZPC{Z.\ Phys.\ C}
 
\newif\ifWritingAuxFile
\newwrite\auxfile
\def\SetUpAuxFile{%
\xdef\auxfileName{\jobname.aux}%
\inputAuxIfPresent{\auxfileName}%
\WritingAuxFiletrue%
\immediate\openout\auxfile=\auxfileName}

\def\L{\left(}\def\R{\right)}
\def\LP{\left.}\def\RP{\right.}
\def\LB{\left[}\def\RB{\right]}

\def\LV{\left|}\def\RV{\right|}
 
\def\bye{\par\vfill\supereject%
\ifAnyCounterChanged\linemessage{
Some counters have changed.  Re-run tex to fix them up.}\fi%
\end}


\SetUpAuxFile

\def\L{\left(}
\def\R{\right)}

\def\c{\mskip 1mu\cdot\mskip 1mu }

\def\eps{\epsilon}

\def\LP{\left.}\def\RP{\right.}

\def\l{\left}

\def\pol{\varepsilon}

\def\dl^#1_#2{\delta^{#1}{}_{#2}}

\def\F#1#2{\,{{\vphantom{F}}_{#1}F_{#2}}}
\def\Li{\mathop{\rm Li}\nolimits}

\def\Ord{{\cal O}}

\catcode`@=11  
\def\meqalign#1{\,\vcenter{\openup1\jot\m@th
   \ialign{\strut\hfil$\displaystyle{##}$ && $\displaystyle{{}##}$\hfil
             \crcr#1\crcr}}\,}
\catcode`@=12  

 
\baselineskip 15pt
\overfullrule 0.5pt

 

\def\pol{\varepsilon}

\def\c{\,\cdot\,}

\def\L{\left(}\def\R{\right)}
\def\LP{\left.}\def\RP{\right.}
\def\spa#1.#2{\left\langle#1\,#2\right\rangle}
\def\spb#1.#2{\left[#1\,#2\right]}
\def\lor#1.#2{\left(#1\,#2\right)}
\def\sand#1.#2.#3{%
\left\langle\smash{#1}{\vphantom1}^{-}\right|{#2}%
\left|\smash{#3}{\vphantom1}^{-}\right\rangle}
\def\sandp#1.#2.#3{%
\left\langle\smash{#1}{\vphantom1}^{-}\right|{#2}%
\left|\smash{#3}{\vphantom1}^{+}\right\rangle}
\def\sandpp#1.#2.#3{%
\left\langle\smash{#1}{\vphantom1}^{+}\right|{#2}%
\left|\smash{#3}{\vphantom1}^{+}\right\rangle}
\catcode`@=11  
\def\meqalign#1{\,\vcenter{\openup1\jot\m@th
   \ialign{\strut\hfil$\displaystyle{##}$ && $\displaystyle{{}##}$\hfil
             \crcr#1\crcr}}\,}
\catcode`@=12  

 
\loadfourteenpoint
\nopagenumbers\hsize=\hstitle\vskip1in
\overfullrule 0pt
\hfuzz 35 pt
\vbadness=10001
%
%

\def\half{{1\over2}}

\def\p#1{{\partial\over\partial\al{#1}}}

\def\al#1{\alpha_{#1}}

\def\ga#1{\gamma_{#1}}  

\def\Li{\mathop{\rm Li}\nolimits}
\def\l{\lambda}
\def\Det{\hat\Delta}
\def\e{\epsilon}
\def\del{\partial}
\def\hf{{\textstyle{1\over2}}}
\def\quarter{{\textstyle{1\over4}}}
\def\prodalphal{\Bigl( \prod \alpha_\ell \Bigr)}
\def\Basic{{\hat I}}
\def\Loop{{\cal I}}
\def\Int{I}\def\Poly#1{\LB #1\RB}
\def\generalPoly{P(\{a_i\})}
\def\TotalDeriv{J}
\def\rg{r_\Gamma}
\def\onemass{{\rm 1m}}
\def\twomass{{\rm 2m}}
\def\threemass{{\rm 3m}}
\def\detprime{{\rm det}^\prime}
\def\ordereps{{\cal O}(\eps)}

\def\m#1{\hat m^2_{#1}}
\def\coeff#1#2{{\textstyle{#1\over#2}}}
%
 
\noindent  hep-ph/9306240 \hfill {SLAC--PUB--5947}
\rightline{SPhT/92--048}
\rightline{UCLA--92--043}
\rightline{May, 1993}
\rightline{(T)}
 
\vskip -1.1 in
\Title{Dimensionally Regulated Pentagon Integrals%
{\tenpoint \raise4pt\hbox{${}^{\star}$}}
}
 
\vskip -.3 in
\centerline{Zvi Bern}
\baselineskip12truept
\centerline{\it Department of Physics}
\centerline{\it University of California, Los Angeles}
\centerline{\it Los Angeles, CA 90024}
\centerline{\tt bern@physics.ucla.edu}
 
\smallskip\smallskip
 
\baselineskip17truept
\centerline{Lance Dixon}
\baselineskip12truept
\centerline{\it Stanford Linear Accelerator Center}
\centerline{\it Stanford, CA 94309}
\centerline{\tt lance@slacvm.slac.stanford.edu}
 
\smallskip \centerline{and} \smallskip
 
\baselineskip17truept
\centerline{David A. Kosower}
\baselineskip12truept
\centerline{\it Theory Division}
\centerline{\it CERN}
\centerline{\it CH-1211 Geneva 23}
\centerline{\it Switzerland}
\vskip 3pt\centerline{and}\vskip 3pt
\centerline{\it Service de Physique Th\'eorique de Saclay}
\centerline{\it Centre d'Etudes de Saclay}
\centerline{\it F-91191 Gif-sur-Yvette cedex, France}
\centerline{\tt kosower@amoco.saclay.cea.fr}
 
\vskip 0.2in\baselineskip13truept
 
\centerline{\bf Abstract}
 
\ifdraftmode
\vskip 5pt
\centerline{{\bf Draft}\hskip 10pt\TimeStamp}
\vskip 5pt
\centerline{{\bf NOT for reproduction}}
\fi
 
{\narrower 
We present methods for evaluating the Feynman parameter
integrals associated with the pentagon diagram in $4-2\e$ dimensions,
along with explicit results for the integrals with all masses
vanishing or with one non-vanishing external mass.  The scalar
pentagon integral can be expressed as a linear combination of box
integrals, up to $\Ord(\e)$ corrections, a result which is the
dimensionally-regulated version of a $D=4$ result of Melrose, and of
van Neerven and Vermaseren.  We obtain and solve differential
equations for various dimensionally-regulated box integrals with
massless internal lines, which appear in one-loop $n$-point
calculations in QCD.  We give a procedure for constructing the tensor
pentagon integrals needed in gauge theory, again through
$\Ord(\eps^0)$.  }
 
\vskip 0.1in
\baselineskip17pt
 
 
\vfill
\vskip 0.1in
\noindent\hrule width 3.6in\hfil\break
\baselineskip13truept
\noindent
${}^{\star}$Research supported in part by the Department of
Energy under contract DE-AC03-76SF00515 (SLAC),
by the Texas National Research Laboratory
Commission under grant FCFY9202 (Z.B.),
and by the {\it Direction des Sciences de la Mati\`ere\/}
of the {\it Commissariat \`a l'Energie Atomique\/} of France
(Saclay).
 
\Date{}
\baselineskip17pt
 

\section{Introduction}
 
The search for new physics at current and future hadron colliders
demands that we first refine our understanding of events
originating in known physics, most importantly in QCD and QCD-associated
processes.  To date, the matrix elements for all pure QCD processes
with up to seven external legs are known exactly at
tree-level~[\ref\TreeLevel{F. Berends, W. Giele, H. Kuijf,
Phys.\ Lett.\ B232:266 (1989); Nucl.\ Phys.\ B333:120 (1990)\semi
M.\ Mangano, S. J. Parke, and Z.\ Xu, Nucl.\ Phys.\ B298:653 (1988)\semi
M.\ Mangano and S. J.\ Parke, Nucl.\ Phys.
B299:673 (1988)}]
allowing the
computation of events with up to five jets in the final state.
(Various techniques [\ref\Approximations{C. J.\ Maxwell, Phys.\ Lett.
B192:190 (1987); Nucl.\ Phys.\ B316:321 (1989)\semi
Z. Kunszt and W. J.\ Stirling, Phys.\ Rev.\ D37:2439 (1988)\semi
F. A.\ Berends and W. T.\ Giele, Nucl.\ Phys.\ B313:595 (1989)}]
allow one to
approximate cross sections with more jets.)
Because the perturbation expansion for jet physics
in QCD is not an expansion strictly in the coupling constant, but is rather
an expansion
in the coupling constant times various infrared logarithms, radiative
corrections play an important role in matching theoretical expectations
to experimental data.  The calculation of radiative corrections requires
of course the computation of loop corrections to the basic tree-level
partonic processes.  Thus far, the one-loop corrections are known only
for the most basic processes, matrix elements with four external
partons~[\ref\ES{R. K. Ellis and J. C. Sexton,
Nucl.\ Phys.\ B269:445 (1986)}].
To go beyond these basic processes in the computation of radiative
corrections in pure QCD (for example, to calculate the next-to-leading order
corrections to three-jet production at hadron colliders),
one must calculate five-point one-loop amplitudes in
a theory with massless particles; and
these in turn require the computation of one-loop Feynman parameter
integrals with five external legs, within the dimensional regularization
method.
To discuss one-loop corrections to five-point amplitudes with external
$W$ and $Z$ bosons, at least one of the external legs must be massive.
In the present paper we address the computation of
such dimensionally-regulated pentagon (and higher-point) integrals.
Recently the techniques described in this paper have been used in the
calculation of the one-loop helicity amplitudes for five external
gluons~[\ref\OurFiveGluon{Z. Bern, L. Dixon and D. A. Kosower,
Phys. Rev. Lett. 70:2677 (1993)}].
 
Various authors~[%
\ref\Melrose{D. B. Melrose, Il Nuovo Cimento 40A:181 (1965)},%
\ref\HVintegrals{G. 't Hooft and M. Veltman, \NPB{153:365 (1979)}},%
\ref\vNV{W. van Neerven and J. A. M. Vermaseren,
\PL{137B:241 (1984)}},\ref\EarlyPentagon{
G. J. van Oldenborgh and J. A. M. Vermaseren, \ZPC{46:425 (1990)}\semi
G. J. van Oldenborgh, PhD thesis, University of Amsterdam (1990)\semi
A. Aeppli, PhD thesis, University of Zurich (1992)}]
have discussed the computation
of pentagon integrals that can be carried out in dimension $D=4$
(i.e. that have neither soft nor collinear infrared divergences).
In particular, Melrose~[\Melrose] and independently
van~Neerven and Vermaseren~[\vNV] were able to express
pentagon integrals as linear combinations
of five different loop integrals with four external legs.
Such box integrals (which, with external masses but no internal masses,
are also required in radiative calculations in QCD) can be
calculated readily in dimensional regularization, by direct integration
or in terms of hypergeometric functions, if the number of masses is
not too large.
 
The techniques of Melrose and of van~Neerven and Vermaseren do not
apply directly to
dimensionally-regulated integrals, however, and the required
pentagon integrals have not yet been presented in a closed and
useful form, which is to say with all poles in $\eps = (4-D)/2$ manifest,
and with all functions of the kinematic invariants expressed in terms of
logarithms and polylogarithms.\footnote{${}^*$}{
   We have been informed that R. K. Ellis,
   W. T. Giele, and E. Yehudai~[\ref\EGY{%
R. K. Ellis, W. T. Giele, and E. Yehudai, to appear}] have recently
   evaluated the pentagon integrals by an independent technique.}
Here we will provide such an expression for the basic scalar integral.
We employ a set of equations derived in a separate paper~[\ref\OurAllN{%
Z. Bern, L. Dixon and D. A. Kosower, Phys. Lett. B302:299 (1993)}].
These equations actually apply more generally to
dimensionally-regulated one-loop $n$-point integrals; they can be used
as a starting point for the reduction of an ($n\geq5$)-point integral
to a linear combination of boxes.
For the pentagon integral ($n=5$), the equations are the
dimensionally-regulated analogs of equations derived in
references~[\Melrose,\vNV].
In this paper we will use the equations to obtain explicit expressions
for the pentagon with all lines massless,
and for the pentagon with one massive external line, up to ${\cal
O}(\e)$ corrections.
Such integrals are of use in the calculation
of next-to-leading-order contributions to processes such as
$gg\to ggg$ and $Z\to q\bar q gg$.
 
Besides the scalar pentagon integral, in QCD one requires
tensor integrals --- loop integrals with up to five powers of the
loop momentum inserted.
In the string-based technique
[\ref\SMT{
Z. Bern and D. A. Kosower, Phys.\ Rev.\ Lett.\ 66:1669 (1991)\semi
Z. Bern and D. A. Kosower, Nucl.\ Phys.\ B379:451, 1992\semi
Z. Bern and D. A. Kosower, in {\it Proceedings of the PASCOS-91
Symposium}, eds.\ P. Nath and S. Reucroft\semi
Z. Bern, D. C. Dunbar, Nucl.\ Phys.\ B379:562, 1992\semi
Z. Bern, UCLA/93/TEP/5, hep-ph/9304249},\use\OurFiveGluon]
for evaluating QCD amplitudes, one obtains
directly integrals over Feynman parameters rather than loop momenta.
Tensor integrals correspond in this framework to the insertion of
polynomials in the Feynman parameters into the numerator of the
integrand.
In order to construct an integral table that meshes
well with this technique, we choose to work in terms of the
Feynman-parametrized integrals.
This approach also lets us take advantage of an observation that
appropriate derivatives of the scalar pentagon insert
Feynman parameters into the numerator of the integrand.
Thus the scalar pentagon may
be used as a generating function for all the tensor integrals.
 
In the more usual momentum-space approach to tensor integrals,
one performs a
Brown-Feynman~[\ref\BrownFeynman{L. M. Brown and R. P. Feynman,
Phys.\ Rev.\ 85:231 (1952)}] or
Passarino-Veltman~[\ref\Passarino{G. Passarino and M. Veltman,
 \NPB{160:151 (1979)}}] reduction, solving a system of algebraic
equations for the tensor integrals.  For example, integrals with
just one loop-momentum inserted in the numerator are reduced
to a linear combination of scalar integrals~[\ref\Stuart{R. G. Stuart,
Comp.\ Phys.\ Comm.\ 48:367 (1988)\semi
R. G. Stuart and A. Gongora, Comp.\ Phys.\ Comm.\ 56:337 (1990)}].
The counterparts of these equations exist for Feynman parameter
integrals.  In particular, integrals with just one Feynman parameter
inserted in the numerator can be expressed as a linear
combination of scalar integrals.
If one now equates these expressions to the above-mentioned
derivative representations of the same one-parameter tensor integrals,
one obtains a set of first-order
partial differential equations for the scalar integral.
Thus an alternate approach to determining the scalar pentagon
is to solve a set of differential equations.
The differential equations are also an efficient way to obtain
various infrared divergent scalar box integrals, with massless internal
lines but with 0, 1, 2 or 3 external masses.
(Most of these box integrals have been obtained previously by other
techniques.)  Together with the infrared
finite box integral with four external masses~[\HVintegrals],
for which a compact form has recently been provided by
Denner, Nierste, and Scharf~[\ref\NewFourPoint{%
A. Denner, U. Nierste, and R. Scharf, \NPB{367:637 (1991)}}],
these constitute the set of box integrals required for computing
one-loop $n$-point amplitudes in QCD without quark masses, for any $n$.
(These box integrals will appear both in the recursive determination
of higher-point diagrams~[\Melrose,\vNV,\OurAllN],
and as diagrams in their own right.)
 
The partial differential equation approach just described is
reminiscent of similar procedures for performing two-loop and
higher-loop integrals (usually with fewer external
legs)~[\ref\TwoLoop{K. G. Chetykrin, A. L. Kataev and F. V. Tkachov,
  Nucl. Phys. B174:345 (1980)\semi
  F. V. Tkachov, Phys. Lett. B100:65 (1981)\semi
  K. G. Chetykrin and F. V. Tkachov, Nucl. Phys. B192:159 (1981)\semi
  D. I. Kazakov, Phys. Lett. B133:406 (1983); Theor. Math. Phys. 58:223
  (1984); Theor. Math. Phys. 62:84 (1985)\semi
  C. Ford and D. R. T. Jones, Phys. Lett. B274:409 (1992), erratum
  B285:399 (1992)}].
However, the latter manipulations have generally been carried out
in terms of either a momentum-space or a configuration-space
representation of the integrals, in contrast to the Feynman parameter
representation used here.
 
The rest of the paper is organized as follows:
in section~\use\Properties, we introduce the Feynman-parametrized
$n$-point integrals, in particular the
pentagon and box integrals, and we make a change of integration variables
and kinematic variables that
allows the tensor integrals to be expressed as derivatives of the
basic scalar integral.
In section~\use\AllNSection\ we present an alternative derivation
of the set of algebraic equations derived in ref.~[\OurAllN].
One of these equations can be used to determine the general
$n$-point scalar one-loop integral recursively, as a linear
combination of $(n-1)$-point integrals.
(For $n\geq7$ there are some subtleties, as explained in
appendix~\use\HigherPointScalarIntegrals.)
The other two equations are useful in the calculation of tensor
integrals, given the scalar integral.  Also, in combination with the
results of section~\use\Properties\ they give partial differential
equations for the scalar integral.
In section~\use\PDESection\ we begin by illustrating the general
derivation of the partial differential equation in
section~\use\AllNSection, using the simple example of
the box integral with all massless external legs.
We then solve the differential equations for box integrals
with 0, 1, 2 or 3 massive external legs.
In section~\use\AlgebraicSection\ we use one of the algebraic
equations derived in section~\use\AllNSection\ to obtain explicit
formulae for the pentagon with all massless external legs, and
with one massive external leg.
In section~\use\GeneratingFunctionSection,
we describe how to obtain the
(tensor) pentagon integrals with Feynman parameters in the numerator,
through ${\cal O}(\eps^0)$.
 
For the  reader's convenience, we  have collected our results for the
scalar box integrals  and for the scalar  and tensor massless pentagon
integrals in appendix~\use\CollectionAppendix.
In appendix~\use\VNVAppendix\ we show that when the
integrals are infrared finite, our results for the scalar pentagon
integral reduce to the non dimensionally-regulated result of
van Neerven and Vermaseren~[\use\vNV].
Appendix~\use\DFiveDrop\ presents an argument
(verifying  an  observation  of Ellis, Giele and Yehudai) which shows
that the approach of section~\use\GeneratingFunctionSection\
generates all tensor pentagon integrals needed in gauge theory
calculations.
In appendix~\use\TwoMassBoxConstantIntegral, we
compute an integration constant for two- and three-mass boxes.
In appendix~\use\AllMassTriangle, as another
illustration of the partial differential equation technique, we obtain
a manifestly symmetric expression for the  triangle integral with all
three external legs massive, to all orders in $\e$.  (To ${\cal
O}(\e^0)$, such a formula has been obtained in ref.~[\ref\HungJung{ H.
J. Lu and C. A. Perez, preprint SLAC--PUB--5809, 1992}].)
In appendix~\use\HigherPointScalarIntegrals, we discuss subtleties that
arise in obtaining scalar integrals for $n\geq7$,
and in appendix~\use\FPinHigherPointIntegrals, we derive and discuss
formul\ae\ for tensor integrals for both the pentagon and
hexagon diagrams.

 
\section{Properties of Feynman Parameter Integrals}
\tagsection\Properties
In this section, we shall show that Feynman parameter integrals
with Feynman parameters inserted in the numerator of the integrand
(which arise from tensor integrals) are given by
appropriate derivatives of the basic scalar integral.
We present the particular cases of the massless box and pentagon
integrals in more detail.
 
For convenience, we assume here that the masses for all internal lines
vanish.  (The extension to nonvanishing internal masses is entirely
straightforward~[\OurAllN].)
Then the $n$-point scalar one-loop integral in $4-2\eps$
dimensions is
$$
\Loop_n\ =\ \mu^{2\eps} \int{d^{4-2\eps}p \over (2\pi)^{4 -2\eps} } {1 \over
p^2 (p-k_1)^2 (p-k_1 - k_2)^2 \cdots (p-k_1-k_2 - \cdots - k_{n-1})^2}
\ ,
\eqn\NPointLoopIntegral
$$
where $k_i$, $i=1,\ldots,n$, are the external momenta and $\mu$ is
the usual dimensional regularization scale parameter.
Performing the usual Feynman parametrization, and integrating out
the loop momentum, we obtain
$$
\Int_n \Poly{1}\ =\ \Gamma(n-2+\e)\int_0^1 d^na_i\
   \delta (1 - {\textstyle \sum_i} a_i)
     { 1 \over
      \LB \sum_{i,j=1}^n S_{ij} a_i a_j - i\varepsilon\RB^{n-2+\eps}}\ ,
\eqn\NPointParameter
$$
where
$$
\Int_n\Poly{1}\ \equiv\ (-1)^{n+1}\, i\, \L4\pi\R^{2-\eps} \mu^{-2\eps} \,
\Loop_n
\eqn\MomentumParameter
$$
is the basic $n$-point parameter integral,
the symmetric matrix $S_{ij}$ is defined by
$$
  S_{ij}\ =\ -\hf (k_i + \cdots + k_{j-1})^2, \qquad i\neq j;
  \qquad\quad  S_{ii}\ =\ 0;
  \qquad\quad  \hbox{($i,j$ are mod $n$)};
\eqn\SDefinition
$$
and where we have put in the $i\varepsilon$ explicitly.
The poles in $\Int_n$
produced by the $\Gamma$ function prefactor are ultraviolet ones; the
remaining poles represent infrared divergences.  In explicit
calculations of cross-sections, they will ultimately cancel corresponding
poles arising from soft and collinear emission of particles in
$(n+1)$-point tree-level processes.
 
We shall use the notation $\Int_n\Poly{\generalPoly}$ to denote an integral
in which the polynomial $P$ appears in the numerator of the integrand,
$$
\Int_n\Poly{\generalPoly}\ =\ \Gamma(n-2+\eps) \int_0^1 d^na_i\
   \delta (1 - {\textstyle \sum_i} a_i)
    {\generalPoly \over
   \LB \sum_{i,j=1}^n S_{ij} a_i a_j - i\varepsilon\RB^{n-2+\eps}}\ .
\eqn\GeneralForm
$$
In QCD calculations, one encounters integrals of this form,
where the degree of $P$ is less than or equal to $n$.
 
For the box (four-point) integral, the ``scalar denominator'' is
$$
\sum_{i,j=1}^4 S_{ij} a_i a_j\ =\ - s a_1 a_3 - t a_2 a_4
 - m_1^2 a_1 a_2 - m_2^2 a_2 a_3 - m_3^2 a_3 a_4 - m_4^2 a_4 a_1\ ,
\eqn\FourDenominator
$$
where $s\equiv (k_1+k_2)^2$ and $t\equiv (k_2+k_3)^2$, and
$m_i^2$ are the masses of the external legs (some or all of which
may vanish).
For the all-massless pentagon integrals,
$$
\sum_{i,j=1}^5 S_{ij} a_i a_j\ =\
   - s_{12} a_1 a_3 - s_{23} a_2 a_4 - s_{34} a_3 a_5
   - s_{45} a_4 a_1 - s_{51} a_5 a_2\ ,
\eqn\FiveDenominator
$$
where $s_{i,i+1} \equiv (k_i+k_{i+1})^2$.
A nonzero mass for external leg 5
would add a term $-m_5^2 a_5 a_1$ to~(\use\FiveDenominator).
All external indices are understood to be taken mod~$n$ for the
$n$-point function.
We will present our results for kinematics in
the Euclidean region, where all momentum invariants
$s_{ij}$, $m_i^2$ are negative.
In this region, the scalar denominator is
always positive, and the integrals are purely real,
which simplifies the resulting expressions.
We define the integral for physical values by analytic continuation
from the Euclidean region; the analytic continuation back to the
physical region should be understood implicitly
in all formul\ae\ presented below, and we shall
henceforth leave the $i\varepsilon$ implicit.
 
Following 't~Hooft and Veltman~[\HVintegrals],
we make the change of integration variables in~(\use\NPointParameter),
$$
\eqalign{
a_i\ &=\ {\al{i} u_i\over \sum^n_{j=1} \al{j} u_j}\ ,\qquad
       {\rm\ no\ sum\ on\ }i,\cr
a_n\ &=\ {\al{n} \L 1-\sum^{n-1}_{j=1} u_j\R\over \sum^n_{j=1} \al{j} u_j}
  \ .\cr
}
\eqn\ChangeOfVar
$$
Assuming that all $\al{i}$ are real and positive, the integral becomes
$$
\Int_n\Poly{1}\ =\ \Gamma(n-2+\e)
   \int_0^1 d^nu\; {\delta\L 1 - \sum u_i \R\,
       \L \prod_{j=1}^n \al{j}\R\,
       \L\sum^n_{j=1} \al{j} u_j\R^{n-4+2\eps}
         \over
        \LB \sum_{i,j} S_{ij} \al{i} \al{j} u_i u_j\RB^{n-2+\eps}}\ .
\eqn\NPointUForm
$$
This form for the integral is most useful if we can also define the
$\alpha_i$ in such a way that all of the dependence on the
$\alpha_i$-variables is scaled out of the denominator.
Let us define the $\alpha_i$, and simultaneously a matrix $\rho$,
through
$$
  S_{ij}\ =\ {\rho_{ij}\over\alpha_i\alpha_j}\ .
\eqn\rhodefn
$$
The elements of the matrix $\rho_{ij}$ are to be thought of as additional
kinematic variables, independent of the $\alpha_i$.  (In specific cases
many of the
elements $\rho_{ij}$ may be taken to be pure numbers.)
 
For the four-point integral described by the
denominator~(\FourDenominator), we can choose
$$
  s\ =\ -{1\over\alpha_1\alpha_3}\ ,\quad
  t\ =\ -{1\over\alpha_2\alpha_4}\ ,\quad
  m_1^2 \ =\ - {\m 1 \over \alpha_1 \alpha_2},\quad
  m_2^2 \ =\ - {\m 2 \over \alpha_2 \alpha_3},\quad
  m_3^2 \ =\ - {\m 3 \over \alpha_3 \alpha_4},\quad
  m_4^2 \ =\ - {\m 4 \over \alpha_4 \alpha_1}.\quad
\eqn\GenBoxAlphas
$$
(Other choices are also possible; see section~\use\PDESection.)
Equations~(\GenBoxAlphas)
do not have a unique solution in terms of the $\alpha_i$.
One simple solution is $\al1 = \al3 =  1/\sqrt{-s}$,
$\al2 = \al4 =  1/\sqrt{-t}$.
However, we would like all four $\alpha_i$
variables to be independent of each other, so that we can use
the scalar integral as a generating function for integrals with
insertions of all four Feynman parameters $a_i$.
Therefore we consider the $\alpha_i$ to be general solutions to
equations~(\use\GenBoxAlphas), with no other constraints on them.
The set of independent kinematic variables corresponding to the
choice~(\GenBoxAlphas) is then $\{\alpha_i;\m i\}$.
 
For the all-massless pentagon integral, the unique solution to
$$
s_{i, i+1}\ =\ -{1\over \al{i} \al{i+2} }
\eqn\stoalpha
$$
is
$$\eqalign{
\al1\ &=\ \sqrt{-{s_{23} s_{34}\over s_{45}s_{51}s_{12} }}\ ,\qquad
\al2\ =\ \sqrt{-{s_{34} s_{45}\over s_{51}s_{12}s_{23} }}\ ,\qquad
\al3\ =\ \sqrt{-{s_{45} s_{51}\over s_{12}s_{23}s_{34} }}\ ,\qquad\cr
\al4\ &=\ \sqrt{-{s_{51} s_{12}\over s_{23}s_{34}s_{45} }}\ ,\qquad
\al5\ =\ \sqrt{-{s_{12} s_{23}\over s_{34}s_{45}s_{51} }}\ .\qquad
}
\eqn\PentagonAlphas
$$
Because we have taken the $s_{ij}$ to be negative, the $\al{i}$ are
real.  No additional kinematic variables are necessary for the
massless pentagon.
 
With these choices of $\al{i}$, the four and five point scalar integrals
become
$$
\eqalign{
\Int_4\Poly{1}\ &=\ \Gamma(2+\e) \biggl( \prod_{j=1}^4 \al{j} \biggr)
     \int_0^1 d^4u\; {\delta\L 1 - \sum u_i \R\,
                 \L\sum^4_{j=1} \al{j} u_j\R^{2\eps}
   \over \LB u_1u_3 + u_2u_4 + \m 1 u_1u_2 + \m 2 u_2u_3
                             + \m 3 u_3u_4 + \m 4 u_4u_1
         \RB^{2+\eps}} \ ,\cr
\Int_5\Poly{1}\ &=\ \Gamma(3+\e) \biggl( \prod_{j=1}^5 \al{j} \biggr)
      \int_0^1 d^5u\; {\delta\L 1 - \sum u_i \R\,
                 \L\sum^5_{j=1} \al{j} u_j\R^{1+2\eps}
   \over \LB u_1 u_3 + u_2 u_4 + u_3 u_5 + u_4 u_1 + u_5 u_2\RB^{3+\eps}}
\ .\cr}
\eqn\URepresentation
$$
 
Further examples of the $\{\alpha_i;\rho_{ij}\}$ change of variables
are to be found in sections~\use\PDESection,\use\AlgebraicSection\
and appendix~\use\AllMassTriangle.
 
It will be helpful to define the {\it reduced integrals}
$$
\Basic_n\Poly{\generalPoly}\ =\ \biggl(\prod_{j=1}^n \al{j}\biggr)^{-1}
   \Int_n\Poly{P(\{a_i/\al{i}\})}\ .
\eqn\ReducedIntegral
$$
As we will see,
dividing out the factors of $\al{i}$ connects the tensor integrals
more simply to the scalar integral.
For the scalar integrals (that is when the polynomial is simply $1$),
we will use the abbreviated notation $\Basic_n \equiv \Basic_n\Poly{1}$.
 
We can use the 't~Hooft-Veltman form of the tensor
integrals~(\use\ReducedIntegral) to obtain derivative relations for
them.
Let $P_m(\{a_i\})$ denote a homogeneous polynomial of degree $m$.
Then for the massless box integral, the change of
variables~(\use\ChangeOfVar)
gives
$$
  \Basic_4 \Poly{P_m(\{a_i\})}\ =\ \Gamma(2+\e)
  \int_0^1 d^4 u \; \delta (1 - {\textstyle \sum_i} u_i)\
  {P_m(\{u_i\}) \L \sum_{j=1}^4 \al{j} u_j \R^{-m+2\eps} \over
   [u_1 u_3 + u_2 u_4]^{2 +\eps} }\ ,
\anoneqn
$$
which we can rewrite in terms of derivatives acting on $\Basic_4$,
$$
  \Basic_4 \Poly{P_m(\{a_i\})}\ =\
   {\Gamma(1-m+2\eps) \over \Gamma(1+2\eps)}\
    P_m \L \left\{ \p{i} \right\} \R
    \, \Basic_4\Poly{1}\ .
\eqn\BasicBox
$$
Similarly, the change of variables~(\use\ChangeOfVar) leads in the
five-point case to
$$
\Basic_5\Poly{P_m(\{a_i\})}\ =\ \Gamma(3+\e)
    \int_0^1 d^5u\; \delta (1 - {\textstyle \sum_i} u_i)\
    {P_m(\{u_i\})\, \L\sum^5_{j=1} \alpha_j u_j\R^{1-m+2\eps}
 \over \LB u_1 u_3 + u_2 u_4 + u_3 u_5 + u_4 u_1 + u_5 u_2\RB^{3+\eps}}
 \ ,
\anoneqn
$$
which we can write as follows,
$$
\Basic_5\Poly{P_m(\{a_i\})}\ =\ {\Gamma(2-m+2\eps)\over\Gamma(2+2\eps)}
         \, P_m\L\left\{ \p{i} \right\}\R \,\Basic_5\Poly{1}\ .
\eqn\BasicPentagon
$$
These equations hold when there
are external masses as well, provided that one holds fixed the matrix
$\rho$ defined in~(\use\rhodefn) when differentiating with respect to
$\alpha_i$.  The result for the general $n$-point integral is
$$
\Basic_n\Poly{P_m(\{a_i\})}\ =\
   {\Gamma(n-3-m+2\eps)\over\Gamma(n-3+2\eps)}
         \, P_m\L\left\{ \p{i} \right\}\R \,\Basic_n\Poly{1}\ .
\eqn\BasicNPoint
$$
Equation~(\use\BasicNPoint) allows one to obtain tensor integrals
by differentating the basic scalar integral.  Certain
subtleties do arise in this approach; they will be dealt with in
section~\use\GeneratingFunctionSection\ and in
appendix~\FPinHigherPointIntegrals.
 
Using equations such as~(\use\BasicBox), (\use\BasicPentagon),
and~(\use\BasicNPoint), one can translate an algebraic
system of equations for integrals with Feynman parameters
inserted, into a system of partial
differential equations for the basic scalar integral;
in principle one can then solve the equations for the latter quantity.
This effectively turns a problem of definite integration into one
of indefinite integration (in a different set of variables).  We shall
use this approach to give concrete expressions for all the box integrals.
It is also possible, as we shall see in the next section, to derive
a purely algebraic set of equations for the $n$-point integrals
$\Basic_n$, in which a new unknown quantity enters only at
${\cal O}(\eps)$.
 
 
\section{Algebraic Equations for $n$-Point One-Loop Integrals}
\tagsection\AllNSection
 
In this section we will derive a set of algebraic equations for
the general $n$-point one-loop integrals.  Some of the equations
are of use in the partial differential equation approach of
section~\use\PDESection;
others can be used to determine the $n$-point scalar integrals for $n\geq5$
in terms of box integrals, in an entirely algebraic fashion
(subject to some subtleties for $n\geq7$, which are explained in
appendix~\use\HigherPointScalarIntegrals).
The equations have been derived in ref.~[\use\OurAllN] using a
momentum-space representation of the loop integrals.  Here
we will derive the same general equations using the Feynman parameter
representation; in this derivation the equations arise from
the consideration of integrals of total
derivatives of the Feynman parameters.\footnote{${}^*$}{
The motivation for considering such objects arose from the observation
that the field-theory limit of integrals of total derivatives
in string theory yields expressions
that are sums of loop integrals with differing numbers of external legs
(multiplied by various coefficients); these sums must necessarily vanish
because the world-sheets in the string loop expansion have no boundaries,
when appropriate analytic continuations of the external momenta
are used.}
For a specific, simple example of the following general derivation,
we refer the reader to the beginning of section~\use\PDESection.
 
The total derivatives we will consider are
$$
\eqalign{
J_{n;m}\ &\equiv\ \Gamma(n-3+\e)
\int_0^1 da_{n-1} \int_0^{1-a_{n-1}} da_{n-2} \,\cdots\,
\int_0^{1-a_1-a_2-\cdots-\widehat{a_m}-\cdots- a_{n-1}} da_m \cr
  &\qquad\qquad \times {d\over d a_m}
{1 \over \LB \sum_{i,j=1}^n S_{ij} a_i a_j\RB^{n-3+\eps} }
    \Biggr\vert_{ a_n = 1 - a_1 - a_2 - \cdots - a_{n-1} }
  \ . \cr}
\eqn\Jni
$$
There are two ways to evaluate $J_{n;m}$.  First, one can carry
out the differentiation with respect to $a_m$, to get
$$
\eqalign{
  J_{n;m}\ &=\ -2 \, \Gamma(n-2+\e) \int d^na_i\
  \delta (1 - {\textstyle \sum_i} a_i)
   { \sum_{j=1}^n (S_{mj}a_j - S_{nj}a_j)
     \over \LB \sum_{i,j=1}^n S_{ij} a_i a_j\RB^{n-2+\eps} } \cr
  \ &=\ -2 \L \prod_{\ell=1}^n \alpha_\ell\R
    \sum_{j=1}^n
     \Basic_{n}\bigl[ (S_{mj} - S_{nj}) \, \alpha_j \, a_j \bigr]
  \ . \cr}
\eqn\Jnione
$$
Second, one can perform the integral over $a_m$.
At the lower integration endpoint, $a_m$ is set to 0. The remaining
$(n-1)$-point integral corresponds to removing
the propagator parametrized by $a_m$ ---
i.e., the propagator between lines $(m-1)$ and $m$ --- from the
original $n$-point (scalar) integral;
we denote such a ``daughter'' integral
of $I_n$ ($\Basic_n$) by $I_{n-1}^{(m)}$ ($\Basic_{n-1}^{(m)}$).%
\footnote{${}^\dagger$}{For more explicit examples of this notation,
see the beginning of subsection~\use\SingleMassBoxSubsection.}
Similarly, at the upper integration endpoint $a_n$ is set to 0,
yielding the $(n-1)$-point integral $I_{n-1}^{(n)}$.
It is always possible to choose the $\alpha_i$ variables for the
integrals $\Basic_{n-1}^{(j)}$ so that they are the same as those for the
parent integral $\Basic_n$.  Having made this choice, the second
evaluation of $J_{n;m}$ gives
$$
 J_{n;m}\ =\ I_{n-1}^{(n)}[1] - I_{n-1}^{(m)}[1]
  \ =\ \L \prod_{\ell=1}^n \alpha_\ell\R
  \LB {\Basic_{n-1}^{(n)}\over\al{n}} - {\Basic_{n-1}^{(m)}\over\al{m}}
  \RB\ .
\eqn\Jnitwo
$$
Equating~(\use\Jnione) and~(\use\Jnitwo), using
$S_{ij} = \rho_{ij}/(\alpha_i\alpha_j)$ and the
definitions~(\use\ReducedIntegral) of the reduced integrals, and
relabelling the index $m\to i$, we have
$$
  \sum_{j=1}^n
  \L {\rho_{ij}\over\alpha_i} - {\rho_{nj}\over\alpha_n} \R
  \Basic_n[a_j]\ =\
  {1\over2} \LB {\Basic_{n-1}^{(i)}\over\al{i}}
              - {\Basic_{n-1}^{(n)}\over\al{n}} \RB\ ,
\qquad\qquad i=1,2,\ldots,n-1.
\eqn\allnmost
$$
We would like to solve for the $n$ one-parameter integrals
$\Basic_n[a_j]$.  To do so we supplement the $n-1$
equations~(\use\allnmost) with the equation that follows from
the constraint on the Feynman parameters, $\sum_{j=1} a_j = 1$,
namely
$$
  \sum_{j=1}^n \alpha_j \, \Basic_n[a_j]\ =\ \Basic_n[1]
   \ =\ \Basic_n.
\eqn\allnlast
$$
 
Before solving equations~(\use\allnmost), (\use\allnlast),
we introduce a little more notation and some ``kinematic''
results from ref.~[\use\OurAllN].
We define the Gram determinant of the $(n-1)$-vector
system associated with the $n$-point integral by
$$
  \Delta_n\ \equiv\ \detprime(2k_i\c k_j),
\eqn\unreducednGram
$$
where the prime signifies that one of the $n$ vectors $k_i$ is to be
omitted before taking the determinant; due to momentum conservation,
$\sum k_i = 0$, any one of the vectors may be
omitted.\footnote{${}^\ddagger$}{
  The notation for, and normalization of, the Gram determinant in
  equation~(\use\unreducednGram) differ from other conventions in
  the literature, e.g. references~[\Melrose,\ref\ByK{E. Byckling and K.
Kajantie, {\tenit Particle Kinematics} (Wiley) (1973)}].}
Next we introduce the {\it rescaled} Gram determinant,
$$
  \Det_n\ \equiv\
  \Bigl( \prod_{\ell=1}^n \alpha_\ell^2 \Bigr) \Delta_n,
\eqn\nGramdefn
$$
which has a simple bilinear representation in
terms of the variables $\alpha_i$:
$$
  \Det_n\ =\ \sum_{i,j=1}^n \eta_{ij}\alpha_i\alpha_j.
\eqn\etadefn
$$
Here $\eta_{ij}$ is independent of the $\alpha_i$; in fact
$\eta$ is proportional~[\use\OurAllN] to the inverse of the
matrix $\rho$ defined in equation~(\use\rhodefn):
$$
  \rho\ =\ N_n\ \eta^{-1},\qquad
  \eta\ =\ N_n\ \rho^{-1},\qquad\qquad
  N_n\ \equiv\ 2^{n-1}\det\rho.
\eqn\Nnetarhoeqn
$$
We also define the variables $\gamma_i$ by
$$
  \gamma_i\ \equiv\ \sum_{j=1}^n \eta_{ij} \alpha_j
   \ =\ {1\over2}{\del\Det_n\over\del\alpha_i}
  \biggr\vert_{\rho_{ij}\ {\rm fixed}}\ .
\eqn\gammadefn
$$
They are in a sense conjugate to the $\al{i}$ variables:
$$
\sum_{j=1}^n \rho_{ij} \gamma_j\ =\ N_n\al{i} \, .
\eqn\gammaconj$$
 
If we define
$$
R_{ki}\ =\ \eta_{ki} - {\gamma_k\gamma_i\over\Det_n}\ ,
\eqn\Rdef
$$
then we may note the following identity,
$$
\sum_{i=1}^n \al{i} R_{ki}\ =\ 0\ ,
\eqn\Rident$$
which follows from equations~(\use\etadefn) and~(\use\gammadefn).
Thus if we multiply both sides of equation~(\allnmost) by $\al{i} R_{ki}$,
and then sum over $i$, the terms not containing $\al{i}$ will drop out,
leaving us with
$$
\sum_{i,j=1}^n
  \L \eta_{ki}\rho_{ij}- {\gamma_k\over\Det_n} \gamma_i\rho_{ij}\R\,
     \Basic_n[a_j]\ =\ {1\over2}
\sum_{i=1}^n R_{ki} \Basic_{n-1}^{(i)}\ ,
\anoneqn$$
or
$$
N_n \Basic_n[a_k]\ =\ {1\over2}
\sum_{i=1}^n R_{ki} \Basic_{n-1}^{(i)}
+ {\gamma_k\over\Det_n}N_n\sum_{j=1}^n \al{j}\Basic_n[a_j]\ .
\anoneqn$$
 
Performing the sum on the right-hand side with the help
of~(\use\allnlast), dividing by $N_n$, and
writing out the definition of $R_{ki}$, we obtain
$$
 \Basic_n[a_i]
  \ =\ {1\over 2 N_n} \sum_{j=1}^n
  \Bigl( \eta_{ij} - {\gamma_i\gamma_j\over\Det_n} \Bigr)
   \Basic_{n-1}^{(j)} \ +\ {\gamma_i\over\Det_n} \Basic_n\ .
\eqn\reducedInaeqn
$$
Combining this set of equations with the derivative
representation~(\use\BasicNPoint) for $m=1$, we obtain a
system of partial differential equations for the $n$-point scalar
integral,
$$
 {1\over n-4+2\e}{\del\Basic_n\over\del\alpha_i}
  \ =\ {1\over 2 N_n} \sum_{j=1}^n
  \Bigl( \eta_{ij} - {\gamma_i\gamma_j\over\Det_n} \Bigr)
   \Basic_{n-1}^{(j)} \ +\ {\gamma_i\over\Det_n} \Basic_n\ .
\eqn\NPointPDEs
$$
Section~\use\PDESection\ is devoted to solving these equations for
various scalar box integrals.
 
In ref.~[\OurAllN] a momentum-space representation was used to
derive an algebraic equation that
involved only scalar integrals, at the expense of introducing a
new object, $\Basic_n^{D=6-2\e}$.  The object $\Basic_n^{D=6-2\e}$
comes from an integral in $D=4-2\e$ with two loop-momenta inserted
in the numerator, but it can also be interpreted as the $n$-point
scalar integral in two higher dimensions.
The latter interpretation is helpful for understanding the properties of
$\Basic_n^{D=6-2\e}$ as $\e\to0$, which are needed in order to
use the ``dimension-changing'' equation to obtain
$D=4-2\e$ scalar integrals through $\Ord(\e^0)$.
We shall now re-derive this equation using
Feynman parameter representations of the integrals.
 
The integral $I_n^{D=6-2\e}[1]$ is most
easily obtained from the $D=4-2\e$ equation~(\use\NPointParameter)
by letting $\e \to \e - 1$,
$$
I_n^{D=6-2\e}[1]\ =\ \Gamma(n-3+\e)\int_0^1 d^na_i\
   \delta (1 - {\textstyle \sum_i} a_i)
     { 1 \over \LB \sum_{i,j=1}^n S_{ij} a_i a_j \RB^{n-3+\eps}}\ ,
\eqn\NPointDeqsix
$$
It may also be obtained by inserting one power of the scalar denominator
of the $D=4-2\e$ integral into the numerator (summations are implicit
in the following derivation):
$$
 I_n^{D=6-2\e}[1]
 \ =\ {\Gamma(n-3+\e)\over\Gamma(n-2+\e)}
 I_n^{D=4-2\e}\LB S_{ij} a_ia_j \RB
 \ =\ {1\over n-3+\e}\ \rho_{ij}\ I_n^{D=4-2\e}
   \LB (a_i/\alpha_i)(a_j/\alpha_j) \RB\ .
\eqn\Dchangeone
$$
In terms of the reduced integrals~(\use\ReducedIntegral),
and using the derivative representation~(\use\BasicNPoint), we have
$$
  \Basic_n^{D=6-2\e}\ =\
  {1\over (n-3+\e)(n-4+2\e)(n-5+2\e)}
  \ \rho_{ij} {\del^2 \Basic_n \over \del\alpha_i\del\alpha_j}\ .
\eqn\Dchangetwo
$$
 
We may now evaluate the right-hand-side of~(\use\Dchangetwo) using
equation~(\use\NPointPDEs) to replace the derivatives, and also using
the relations between $\rho$, $\eta$ and $\gamma$ to simplify the
expressions:
$$
\eqalign{
  {1\over n-4+2\eps}&\rho_{ij}
   {\del^2 \Basic_n \over \del\alpha_i\del\alpha_j}
  \ =\
\rho_{ij}\p{i}\left(
  {1\over 2N_n}\LB \eta_{jk}-{\ga{j}\ga{k}\over\Det_n}\RB
  \,\Basic_{n-1}^{(k)} + {\ga{j}\over\Det_n}\Basic_n \right) \cr
&=\ {1\over 2 N_n}\rho_{ij}\LB 2 {\ga{i}\ga{j}\ga{k}\over\Det_n^2}
   - {\L\eta_{ij}\ga{k} + \eta_{ik}\ga{j}\R\over\Det_n}\RB
      \,\Basic_{n-1}^{(k)}
  + {1\over2 N_n} \LB\eta_{jk}-{\ga{j}\ga{k}\over\Det_n}\RB \rho_{ij}
   {\del\Basic_{n-1}^{(k)} \over \del\alpha_i} \cr
&\qquad + {\rho_{ij}\over\Det_n} \LB \eta_{ij}
                   - 2{\ga{i}\ga{j}\over\Det_n}\RB\, \Basic_n
   + (n-4+2\e){\rho_{ij}\ga{j}\over\Det_n}
    \L {1\over2 N_n}\LB\eta_{ik}-{\ga{i}\ga{k}\over\Det_n}\RB
            \,\Basic_{n-1}^{(k)} + {\ga{i}\over\Det_n}\Basic_n\R \cr
&=\
  - {n-1\over 2} {\ga{k}\over\Det_n} \, \Basic_{n-1}^{(k)}
  + {1\over2}\LB \p{k} - {\ga{k}\over\Det_n}\alpha_i\p{i} \RB
    \Basic_{n-1}^{(k)}
  + \Bigl( (n-2) + (n-4+2\e) \Bigr)
          {N_n\over\Det_n} \, \Basic_n\ . \cr
}
$$
Now $\Basic_{n-1}^{(k)}$ is actually independent of $\alpha_k$
(since $a_k$ has been set to 0 in $\Basic_{n-1}^{(k)}$); also
$$
\al{i}\p{i}\Basic_{n-1}^{(k)}\ =\ (n-5+2\eps)\ \Basic_{n-1}^{(k)} \, .
\anoneqn
$$
So we obtain
$$
{1\over n-4+2\eps}\rho_{ij}
   {\del^2 \Basic_n \over \del\alpha_i\del\alpha_j}
  \ =\
(n-3+\e)\left[ - \sum_{k=1}^n {\gamma_k\over \Det_n} \Basic_{n-1}^{(k)}
  + {2N_n\over\Det_n} \Basic_n \right]\ ,
\anoneqn
$$
which can be solved for $\Basic_n$ using equation~(\use\Dchangetwo),
$$
  \Basic_n\ =\ {1\over 2N_n} \Biggl[ \sum_{i=1}^n \gamma_i \,
   \Basic_{n-1}^{(i)}\ +\ (n-5+2\e) \,\Det_n \,
   \Basic_n^{D=6-2\e} \Biggr]\ .
\eqn\reducedallNeqn
$$
In ref.~[\OurAllN] it is shown how to use this equation
to obtain $n$-point integrals with $n\geq6$.
However, for $n\geq7$ there are some complications, which are discussed
in appendix~\use\HigherPointScalarIntegrals.
In this paper our main interest is the pentagon integral ($n=5$).
For the scalar pentagon integral
it suffices to note that the integral $\Basic_5^{D=6-2\e}$ is finite
as $\e\to0$, because the $D=6$ scalar pentagon integral possesses
neither ultraviolet nor infrared divergences (soft or collinear),
and also that the coefficient of $\Basic_5^{D=6-2\e}$ in
equation~(\use\reducedallNeqn) is of order $\e$.
Therefore to ${\cal O}(\e^0)$ the general scalar
pentagon integral is given by the sum of five scalar box integrals,
$$
  \Basic_5\ =\ {1\over 2N_5}\sum_{i=1}^5 \gamma_i \, \Basic_4^{(i)}
  \ +\ {\cal O}(\e).
\eqn\reducedNeqfiveeqn
$$
A schematic depiction of this equation, with the coefficients
suppressed, is given in \fig\vNVEquationFigure.
For the tensor pentagon integrals we have to keep the
$\Basic_5^{D=6-2\e}$ term around a while longer (see
section~\use\GeneratingFunctionSection).
 
One further equation for all $n$ can be obtained by eliminating
$\Basic_n$ from equation~(\use\reducedInaeqn) using
equation~(\use\reducedallNeqn), with the result
$$
  {1\over n-4+2\e}\ {\del\Basic_n\over\del\alpha_i}\ =\ \Basic_n[a_i]
  \ =\ {1\over 2 N_n} \Biggl[ \sum_{j=1}^n
   \eta_{ij}\ \Basic_{n-1}^{(j)}
   \ +\ (n-5+2\e)\,\gamma_i\ \Basic_n^{D=6-2\e} \Biggr]\ .
\eqn\reducedInamixedeqn
$$
Since the $D=6$ scalar box is also finite, setting $n=4$ in
equation~(\use\reducedInamixedeqn) yields a simple set of partial
differential equations for the box integrals, through ${\cal O}(\e)$:
$$
  {\del\Basic_4\over\del\alpha_i}
  \ =\ {\e \over N_4} \Biggl[
  \sum_{j=1}^4 \eta_{ij}\ \Basic_3^{(j)}
  \ +\ (-1+2\e)\,\gamma_i\ \Basic_4^{D=6-2\e} \Biggr]
  \ =\ {\e\over N_4} \sum_{j=1}^4 \eta_{ij}\ \Basic_3^{(j)}
   \ +\ {\cal O}(\e)\ .
\eqn\boxPDEs
$$
The right-hand-side depends only on the infrared singular pieces of
the triangle integrals.
 
This completes our re-derivation of general all-$n$ results
presented in ref.~[\OurAllN];
we now apply these results to various box and pentagon integrals.

 
\section{Partial Differential Equation Technique}
\tagsection\PDESection
 
In this section, we solve the partial differential
equations~(\use\NPointPDEs), (\use\boxPDEs) for scalar box integrals
with all internal lines massless, but with 0, 1, 2 or 3 massive external
lines.
 
\subsection{The Massless Box Integral}
 
We begin with the box integral with all external lines massless,
$$
\Int_4 \Poly{1}\ =\ \Gamma(2+\e)\int_0^1 d^4a_i\
\delta (1 - {\textstyle \sum_i} a_i)  { 1 \over
\LB -sa_1a_3 - ta_2a_4 \RB^{2+\eps}}\ .
\eqn\NewMasslessBoxDef
$$
This integral is simple enough to perform directly
after the following change of variables [\ref\Karplus{R. Karplus and M.
Neuman, Phys.\ Rev.\ 83:776 (1951)}]
which factorizes the integrand\footnote{${}^\dagger$}{
  J. Vermaseren has pointed out to us that the factorization of
  the integrand in terms of $x,y,z$ arises naturally if one
  combines pairs of propagators using Feynman parameters, and then
  combines the two resulting factors using another Feynman parameter.
  See also ref.~[\Karplus].}
$$
  a_1\ =\ y(1-x),\qquad a_2\ =\ z(1-y),\qquad a_3\ =\ (1-y)(1-z),\qquad
  a_4\ =\ xy.
\eqn\Boxxyz
$$
However, our purpose here is to illustrate the partial differential
equation technique, including the derivation of the equations,
via this simple example.
 
As noted above, algebraic equations for Feynman parameter integrals
can be obtained by considering integrals of total derivatives.
Here we consider the box integral $I_4[1]$,
with the parameter $a_4$ eliminated,
and with the integrand differentiated with respect to $a_1$:
$$
\TotalDeriv_{4;1}\ \equiv\ \Gamma(1+\e)
\int_0^1 da_3 \int_0^{1-a_3} da_2
\int_0^{1-a_2- a_3} da_1  \; {\del\over \del a_1}
{1 \over \LB -s a_1 a_3 - t a_2 (1-a_1-a_2-a_3)\RB^{1+\eps} }\ .
\eqn\FourDerivative
$$
Observe that $\TotalDeriv_{4;1}$ can be evaluated in two ways,
either by explicit differentiation,
or by evaluating the integrand at the boundaries
$a_4=1-a_1-a_2-a_3=0$ and $a_1=0$.
The boundary terms yield
$$
  \Gamma(1+\e) \int_0^1 da_1 da_2 da_3
  {\delta (1 - {\textstyle \sum_{i=1}^3} a_i)
        \over [-s a_1 a_3]^{1+\eps}} \ -\
  \Gamma(1+\e) \int_0^1 da_2 da_3 da_4
  {\delta (1 - {\textstyle \sum_{i=2}^4} a_i)
        \over [-t a_2 a_4]^{1+\eps}}\ ,
\eqn\BoxBoundaryTerm
$$
which is the difference of two triangle integrals,
each with one massive external leg,
as depicted in \fig\OffShellTriangleFigure.  These integrals
are easily evaluated,
$$
  \Int_3^\onemass(s)\ \equiv\ \Gamma(1+\e) \int_0^1 d^3a_i
    {\delta ( 1 - {\textstyle \sum_{i=1}^3} a_i)
        \over [-s a_1 a_3]^{1+\eps}}
  \ =\ {\rg\over\eps^2}  (-s)^{-1-\eps}\ ,
\eqn\OneMassTriangle
$$
where
$$
  \rg\ \equiv\ {\Gamma(1+\eps)\Gamma^2(1-\eps)\over\Gamma(1-2\eps)}
$$
is a ubiquitous prefactor.
Thus
$$
  \TotalDeriv_{4;1}\ =\ {\rg\over\eps^2}
  \Bigl( (-s)^{-1-\eps} - (-t)^{-1-\eps} \Bigr)\ .
\eqn\MasslessBoxRHS
$$
The other way of evaluating equation~(\use\FourDerivative), explicit
differentiation, yields
$$
\eqalign{
  \TotalDeriv_{4;1}\ &=\
  -\Gamma(2+\eps) \int_0^1 d^4a \; \delta(1- {\textstyle \sum_i^4} a_i)
  {-s a_3  + t a_2 \over  \LB-s a_1 a_3 - t a_2 a_4\RB^{2+\eps} }\cr
   &=\ I_4[s a_3 - t a_2]
  \ =\ {1\over2\e}\L \prod_{i=1}^4 \al{i} \R
     \LB -{1\over\al{1}}\p{3} + {1\over\al{4}}\p{2} \RB
     \Basic_4\ ,\cr}
\eqn\DiffEval
$$
where we have used equations~(\use\GenBoxAlphas),
(\use\ReducedIntegral) and (\use\BasicBox) in the last step.
 
Equations~(\use\MasslessBoxRHS) and (\use\DiffEval) together constitute
one differential equation for $\Basic_4$.
In fact, due to the symmetries of the original integral,
total derivatives in other Feynman parameters do not yield independent
equations.
Instead, we recognize at this stage that $\Basic_4$ is really a
function of $s$ and $t$ alone, not of all four $\alpha_i$,
$$
  \Basic_4\ =\ \Basic_4(s,t)
  \ =\ \Basic_4\bigl(-(\al{1}\al{3})^{-1},-(\al{2}\al{4})^{-1}\bigr),
\anoneqn
$$
so that
$$
  {1\over\al{1}} {\del\Basic_4\over\del\al{3}}
  \ =\ s^2 {\del\Basic_4\over\del s}\ , \qquad
  {1\over\al{4}} {\del\Basic_4\over\del\al{2}}
  \ =\ t^2 {\del\Basic_4\over\del t}\ .
\eqn\stDeriv
$$
Combining equations~(\use\MasslessBoxRHS), (\use\DiffEval) and
(\use\stDeriv), we see that $\Basic_4(s,t)$ satisfies
the partial differential equation
$$
  s^2 {\del\Basic_4\over\del s}\ -\ t^2 {\del\Basic_4\over\del t}
\ =\ -{2\rg\over\e} st \LB (-s)^{-1-\e}\ -\ (-t)^{-1-\e} \RB\ .
\eqn\PDEFourst
$$
We still need one additional equation, which comes from the
fact that the dimension of $\Basic_4$ is equal to
$-\eps\times{\rm dimension}(s,t)$,
so that
$$
  s {\del\Basic_4\over\del s}\ +\ t {\del\Basic_4\over\del t}
\ =\ -\e \, \Basic_4\ .
\eqn\PDEFourDimension
$$
Equations~(\use\PDEFourst) and~(\use\PDEFourDimension)
form a complete set of partial differential equations.
 
\def\BasicH{\Basic^0}
If we consider instead of $\Basic_4$ the dimensionless quantity
$\BasicH_4$, defined by
$$
\BasicH_4(s,t) \equiv \L -{s+t\over st} \R^{-\eps}
\Basic_4(s,t)\ ,
\anoneqn
$$
we see that it is a function only of the ratio $\chi \equiv t/s$,
and that
$$
  s^2 {\del\BasicH_4\over\del s}\ -\ t^2 {\del\BasicH_4\over\del t}
    \ =\ -t (1+\chi) {d\BasicH_4\over d\chi}\ .
\anoneqn
$$
In terms of $\chi$, the first equation~(\use\PDEFourst) becomes
$$
{d\BasicH_4\over d\chi}\ =\ -{2\rg\over\eps}\,
    {(\chi^{\eps} - \chi^{-1})\over \L 1+\chi\R^{1+\eps}}\ .
\eqn\PDEFourFinal
$$
 
One can solve this differential equation to all orders in $\eps$
as follows.
We observe that the transformation $\chi\rightarrow \chi^{-1}$
interchanges the two terms on the right-hand side.  Taking the
second term, shifting $\chi\rightarrow \chi-1$, using the
hypergeometric function formul\ae
$$
\int dz\; z^c\ \F{p}{q}\L\{a_i\};\{b_i\};z\R
 \ =\ {z^{c+1}\over c+1}
 \ \F{p+1}{q+1}\L\{a_i\},c+1;\{b_i\},c+2;z\R
\anoneqn
$$
and
$$
\F{1}{0}(\xi ; z)\ =\ (1 - z)^{-\xi}\;,
\anoneqn
$$
the hypergeometric function identity
$$
\F21(1, -\eps; 1-\eps; 1+\chi)\ =\
(-\chi)^\eps \F21(-\eps, -\eps; 1-\eps; 1+\chi^{-1})\ ,
\anoneqn
$$
and using the interchange of $\chi$ and $\chi^{-1}$ to furnish the first
term, we obtain (note that $\chi$ should be thought of as having a
small imaginary part in order to avoid difficulties with branch cuts)
$$\eqalign{
\BasicH_4 &= {2\rg\over\eps^2} \LB \L1+\chi^{-1}\R^{-\eps}
         \F21\L 1,-\eps;1-\eps;1+\chi^{-1}\R
         +(1+\chi)^{-\eps} \F21\L 1,-\eps;1-\eps;1+\chi\R\RB\cr
   &=
   {2\rg\over\eps^2}\,\L 1+\chi^{-1}\R^{-\eps}
   \LB (-\chi^{-1})^{\eps} \F21\L -\eps,-\eps;1-\eps;1+\chi\R
   +(-\chi)^{\eps}\chi^{-\eps}\F21\L -\eps,-\eps;1-\eps;1+\chi^{-1}\R\RB
   \ .\cr
}\eqn\BasicHFullSolution
$$
The constant of integration may be determined by evaluating the
integral~(\use\NewMasslessBoxDef) directly at some convenient value
of $\chi$, say $\chi = t/s = 1$ ($s=t=-1$),
$$
\eqalign{
 \Basic_4^0(\chi=1)\ =\ 2^{-\e} \, I_4[1]
 \ &=\ 2^{-\e} \Gamma(2+\e) \int_0^1 d^4a_i\
  { \delta(1-\sum a_i) \over [a_1a_3+a_2a_4]^{2+\e} } \cr
   &=\ 2^{-\e} \Gamma(2+\e) \int_0^1 dx \int_0^1 dy \int_0^1 dz\
   y^{-1-\e} (1-y)^{-1-\e} \bigl[ (1-x)(1-z) + xz \bigr]^{-2-\e}\ ,
   \cr}
\eqn\MasslessBoxConstInt
$$
where we have made the change of variables~(\use\Boxxyz).
The $y$ and $z$ integrals are elementary and leave us with a standard
hypergeometric integral,
$$
\eqalign{
  \Basic_4^0(\chi=1)\ &=\ -{2^{1-\e} \rg \over \e}
   \int_0^1 dx\ { x^{-1-\e} - (1-x)^{-1-\e} \over 1-2x} \cr
   &=\ -{2^{1-\e} \rg \over \e}
     {\Gamma(-\e)\Gamma(1) \over \Gamma(1-\e)}
     \lim_{\delta\to0} \Bigl(
        \F21(1, -\eps; 1-\eps; 2+i\delta)
      - \F21(1, 1; 1-\eps; 2+i\delta) \Bigr) \cr
   &=\ {2^{1-\e} \rg \over \e^2}
     \lim_{\delta\to0} \Bigl(
      \F21(1, -\eps; 1-\eps; 2+i\delta)
    + \F21(1, -\eps; 1-\eps; 2-i\delta) \Bigr)\ . \cr}
\eqn\MasslessBoxConstIntTwo
$$
Comparing with the first line of equation~(\use\BasicHFullSolution),
we see that the constant of integration vanishes.
 
Alternatively, we may solve equation~(\use\PDEFourFinal) order by order
in $\eps$.   Observe that $\BasicH_4$ must
contain $1/\eps^2$ poles from the overlap of collinear and
soft singularities.  As the right-hand side of the differential
equation only contains a single power of $1/\eps$, this leading
pole should be multiplied by something to the $\pm\eps$ power, so
that one power of $\eps$ is cancelled upon differentiation.
 Through $\Ord(\eps^0)$, we then have
$$
\eqalign{
\BasicH_4\ &=\ \rg\left\{ {2\over \eps^2 } \LB (1+\chi)^{-\eps}
+ (1+\chi^{-1})^{-\eps} \RB
\ -\ \ln^2\chi - \pi^2 \right\}\ +\ \Ord(\eps) \cr
&=\ \rg\left\{ {2\over \eps^2 } \LB \Bigl({s+t\over s} \Bigr)^{-\eps}
+ \Bigl({s+t \over t} \Bigr)^{-\eps} \RB
\ -\ \ln^2 \Bigl( {t\over s} \Bigr) - \pi^2 \right\} \ +\ \Ord(\eps)\ ,\cr
}\eqn\BasicHSolution
$$
where the constant of integration can be fixed as in the all-orders
solution.
 
Restoring the prefactor $(-(s+t)/st)^\e$, and
expressing the result in terms of the $\alpha_i$, we have
$$
\eqalign{
 \Basic_4[1]\ &=\ {2\rg\over\e^2} \LB (-\al2\al4)^\e
   \F21\L -\eps,-\eps;1-\eps; 1+{\al1\al3\over\al2\al4}\R
      \ +\ (-\al1\al3)^\e
   \F21\L -\eps,-\eps;1-\eps; 1+{\al2\al4\over\al1\al3}\R \RB
   \cr
   &=\ \rg \LB {2\over\e^2} \L (\al1\al3)^\e + (\al2\al4)^\e \R
     \ -\ \ln^2\L {\al1\al3\over\al2\al4} \R -\pi^2 \RB \ +\ \Ord(\e)
 \ .\cr}
\eqn\BoxFinalForm
$$
In this form, the differentiation formula~(\use\BasicBox)
may be applied to the scalar integral $\Basic_4$
to obtain the integrals with arbitrary Feynman parameter
polynomials inserted.
Because of the $\Gamma(1-m+2\e)$ prefactor in~(\use\BasicBox),
the $\Ord(\e)$ terms in $\Basic_4$ contribute to
the polynomial integrals at $\Ord(\e^0)$.
Instead of displaying the $\Ord(\e)$ terms in $\Basic_4$ explicitly,
we quote the reduced integrals with one parameter inserted,
$\Basic_4[a_i]$, to $\Ord(\e^0)$:
$$
\eqalign{
  \alpha_1 \, \Basic_4[a_1]\ &=\ \alpha_3 \, \Basic_4[a_3]\ =\
    \rg\, \left\{ {1\over\e^2}{(\al2\al4)^\e}
    - {1\over2}\L {\al1\al3 \over \al1\al3+\al2\al4} \R
    \LB \ln^2\L {\al1\al3\over\al2\al4} \R + \pi^2 \RB \right\}
    \ +\ \Ord(\e),\cr
  \alpha_2 \, \Basic_4[a_2]\ &=\ \alpha_4 \, \Basic_4[a_4]\ =\
    \rg\, \left\{ {1\over\e^2}{(\al1\al3)^\e}
    - {1\over2}\L {\al2\al4 \over \al1\al3+\al2\al4} \R
    \LB \ln^2\L {\al1\al3\over\al2\al4} \R + \pi^2 \RB \right\}
    \ +\ \Ord(\e).\cr
}\eqn\MasslessBoxOneA
$$
The latter integrals
may be differentiated further to obtain through $\Ord(\e^0)$
 the integral with any polynomial of the Feynman parameters inserted.
 
As mentioned previously, the branch cuts can be obtained
by inserting the $i \varepsilon$ associated with each kinematic
variable,
$$\eqalign{
(-s)^{-\eps}\ &\rightarrow\ |s|^{-\eps} e^{+i\pi\eps\Theta(s)}\;,\cr
\ln(-s)\ &\rightarrow\ \ln |s| - i\pi\Theta(s)\;,\cr
}
\anoneqn
$$
where $\Theta(x)$ is the usual
Heavyside function: $\Theta(x) = 1$ for $x>0$ and $\Theta(x) = 0$
for $x<0$.  For the massless scalar box we therefore obtain
$$
\eqalign{
{\cal I}_4 [1]\ & =\ i {\rg \over (4 \pi)^2 } \, {1\over s t} \biggl\{
{2 \over \eps^2} \biggl[ \Bigl( {|s| \over 4 \pi \mu^2} \Bigr)^{-\eps}
e^{i \pi \eps \Theta(s)}
+  \Bigl( {|t| \over 4 \pi \mu^2} \Bigr)^{-\eps}
e^{i \pi \eps \Theta(t)} \biggr]  \cr
& \hskip 0.5 cm
- \ln^2\LV {s \over t} \RV +
2 \pi i \L\Theta(s) - \Theta(t)\R \ln \LV{s \over t} \RV
- \pi^2  \LB 1 - \L\Theta(s) - \Theta(t)\R^2 \RB
\biggr\} + {\cal O}(\eps) \ ,  \cr}
\anoneqn
$$
where $s$ and $t$ are the Mandelstam variables defined below
equation~(\use\FourDenominator).
 
 
\subsection{The Box Integral with One External Mass}
\tagsubsection\SingleMassBoxSubsection
 
Following the same techniques, we can obtain partial differential
equations for boxes with one external massive leg (or equivalently,
one external leg off-shell),
$$
\Int_4^\onemass(s_1,s_2,m_4^2)\ =\ \Gamma(2+\e)
\int_0^1 d^4a\ \delta(1 - {\textstyle \sum_i} a_i)
{1\over \LB -s a_1 a_3 - t a_2 a_4 - m_4^2 a_4 a_1\RB^{2+\eps}}\;.
\anoneqn
$$
(This integral could also be evaluated using the same
change of variables~(\use\Boxxyz) as for the massless box.)
By analogy with equation~(\use\BoxBoundaryTerm),
such integrals will clearly arise in
the consideration of massless pentagon integrals.
Following the conventions of section~\use\AllNSection,
we label these boxes by $\Int_4^{(i)}$ when the
momentum invariant $s_{i-1,i}$ for the adjacent legs $(i-1)$ and $i$
of the pentagon diagram serves as the ``mass'' of the massive leg
of the box.  For example,
$$\eqalign{
  \Int_4^{(5)}\Poly{1}\ &=\ \Int_4^\onemass(s_{12},s_{23},s_{45})\cr
  &=\ \Gamma(2+\e) \int_0^1 d^4 a \, \delta(1 - {\textstyle \sum_i} a_i) \,
 {1\over \LB -s_{12} a_1 a_3 - s_{23} a_2 a_4 - s_{45} a_4 a_1\RB^{2+\eps}}\cr
}
\eqn\FourFiveBox
$$
is the box integral arising from the diagram depicted in
\fig\MassiveBoxFigure, in which a tree with external legs~4 and~5 is
attached to a four-point loop.
Note that the scalar denominator for the integral~(\use\FourFiveBox)
can be obtained from the massless pentagon denominator by setting
the parameter $a_5$ to zero.  Similarly, $\Int_4^{(i)}$ can be
obtained by setting $a_i\to0$ in the massless pentagon.
 
{}From these remarks it is clear that the change of
integration variables described earlier for the pentagon
can be used here to remove the kinematic factors from the
denominator of the box integral,
$$
\eqalign{
  \Int_4^{(5)}\Poly{1}\ &=\ \Gamma(2+\e)
     \biggl( \prod_{j=1}^4 \al{j} \biggr)
     \int_0^1 d^4u\; {\delta\L 1 - \sum u_i \R\,
                 \L\sum^4_{j=1} \al{j} u_j\R^{2\eps}
   \over \LB u_1 u_3 + u_2 u_4 + u_4 u_1\RB^{2+\eps}}\ ,\cr
}\anoneqn
$$
where $\al{i}$ are given by equation~(\use\PentagonAlphas).
(These variables $\al{i}$ should not be confused with the
corresponding $\al{i}$ for the massless box.)
The other integrals that will arise,
$$
\eqalign{
\Int_4^{(1)} &= \Int_4^\onemass(s_{23},s_{34},s_{51}),\qquad
\Int_4^{(2)} = \Int_4^\onemass(s_{34},s_{45},s_{12}),\cr
\Int_4^{(3)} &= \Int_4^\onemass(s_{45},s_{51},s_{23}),\qquad
\Int_4^{(4)} = \Int_4^\onemass(s_{51},s_{12},s_{34}),
}\anoneqn
$$
can be obtained from $\Int_4^{(5)}$ by cyclic permutation of the
$\al{i}$.
We define the reduced integral, $\Basic_4^{(i)}$
or $\Basic_4^\onemass$, via equation~(\use\ReducedIntegral).
 
We can now apply the general results of section~\use\AllNSection\
to the example of equation~(\FourFiveBox).
The matrix $\rho$ defined in equation~(\use\rhodefn) is now given by
$$
  \rho^{\rm 1m}\ =\
  {1\over2} \L\matrix{  0 & 0 & 1 & 1\cr
                        0 & 0 & 0 & 1\cr
                        1 & 0 & 0 & 0\cr
                        1 & 1 & 0 & 0\cr   }\R\ ,
\eqn\onemassrho
$$
so that $N_4^{\rm 1m} = {1\over2}$, and the rescaled Gram determinant
is given, using eqs.~(\use\etadefn) and (\use\Nnetarhoeqn), by
$$
  \Det_4^\onemass = 2(\al1\al3+\al2\al4-\al2\al3)\,.
\eqn\onemassgramdet
$$
Using equations~(\use\etadefn) and (\use\gammadefn),
the explicit values of the quantities $\ga{i}^\onemass$ and
$\eta_{ij}^\onemass$ can be read off from~(\use\onemassgramdet):
$$
\eqalign{
 \ga1^\onemass\ &=\ \al3,\quad \ga2^\onemass\ =\ \al4-\al3,
 \quad \ga3^\onemass\ =\ \al1-\al2, \quad \ga4^\onemass\ =\ \al2, \cr
 \eta_{13}^\onemass\ &=\ \eta_{24}^\onemass\ =\ -\eta_{23}^\onemass\ =\
 \eta_{31}^\onemass\ =\ \eta_{42}^\onemass\ =\ -\eta_{32}^\onemass\ =\ 1,
  \qquad {\rm remaining}\ \eta_{ij}^\onemass\ =\ 0. \cr
}\anoneqn
$$
In terms of these quantities, the differential
equations~(\use\reducedInaeqn) read
$$
\eqalign{
{\partial\Basic_4^\onemass\over\partial\al{i}}\ &=\ 2\e\,
 \LB \sum_{j=1}^4 \L \eta_{ij}^\onemass-{\ga{i}^\onemass\ga{j}^\onemass\over
                                 \Det_4^\onemass}\R\,\Basic_3^{(j)}
 + {\ga{i}^\onemass\over\Det_4^\onemass} \Basic_4^\onemass\RB\cr
\ &=\ 2\eps\, \LB
   \sum_{j=1}^4 \sqrt{\Det_4^\onemass} {\partial^2 \sqrt{\Det_4^\onemass}\over
           \partial\al{i}\partial\al{j}}\,\Basic_3^{(j)}
 + {1\over\sqrt{\Det_4^\onemass}} {\partial\sqrt{\Det_4^\onemass}\over
                            \partial\al{i}}\,\Basic_4^\onemass\RB \; . \cr
}\eqn\OneMassDifferentialEqn
$$
The triangle integrals appearing on the right-hand-side
of~(\use\OneMassDifferentialEqn) include both the triangle integral
with one external massive leg, $\Int_3^\onemass$, defined in
equation~(\use\OneMassTriangle), and the triangle with two external
masses,
$$
\Int_3^\twomass(s_1,s_2)\ =\ {\rg\over\e^2}
  {(-s_1)^{-\eps}-(-s_2)^{-\eps}\over (-s_1)-(-s_2)}\ .
\eqn\TwoMassTriangle
$$
Explicitly, the following reduced triangle integrals appear
(defined again via~(\use\ReducedIntegral)):
$$
\eqalign{
\Basic_3^{(1)}\ &=\
 {1\over\al2\al3\al4}\Int_3^\onemass\L{-1\over\al2\al4}\R
 \ =\ {\rg\over\eps^2} {\al4^\e \al2^\e \over \al3}
 \;,\cr
\Basic_3^{(2)}\ &=\
 {1\over\al1\al3\al4} \Int_3^\twomass\L{-1\over\al1\al3},{-1\over\al4\al1}\R
 \ =\ -{\rg\over\eps^2}\al1^\e {\al4^\eps-\al3^\eps \over \al4-\al3}
 \;,\cr
\Basic_3^{(3)}\ &=\
 {1\over\al1\al2\al4}\Int_3^\twomass\L{-1\over\al2\al4},{-1\over\al4\al1}\R
 \ =\ -{\rg\over\eps^2}\al4^\e {\al1^\eps-\al2^\eps \over \al1-\al2}
 \;,\cr
\Basic_3^{(4)}\ &=\
 {1\over\al1\al2\al3} \Int_3^\onemass\L{-1\over\al1\al3}\R
 \ =\ {\rg\over\eps^2} {\al1^\e \al3^\e \over \al2}
 \; .\cr
}\anoneqn
$$
 
The differential equations~(\use\OneMassDifferentialEqn)
have the solution,
$$
\eqalign{
\Basic_4^\onemass
\ &=\ {2\rg\over\eps^2} \; \Biggl[
     \L -\al3(\al1-\al2) \R^\eps
   \F21\L-\eps,-\eps;1-\eps;
   {\al1\al3+\al2\al4-\al2\al3\over\al3(\al1-\al2)}\R \cr
&\hskip 7mm
    +\L -\al2(\al4-\al3) \R^{\eps} \F21\L -\eps,-\eps;1-\eps;
          {\al1\al3+\al2\al4-\al2\al3\over \al2(\al4-\al3)}\R\cr
&\hskip 7mm -\L (\al1-\al2)(\al4-\al3) \R^\eps
  \F21\L -\eps,-\eps;1-\eps;
     -{\al1\al3+\al2\al4-\al2\al3\over (\al1-\al2) (\al4-\al3)}\R
     \Biggr]\cr
&=\ {2\rg\over\eps^2} \; \Biggl[ \L-\ga1^\onemass\ga3^\onemass\R^\eps
 \F21\L-\eps,-\eps;1-\eps; {\Det_4^\onemass\over2\ga1^\onemass\ga3^\onemass}\R
        \cr
&\hskip 7mm   +\L-\ga2^\onemass\ga4^\onemass\R^\eps
\F21\L -\eps,-\eps;1-\eps; {\Det_4^\onemass\over2\ga2^\onemass\ga4^\onemass}\R
      \cr
&\hskip 7mm   -\L\ga2^\onemass\ga3^\onemass\R^\eps
 \F21\L -\eps,-\eps;1-\eps;-{\Det_4^\onemass\over2\ga2^\onemass\ga3^\onemass}\R
      \Biggr]\cr
&=\ 2\rg\, \LB {(\al2\al3)^\e\over\eps^2} + \Li_2\L 1-{\al1\over\al2}\R
   +\Li_2\L1-{\al4\over\al3}\R - {\pi^2\over6}\RB + \Ord(\eps) \; ,\cr
}\eqn\SingleMassBoxSolution
$$
where $\Li_2$ is the dilogarithm~[\ref\Lewin{L. Lewin, {\it
Dilogarithms and Asscociated Functions} (Macdonald) (1958)}],
which satisfies
$$
  {d\over dx}\Li_2(1-x)\ =\ {\ln(x)\over 1-x}\ ,
\eqn\dilogdiff
$$
and also the identity
$$
\Li_2 (1-x) + \Li_2 (1-x^{-1})\ =\ - {1\over 2} \ln^2(x),
\hskip 2 cm x>0.
\eqn\dilogid
$$
In principle, one could also add any solution of the homogeneous
equations, (\use\OneMassDifferentialEqn) with $\Basic_3^{(j)}$ set
to zero.
The coefficient of such a solution vanishes, as may again be
demonstrated by evaluation at a special kinematic point.
 
In terms of momentum invariants, the unreduced integral is
$$
\Int_4^{(5)}\ =\ \rg\,
{2\over s_{12} s_{23} }  {(-s_{12})^{-\eps} (-s_{23})^{-\eps}
\over (-s_{45})^{-\eps} }
\biggl[ {1\over \eps^2 } + \Li_2\Bigl( 1 - {s_{12}\over s_{45}} \Bigr)
+ \Li_2 \Bigl( 1 - {s_{23} \over s_{45}} \Bigr) - {\pi^2 \over 6} \biggr]
+\Ord(\eps),
\eqn\NewOneMass
$$
or alternatively, after using the dilogarithm identity (\use\dilogid) and
rearranging the terms
$$
\eqalign{
  I_4^{(5)} &=\ {\rg \over s_{12} s_{23}} \biggl\{
  {2\over\e^2} \Bigl[ (-s_{12})^{-\e} + (-s_{23})^{-\e}
- (-s_{45})^{-\e} \Bigr] \cr
 &\ -\ 2\ \Li_2\left(1-{s_{45}\over s_{12}}\right)
  \ -\ 2\ \Li_2\left(1-{s_{45}\over s_{23}}\right)
  \ -\ \ln^2\left({s_{12}\over s_{23}}\right)\ -\ {\pi^2\over3} \biggr\}
  \ +\ \Ord(\e). \cr}
\eqn\NewOneMassAlt
$$
This second form is appropriate for studying the limit $s_{45}
\rightarrow 0$ as we do at the end of this section.
 
Including the overall normalization factors appropriate for the
momentum-space integral (\use\NPointLoopIntegral) yields
$$
\Loop_4^{(5)}\ =\ {i \rg\over(4\pi)^2}\,
{2\over s_{12} s_{23} } \Bigl( {-s_{12} \over 4 \pi \mu^2} \Bigr)^{-\eps}
\Bigl({ - s_{23} \over 4 \pi \mu^2 }\Bigr)^{-\eps}
\Bigl({ -s_{45} \over 4 \pi \mu^2 }\Bigr)^{\eps}
\biggl[ {1\over \eps^2 } + \Li_2\Bigl( 1 - {s_{12}\over s_{45}} \Bigr)
+ \Li_2 \Bigl( 1 - {s_{23} \over s_{45}} \Bigr) - {\pi^2 \over 6} \biggr]
+\Ord(\eps),
\anoneqn
$$
in agreement with the results of
refs.~[\ref\SingleMassBox{K. Fabricius and I. Schmitt, Z. Phys.\
C3:51 (1979)\semi
S. Papadopoulos, A. P.\ Contogouris, J. Ralston,
   Phys.\ Rev.\ D25:2218 (1982)}].
The correct analytic continuation to the physical region can
be obtained from this expression by taking
$s_{ij} \rightarrow s_{ij} + i \varepsilon$.
 
As with the massless box, it is useful to quote the integrals
$\Basic_4^\onemass[a_i]$ to $\Ord(\e^0)$; further
differentiation of them will give any desired integral
to $\Ord(\e^0)$ as well.
The $\Basic_4^\onemass[a_i]$ may be read off from
equations~(\use\OneMassDifferentialEqn) and
(\use\SingleMassBoxSolution).
We rewrite them in terms of a combination of dilogarithms and
logarithms that will reappear in the massless pentagon tensor
integrals:
$$
\eqalign{
  \Basic_4^{(5)}[a_1]\ &=\ \rg \left[ -{1\over \e^2}
     { \al4^\e (\al1^\e - \al2^\e) \over \al1 - \al2 }
      + {\al3 \, L_5 \over \al1\al3+\al2\al4-\al2\al3} \right]\ , \cr
  \Basic_4^{(5)}[a_2]\ &=\ \rg
     \left[ {1\over \e^2} \left( {\al1^\e \al3^\e \over \al2}
       + { \al4^\e (\al1^\e - \al2^\e) \over \al1 - \al2 }  \right)
       + {(\al4 - \al3) \, L_5 \over
              \al1\al3+\al2\al4-\al2\al3} \right]\ , \cr
  \Basic_4^{(5)}[a_3]\ &=\ \rg
      \left[ {1\over \e^2} \left( {\al4^\e \al2^\e \over \al3}
       + { \al1^\e (\al4^\e - \al3^\e) \over \al4 - \al3 }  \right)
       + {(\al1 - \al2) \, L_5 \over
             \al1\al3+\al2\al4-\al2\al3} \right]\ , \cr
  \Basic_4^{(5)}[a_4]\ &=\ \rg
      \left[ -{1\over \e^2}
      { \al1^\e (\al4^\e - \al3^\e) \over \al4 - \al3 }
      + {\al2 \, L_5 \over \al1\al3+\al2\al4-\al2\al3} \right]\ , \cr
        }
\eqn\MassBoxOneA
$$
where
$$
  L_i\ \equiv\ \Li_2\L 1-{\al{i+1}\over\al{i+2}} \R
              + \Li_2\L 1-{\al{i-1}\over\al{i-2}} \R
  + \ln\L{\al{i+1}\over\al{i+2}}\R \ln\L{\al{i-1}\over\al{i-2}}\R
                  -{\pi^2 \over 6}\ .
\eqn\LDef
$$
Note that $L_i$ vanishes as
$\al{i+1}\al{i-2}+\al{i+2}\al{i-1}-\al{i+2}\al{i-2} \rightarrow 0$,
so the $\Basic_4^{(i)}[a_j]$ are not singular in that limit.
 
 
\subsection{Box Integrals with Two External Masses}
 
\def\easytwo{{2{\rm m}e}}
\def\hardtwo{{2{\rm m}h}}
 
In order to evaluate the pentagon integral with one external mass,
or the all-massless hexagon integral, one needs box integrals with
two external masses, of which there are two types, which we will call
`easy' and `hard'.  Both of these integrals have been performed
previously~[\ref\TwoMassBoxRef{J. Ohnemus and J. Owens,
  Phys.\ Rev.\ D43:3626 (1991)},\EGY].
The `easy' box, with external masses at diagonally opposite corners,
can be done with the same change of variables~(\use\Boxxyz)
described in section~\use\PDESection.
We will not discuss it further, but merely quote the result,
$$
\eqalign{
  I_4^\easytwo[1]\ &\equiv\ \Gamma(2+\e)
  \int_0^1 d^4a_i\ {\delta\bigl( 1-\sum a_i \bigr) \over
 \bigl[ -sa_1a_3 - ta_2a_4 - m_1^2a_1a_2 - m_3^2a_3a_4 \bigr]^{2+\e} }
    \cr
  &=\  {2\,\rg\over st - m_1^2m_3^2}
  \left\{ {1\over \e^2} \L
  (-s)^{-\e} + (-t)^{-\e} - (-m_1^2)^{-\e} - (-m_3^2)^{-\e}\R
  \ +\ \Li_2\L 1-{m_1^2m_3^2\over st}\R\RP \cr
  &\ \ \LP
  - \Li_2\L 1-{m_1^2\over s}\R
  - \Li_2\L 1-{m_1^2\over t}\R
  - \Li_2\L 1-{m_3^2\over s}\R
  - \Li_2\L 1-{m_3^2\over t}\R
  - {1\over2} \ln^2\L{s\over t}\R \right\} + \ordereps\ . \cr}
\eqn\EasyDouble
$$
 
The `hard' box, with external masses at adjacent corners (legs~3
and~4),
$$
\Int_4^\hardtwo[1]\ =\
  \Gamma(2+\e) \int_0^1 d^4a {\delta(1 - {\textstyle \sum_i} a_i)
 \over \LB -s a_1 a_3 - t a_2 a_4 - m_3^2 a_3 a_4
                                  - m_4^2 a_4 a_1 \RB^{2+\eps}}\;,
\eqn\HardDoubleMassIntegral
$$
cannot be easily done this way;
but it is amenable to the partial differential equation technique.
We change to $\alpha_i$ variables defined by
$$
  s\ =\ -{1\over\alpha_1\alpha_3}\ , \qquad
  t\ =\ -{1\over\alpha_2\alpha_4}\ , \qquad
  m_3^2\ =\ -{1\over\alpha_3\alpha_4}\ , \qquad
  m_4^2\ =\ -{1\over\alpha_4\alpha_1}\ .
$$
Then the matrix $\rho$ is given by
$$
  \rho^\hardtwo\ =\
  {1\over2} \L\matrix{  0 & 0 & 1 & 1\cr
                        0 & 0 & 0 & 1\cr
                        1 & 0 & 0 & 1\cr
                        1 & 1 & 1 & 0\cr   }\R\ ,
\eqn\hardtworho
$$
we have $N_4^\hardtwo = {1\over2}$, and the rescaled Gram determinant
is
$$
\Det_4^\hardtwo = 2(\alpha_1\alpha_3 + \alpha_2\alpha_4
   - \alpha_1\alpha_2 - \alpha_2\alpha_3 + \alpha_2^2)\;,
\eqn\hardtwogramdet
$$
from which $\gamma_i^\hardtwo$ and $\eta_{ij}^\hardtwo$ can be
obtained via equations~(\use\etadefn) and (\use\gammadefn).
 
To obtain the box integral to ${\cal O}(\e^0)$, we can use the
simple partial differential equations~(\use\boxPDEs), which
are sensitive only to the pieces of the triangle integrals that are
singular as $\e\to0$.  In particular, the three-mass triangle
does not contribute, because it is finite as $\e\to0$.
We find that to ${\cal O}(\e^0)$,
$$
\eqalign{
{\del\Basic_4^\hardtwo\over\del\alpha_i}\ &=\
  2\eps \, \sum_{j=1}^4 \eta_{ij}^\hardtwo \, \Basic_3^{(j)} \cr
 &=\ 2\,\rg \left[ {1\over\e} \eta_{i4}^\hardtwo {1\over\alpha_2}
   - \eta_{i1}^\hardtwo {\ln(\alpha_2/\alpha_3)\over\alpha_2-\alpha_3}
   - \eta_{i3}^\hardtwo {\ln(\alpha_1/\alpha_2)\over\alpha_1-\alpha_2}
   + \eta_{i4}^\hardtwo {\ln(\alpha_1\alpha_3)\over\alpha_2} \right]\ .
 \cr}
\eqn\DoubleMassBoxDerivatives
$$
Writing
$$
\Basic_4^\hardtwo\ =\ \rg\,\LB{1\over\eps^2}
              +{1\over\eps} X_{-1} + X_0 + c_0 \RB\ ,
\anoneqn
$$
and solving the differential equations for $X_{-1}$ and $X_0$,
we find
$$
\eqalign{
  X_{-1}\ &=\ 2\ln\alpha_2, \cr
   X_0\ &=\ 2 \Li_2\L 1-{\alpha_1\over\alpha_2}\R
          + 2 \Li_2\L 1-{\alpha_3\over\alpha_2}\R
          + 2 \ln^2\alpha_2, \cr}
\anoneqn
$$
or
$$
\Basic_4^\hardtwo\ =\ \rg\, \alpha_2^{2\eps}\LB{1\over\eps^2}
    + 2 \Li_2\L 1-{\alpha_1\over\alpha_2}\R
    + 2 \Li_2\L 1-{\alpha_3\over\alpha_2}\R
    + c_0 \RB\ +\ {\cal O}(\e).
\eqn\HardDoubleMassResult
$$
 
The constant $c_0$ may be determined
by computing the function at a specific point,
say where all the $\alpha_i$ are equal; the resulting integral
is evaluated explicitly in appendix~\use\TwoMassBoxConstantIntegral,
whence we find $c_0 = 0$.
Finally, rewriting the result~(\use\HardDoubleMassResult) back in
terms of the conventional kinematic variables yields
$$
I_4^\hardtwo [1] = \rg\, {(-m_3^2)^{\eps} (-m_4^2)^{\eps}
\over (-t)^{1+2\eps} (-s)^{1+\eps} }
\LB{1\over\eps^2}
 + 2 \Li_2\L 1-{t\over m_3^2}\R + 2 \Li_2\L 1-{t\over m_4^2}\R \RB\
  +\ {\cal O}(\e).
\eqn\HardTwoMassAnswer
$$
 
Using the dilogarithm identity (\use\dilogid) and rearranging the terms
this can be written in the alternative form
$$
\eqalign{
  I_4^{2{\rm m}h}[1] \ &=\ {\rg \over st} \biggl\{
  {2\over\e^2} \Bigl[ (-s)^{-\e} + (-t)^{-\e}
              - (-m_3^2)^{-\e} - (-m_4^2)^{-\e} \Bigr]
  \ +\ {1\over\e^2}
    { (-m_3^2)^{-\e}(-m_4^2)^{-\e} \over (-s)^{-\e} }  \cr
 &\ -\ 2\ \Li_2\left(1-{m_3^2\over t}\right)
  \ -\ 2\ \Li_2\left(1-{m_4^2\over t}\right)
  \ -\ \ln^2\left({s\over t}\right)\biggr\}\ +\ \Ord(\e), \cr}
\eqn\NewHardTwoMass
$$
which is more convenient for studying the massless limit, as we do
at the end of this section.
 
 
\subsection{The Box Integral with Three External Masses}
\def\threemass{{3{\rm m}}}
 
Here we compute the three-mass scalar box integral,
$$
  I_4^\threemass[1]\ =\ \Gamma(2+\e)
  \int d^4a_i\ {\delta\bigl( 1-\sum a_i \bigr)\over
   \bigl[ -sa_1a_3 - ta_2a_4 - m_2^2a_2a_3 - m_3^2a_3a_4 - m_4^2a_4a_1
   \bigr]^{2+\e}}\ .
\eqn\threemassdef
$$
We again use the partial differential equations~(\use\boxPDEs),
with the change-of-variables
$$
  s\ =\ -{1\over\alpha_1\alpha_3},\quad
  t\ =\ -{1\over\alpha_2\alpha_4},\quad
  m_2^2\ =\ -{\lambda\over\alpha_2\alpha_3},\quad
  m_3^2\ =\ -{1\over\alpha_3\alpha_4},\quad
  m_4^2\ =\ -{1\over\alpha_4\alpha_1},
\eqn\threemasskin
$$
which is the same as that used for the hard two-mass box, except that
now $\lambda\neq0$.
The matrix $\rho$ becomes
$$
  \rho_4^\threemass\ =\ {1\over2} \L\matrix{
        0   &    0    &   1     &   1    \cr
        0   &    0    & \lambda &   1    \cr
        1   & \lambda &   0     &   1    \cr
        1   &    1    &   1     &   0    \cr }\R\ ,
\eqn\rhofour
$$
the normalization factor in~(\use\boxPDEs) is
$N_4^\threemass\ =\ \hf(1-\lambda)^2$,
and the matrix $\eta$ used to construct $\Det_4^{\threemass}$ is
$$
  \eta^\threemass\ =\  \L\matrix{
        2\l  & -1-\l   & 1-\l    & -\l(1-\l)  \cr
       -1-\l &    2    & -(1-\l) & 1-\l       \cr
        1-\l & -(1-\l) &   0     &   0        \cr
   -\l(1-\l) &  1-\l   &   0     &   0        \cr }\R\ .
\eqn\threemassetadefn
$$
 
We expand $\Basic_4^\threemass$ as
$$
  \Basic_4^\threemass\ =\ \rg \Bigl[ {c_1(\lambda)\over\e}\ +\
  X_0(\alpha_i,\lambda)\ +\ c_0(\lambda) \Bigr]\ + \ordereps.
\eqn\epsexpand
$$
(We will see below that there is no $1/\e^2$ singularity.)
To solve the partial differential equations order-by-order in $\e$
we need to first know $c_1(\l)$.
We know that $c_1(\l)$ is independent of
the $\alpha_i$ because the daughter triangles here are of the
two-mass and three-mass varieties;
the two-mass triangle has a $1/\e$ pole, which feeds into
$X_0(\alpha_i,\l)$, while the three-mass triangle is finite and can be
ignored altogether.
So we may compute $c_1(\lambda)$ by doing the integral
$\Basic_4^\threemass$ for the special choice of all $\alpha_i=1$.
We should compute the finite part of the integral
while we're at it, since this result will fix the constant
of integration $c_0(\lambda)$.  This computation
is done in appendix~\use\TwoMassBoxConstantIntegral,
where we find
$$
c_1(\lambda) = {\ln\lambda\over 1-\lambda}\ .
\anoneqn$$
 
Next we solve the partial differential equations~(\use\boxPDEs).
Plug the expansion of $\Basic_4^\threemass$ (equation~(\use\epsexpand))
and the divergent pieces of the 2-mass triangles
$\Basic_3^{(3)}$ and $\Basic_3^{(4)}$,
$$
\eqalign{
  \Basic_3^{(3)}\ &=\ -{\rg\over\e}
  { \ln(\alpha_1/\alpha_2) \over \alpha_1-\alpha_2}
  \ +\ {\cal O}(\eps^0)\ ,\cr
  \Basic_3^{(4)}\ &=\ -{\rg\over\e}
  { \ln(\l\alpha_1/\alpha_2) \over \l\alpha_1-\alpha_2}
  \ +\ \Ord(\eps^0)\ ,\cr}
\eqn\triexpand
$$
into the far right-hand-side of~(\use\boxPDEs) and use the
result~(\use\threemassetadefn) for
$\eta_{ij}$, to get
$$
\eqalign{
  {\del X_0\over\del\alpha_1}\ &=\ {2 \over 1-\l}\biggl[
  -{\ln(\alpha_1/\alpha_2)\over \alpha_1-\alpha_2}
  +\l{\ln(\l\alpha_1/\alpha_2)\over \l\alpha_1-\alpha_2} \biggr]\ ,\cr
  {\del X_0\over\del\alpha_2}\ &=\ {2 \over 1-\l}\biggl[
  {\ln(\alpha_1/\alpha_2)\over \alpha_1-\alpha_2}
  - {\ln(\l\alpha_1/\alpha_2)\over \l\alpha_1-\alpha_2} \biggr]\ ,\cr
  {\del X_0\over\del\alpha_3}\ &=\ 0\ ,\cr
  {\del X_0\over\del\alpha_4}\ &=\ 0\ .\cr}
\eqn\explicitPDEs
$$
Solving these equations for $X_0(\alpha_i,\l)$, and fixing the constant
$c_0(\l)$ using equation~(\use\alphaoneanswer), yields
$$
\eqalign{
 \Basic_4^\threemass\ =\ {\rg \over 1-\l} \biggl[ & {\ln\l\over\e}
  \ +\ 2\ \Li_2\left(1-{\alpha_1\over\alpha_2}\right)
  \ -\ 2\ \Li_2\left(1-{\l\alpha_1\over\alpha_2}\right)
  \ +\ 2\ \Li_2(1-\l)
  \ +\ 2\ln\l\ \ln\alpha_2\ -\ {1\over2}\ln^2\l \biggr]\ \cr
\null & + \ordereps\ . \cr}
\eqn\Basicanswer
$$
Returning to the original kinematic variables, and using dilogarithm
identities~[\Lewin], we get
$$
\eqalign{
  I_4^\threemass(s,t,m_i^2)\ &=\ {\rg \over st-m_2^2m_4^2} \biggl\{
  {2\over\e^2} \Bigl[ (-s)^{-\e} + (-t)^{-\e}
     - (-m_2^2)^{-\e}- (-m_3^2)^{-\e} - (-m_4^2)^{-\e} \Bigr] \cr
  &  \ +\ {1\over\e^2}
     { (-m_2^2)^{-\e} (-m_3^2)^{-\e} \over (-t)^{-\e} }
  \ +\ {1\over\e^2}
    { (-m_3^2)^{-\e}(-m_4^2)^{-\e} \over (-s)^{-\e} } \cr
  &\ -\ 2\ \Li_2\left(1-{m_2^2\over s}\right)
   \ -\ 2\ \Li_2\left(1-{m_4^2\over t}\right)
  \ +\ 2\ \Li_2\left(1-{m_2^2m_4^2\over st}\right)
      \ -\ \ln^2\left({s\over t}\right) \biggr\} + \ordereps\ . \cr }
\eqn\threemassanswer
$$
 
 
\subsection{The Box Integral with Four External Masses}
\def\fourmass{{4{\rm m}}}
 
The four-mass box integral is infrared finite and has been
performed in $D=4$ in ref.~[\use\HVintegrals];
a compact expression is given in ref.~[\use\NewFourPoint].
Amusingly, the partial differential
equations~(\use\boxPDEs) for it are trivial,
because the three-mass triangles
appearing on the right-hand-side are non-singular as $\e\to0$.
In other words, through $\ordereps$, the
reduced four-mass box cannot depend on the $\alpha_i$,
but only on the two other, dimensionless variables,
say $\l_1$ and $\l_2$, where we define
$$
  s\ =\ -{1\over\alpha_1\alpha_3},\quad
  t\ =\ -{1\over\alpha_2\alpha_4},\quad
  m_1^2\ =\ -{\l_1\over\alpha_1\alpha_2},\quad
  m_2^2\ =\ -{\l_2\over\alpha_2\alpha_3},\quad
  m_3^2\ =\ -{1\over\alpha_3\alpha_4},\quad
  m_4^2\ =\ -{1\over\alpha_4\alpha_1}.
\eqn\fourmasskin
$$
One can check that the answer $D_0(s,t,m_i^2)$
given in ref.~[\use\NewFourPoint] does have this property --- when the
integral is divided by $\alpha_1\alpha_2\alpha_3\alpha_4$ it
depends only on $\l_1$ and $\l_2$.  Indeed,
$$
\eqalign{
  \Basic_4^\fourmass\ =\ {D_0\over\prod \alpha_i}\ =\
  {1\over r}\biggl\{ & \Li_2\left(\hf(1-\l_1+\l_2+r)\right)
   \ -\ \Li_2\left(\hf(1-\l_1+\l_2-r)\right) \cr
  &\ +\ \Li_2\left(\textstyle{-1\over2\l_1}(1-\l_1-\l_2-r)\right)
   \ -\ \Li_2\left(\textstyle{-1\over2\l_1}(1-\l_1-\l_2+r)\right) \cr
  &\ +\ {1\over2}\ln\left({\l_1\over\l_2^2}\right)
   \ln\left({ 1+\l_1-\l_2+r \over 1+\l_1-\l_2-r }\right) \biggr\}+\ordereps\ ,
   \cr}
\eqn\fourmassfinal
$$
where
$$
  r\ \equiv\ \sqrt{1 - 2\l_1 - 2\l_2 + \l_1^2 - 2\l_1\l_2 + \l_2^2}\ .
\eqn\rdefinition
$$
 
 
\subsection{The Massless Limit of Massive Boxes}
 
In general there is no reason for the massless limits to be
smooth.  The limit of taking a mass to zero does not necessarily
commute with the $1/\e$ expansion of dimensional regularization, which
has been truncated at $\Ord(\e^0)$.
For $\e < 0$ (as is
required to regulate the infrared divergences in the box integrals),
we see that the single external mass box $I_4^{1{\rm m}}(s,t,m_4^2)$
(given in equation (\use\NewOneMassAlt) with $s=s_{12}$, $t =s_{23}$,
$m_4^2 = s_{45}$) goes over smoothly to the massless box $I_4^{0{\rm
m}}(s,t)$ as $m_4\to0$, and the easy two-mass box $I_4^{2{\rm
m}e}(s,t,m_3^2,m_4^2)$ goes over smoothly to the one-mass box
$I_4^{1{\rm m}}(s,t,m_4^2)$ as $m_3\to0$.  On the other hand, the
limits, $I_4^{2{\rm m}h} \to I_4^{1{\rm m}}$, $I_4^{3{\rm
m}} \to I_4^{2{\rm m}e}$, and $I_4^{3{\rm m}} \to I_4^{2{\rm m}h}$,
are not smooth: in each of these cases there are ``missing''
dilogarithms.
 
The fact that some of the above limits happen to be smooth, with only
the exponentiation of the logarithms $(-s)^{-\eps}, (-t)^{-\eps},
(-m_i^2)^{-\eps}$, can be understood from the representation~(\reducedallNeqn)
(for $n=4$) of the $D=4-2\e$ box integral as the sum of $D=4-2\e$
triangles and a $D=6-2\e$ box integral.  The $D=6-2\e$ box integral is
infrared (and ultraviolet) convergent for any choice of mass, so it
has a smooth limit as any mass goes to zero.
The $D=4-2\e$ triangles appearing in the
representations~(\reducedallNeqn) for $I_4^{0{\rm m}}$,
$I_4^{1{\rm m}}$ and $I_4^{2{\rm m}e}$ have either one or two
nonvanishing external masses; these integrals can be written in closed
form to all order in $\e$ merely by exponentiating logarithms.  (See
equations~(\OneMassTriangle) and (\TwoMassTriangle).)  In contrast,
the representations~(\reducedallNeqn) of the box integrals $I_4^{2{\rm
m}h}$ and $I_4^{3{\rm m}}$ require the triangle with three external
masses, $I_3^{3{\rm m}}$, whose all-orders-in-$\e$
form~(\use\finaltrians) is considerably more complicated, involving
hypergeometric functions.  One should not expect that these latter box
integrals, truncated to $\Ord(\e^0)$, could be made to have smooth
limits simply by exponentiating logarithms.
 
\section{\bf Algebraic Approach to Pentagon Integrals }
\tagsection\AlgebraicSection
 
It is possible to solve the partial differential
equations~(\NPointPDEs) for the massless scalar pentagon through
$\Ord(\e^0)$.
However, a simpler approach, which works equally well for arbitrary
pentagon kinematics, is to use
the general algebraic equation~(\use\reducedallNeqn) derived
in section~\use\AllNSection\ to express the
scalar pentagon integral $\Basic_5$ as a sum of five scalar
box integrals, up to $\Ord(\eps)$ corrections:
$$
  \Basic_5\ =\ {1\over 2N_5} \Biggl[ \sum_{i=1}^5 \gamma_i \,
   \Basic_{4}^{(i)}\ +\ 2\e \,\Det_5 \,
   \Basic_5^{D=6-2\e} \Biggr]\ .
\eqn\PreGeneratingFunction
$$
(See also fig.~\vNVEquationFigure.)
To give an explicit expression for the pentagon, we need only
collect the relevant scalar box integrals from section~\PDESection,
and compute the kinematic coefficients $N_5$, $\Det_5$, $\gamma_i$ and
$\eta_{ij}$.  (The $\eta_{ij}$ are relevant for computing tensor
integrals.)
We now do this for the all-massless pentagon integral,
and for the pentagon with one external mass.
 
 
\subsection{The Massless Pentagon Integral}
 
For the massless pentagon, equation~(\use\FiveDenominator)
for the scalar denominator, with the change of
variables~(\use\stoalpha), leads to a matrix $\rho$ given by
$$
\rho = {1\over2}\L\matrix{
  0&0&1&1&0\cr
  0&0&0&1&1\cr
  1&0&0&0&1\cr
  1&1&0&0&0\cr
  0&1&1&0&0\cr}\R\ .
\anoneqn$$
We find that $N_5=1$, and
$$\eqalign{
  \Det_5\ &=\ \sum_{i=1}^5
     \bigl( \al{i}^2 - 2\al{i}\al{i+1}+2\al{i}\al{i+2} \bigr)\ ,\cr
  \gamma_i\ &=\ \al{i-2}-\al{i-1}+\al{i}-\al{i+1}+\al{i+2}\ ,\cr
  \eta_{ij}\ &=\ 1 - 2\,\delta_{i,j-1}-2\,\delta_{i,j+1}\ .\cr}
\eqn\masslessdetgammaeta
$$
Plugging the one-mass box integrals~(\use\SingleMassBoxSolution)
into equation~(\use\reducedNeqfiveeqn), and using the dilogarithm
identity~(\use\dilogid), we obtain
$$
  \Basic_5[1]\ =\ \rg
  \sum_{j=1}^5 \alpha_j^{1+2\e} \biggl[ {1\over \eps^2}
  + 2\Li_2\Bigl(1 - {\alpha_{j+1} \over\alpha_j}\Bigr)
  + 2\Li_2\Bigl(1- {\alpha_{j-1}\over\alpha_j} \Bigr)
  -{\pi^2\over 6} \biggr]\ +\ \Ord(\eps).
\eqn\PentagonSolution
$$
In terms of momentum invariants, the unreduced integral is
$$
\Int_5\ =\ {\rg\ (-s_{51})^\eps (-s_{12})^\eps
   \over (-s_{23})^{1+\eps} (-s_{34} )^{1+\eps} (-s_{45})^{1+\eps} }
  \LB {1\over \eps^2} + 2\Li_2\Bigl(1- {s_{23} \over s_{51} }\Bigr)
  + 2\Li_2\Bigl( 1- {s_{45}\over s_{12}} \Bigr)
  - {\pi^2 \over 6} \RB\ +\ \hbox{cyclic}\ +\ \Ord(\eps).
\eqn\UnreducedPentagon
$$
 From this expression we can obtain the value in any region by using the
usual $i\varepsilon $ prescription and observing that
$\Int_5$ is manifestly real in the region where all $s_{ij} <0$.
 
For the tensor integrals, we do need some information about the
$\Ord(\e)$ parts of the scalar pentagon.  It turns out that
leaving the six-dimensional pentagon  $\Basic_5^{D=6-2\e}$ in
equation~(\use\PreGeneratingFunction) leaves us with
enough information about these terms
that we {\it can} use the scalar pentagon as
a generating function for the tensor integrals to $\Ord(1)$,
without having to evaluate $\Basic_5^{D=6-2\e}$ explicitly.
(The explicit solution for $\Basic_5^{D=6}$
involves a rather long combination
of $\Li_3$'s, $\Li_2$'s, and logarithms whose arguments
are complicated solutions of various quadratic equations.)
We show how to do so in the next section.

\def\msq{\hat m_5^2}
 
\subsection{The Pentagon Integral with One External Mass}
 
For the pentagon with one external mass, $m_5\neq0$,
we use the same change of variables~(\use\stoalpha),
(\use\PentagonAlphas) as in the massless case, except that
we also define the rescaled mass $\hat m_5^2 \equiv -\al5\al1 m_5^2$,
which is taken to be a variable independent of the $\al{i}$.
We find that the normalization factor is $N_5 = 1- \hat m_5^2$,
while the rescaled Gram determinant is given by
$$
\Det_5^{\rm 1m}
  \ =\ \Det_5^{\rm 0m}\ +\
  \hat m_5^2 (-2\alpha_1\alpha_3 + 2\alpha_2\alpha_3 - 2\alpha_3^2
  -4\alpha_2\alpha_4 + 2\alpha_3\alpha_4 - 2\alpha_3\alpha_5)
  \ +\ \alpha_3^2 (\hat m_5^2)^2\ ,
 \eqn\fonemass
$$
where $\Det_5^{\rm 0m}$ is given in
equation~(\use\masslessdetgammaeta).
Using these values in the general
expression for the scalar pentagon~(\use\reducedNeqfiveeqn),
and collecting the box integrals with one and two external masses
from section~\PDESection, we get
$$
\hskip -15pt\eqalign{
  I_5^{\rm 1m}&[1]\ =\ \rg \prodalphal \Biggl\{
 {1\over\e^2}\biggl[ \alpha_2^{1+2\e} + \alpha_3^{1+2\e}
                   + \alpha_4^{1+2\e}
+ \left[ (\alpha_1-\msq\alpha_2)\alpha_1^{2\e}
  + (\alpha_5-\msq\alpha_4)\alpha_5^{2\e} \right]
  { 1 - (\msq)^{-\e} \over 1-\msq } \biggr]  \cr
&+\ 2\alpha_2\left[  \Li_2\left(1-{\alpha_1\over\alpha_2}\right)
      +  \Li_2\left(1-{\alpha_3\over\alpha_2}\right) \right]
+\ 2\alpha_4\left[  \Li_2\left(1-{\alpha_5\over\alpha_4}\right)
      +  \Li_2\left(1-{\alpha_3\over\alpha_4}\right) \right] \cr
&+\ 2\alpha_3\left[  \Li_2\left(1-{\alpha_2\over\alpha_3}\right)
      +  \Li_2\left(1-{\alpha_4\over\alpha_3}\right) \right]
+\ {\alpha_1-\alpha_2+(1-\msq)\alpha_3-\alpha_4+\alpha_5 \over 1-\msq}
  \Li_2\left(1-\msq\right) \cr
&+\ 2 {\alpha_1-\msq\alpha_2\over 1-\msq}
 \left[  \Li_2\left(1-{\alpha_2\over\alpha_1}\right)
       - \Li_2\left(1-{\alpha_2\over\alpha_1}\msq\right) \right]
+\ 2 {\alpha_5-\msq\alpha_4\over 1-\msq}
 \left[  \Li_2\left(1-{\alpha_4\over\alpha_5}\right)
       - \Li_2\left(1-{\alpha_4\over\alpha_5}\msq\right) \right] \cr
&-\ {\pi^2\over3}\,\alpha_3 \Biggr\}\ +\ \Ord(\e),\cr }
\eqn\onemasspent
$$
or in terms of more conventional kinematic variables
$$
\hskip -.7 cm
\eqalign{
  I_5^{\rm 1m}&[1]
  = - {\rg \over s_{12} s_{23} s_{34} s_{45} s_{51}} \Biggl\{
 {1\over\e^2}\biggl[
  {(-s_{34})^{1+\e} (-s_{45})^{1+\e} \over
      (-s_{51})^\e (-s_{12})^\e (-s_{23})^\e}
  + {(-s_{45})^{1+\e} (-s_{51})^{1+\e} \over
      (-s_{12})^\e (-s_{23})^\e (-s_{34})^\e}
 + {(-s_{51})^{1+\e} (-s_{12})^{1+\e} \over
         (-s_{23})^\e (-s_{34})^\e (-s_{45})^\e } \cr
& \null\hskip 1 cm
+ {s_{45} s_{51}  \over s_{45} s_{51} - m_5^2 s_{23} }
\biggl[ s_{23} s_{34} \left(1 - {m_5^2 \over s_{51}} \right)
{ (- s_{23})^\e (-s_{34})^\e \over (-s_{45})^\e
(-s_{51})^\e (-s_{12})^\e } \cr
& \hskip 2 cm +\ s_{12} s_{23}\left(1- {m_5^2 \over s_{45} }\right)
 {(-s_{12})^\e  (-s_{23})^\e \over
(-s_{34})^\e (-s_{45})^\e (-s_{51})^\e} \biggr]
\biggl( 1 - \Bigl( {m_5^2 s_{23} \over s_{45} s_{51} }\Bigr)^{-\e}
\biggr) \biggr]  \cr
&+\ 2 s_{34} s_{45} \left[ \Li_2\left(1-{s_{23}\over s_{45}}\right)
      +  \Li_2\left(1-{s_{51}\over s_{34}}\right) \right]
+\ 2 s_{51} s_{12} \left[  \Li_2\left(1-{s_{23}\over s_{51}}\right)
      +  \Li_2\left(1-{s_{45}\over s_{12}}\right) \right] \cr
&+\ 2 s_{45} s_{51} \left[  \Li_2\left(1-{s_{34}\over s_{51}}\right)
      +  \Li_2\left(1-{s_{12}\over s_{45}}\right) \right] \cr
&+\ s_{45} s_{51}
{s_{23} s_{34}- s_{34} s_{45} + s_{45}s_{51} -m_5^2 s_{23} - s_{51} s_{12}
+ s_{12}s_{23} \over s_{45} s_{51} - m_5^2 s_{23} }
  \Li_2\left(1- {m_5^2  s_{23} \over s_{45} s_{51}} \right) \cr
&+\ 2 {(s_{51} - m_5^2) s_{23} s_{34} s_{45} \over
                  s_{45}s_{51} - s_{23} m_5^2 }
 \left[  \Li_2\left(1-{s_{45}\over s_{23} }\right)
       - \Li_2\left(1-{m_5^2 \over s_{51} }\right) \right] \cr
& +\ 2 {(s_{45} - m_5^2) s_{51} s_{12} s_{23} \over s_{45} s_{51} -
       s_{23} m_5^2}
 \left[  \Li_2\left(1-{s_{51}\over s_{23}}\right)
       - \Li_2\left(1-{m_5^2 \over s_{45}}\right) \right]
       - {\pi^2\over3}\, s_{45} s_{51}  \Biggr\}\ +\ \Ord(\e).\cr }
\eqn\onemasspentsvar
$$
Observe that it has the expected symmetry under flipping external legs
$1\leftrightarrow 4$ and $2 \leftrightarrow 3$.  The limit of the
expression~(\onemasspentsvar) as $m_5\to0$ does not yield the massless
pentagon integral~(\UnreducedPentagon), for similar reasons as
explained at the end of section~\use\PDESection\ for box integrals.
The single mass pentagon $I_5^{{\rm 1m}}$, which is given through
$\Ord(\e^0)$, should not be expected to have a smooth limit onto the
massless pentagon as $m_5\to0$, because $I_5^{{\rm 1m}}$ incorporates
the box integral $I_4^{2{\rm m}h}$, and through it the triangle
integral $I_3^{3{\rm m}}$ which does not have a smooth limit.

 
\section{Feynman Parameters in the Numerator}
\tagsection\GeneratingFunctionSection
 
In this section, we explain how to use the
scalar pentagon $\Basic_5$, when expressed in terms of box integrals and
$\Basic_5^{D=6-2\e}$ via equation~(\use\PreGeneratingFunction),
as a generating function for the tensor integrals $\Basic_5[P(a_i)]$
through $\Ord(\e^0)$.
The general discussion applies to the pentagon integral with
any number of external (or internal) masses;
we shall also give explicit formul\ae\ for the massless pentagon
at the end of the section.
 
The only complication in applying the differentiation
formula~(\use\BasicPentagon) is the appearance of $\Basic_5^{D=6-2\e}$
and its derivatives at $\Ord(\e^0)$ when the degree of $P(a_i)$ is
two or higher.
It is easy to eliminate the derivatives of $\Basic_5^{D=6-2\e}$ in favor of
$\Basic_5^{D=6-2\e}$ itself and the $D=6-2\e$ scalar box integrals
$\Basic_4^{D=6-2\e\ (j)}$.
We just let $\e\to\e-1$ in equation~(\use\NPointPDEs), whence
$$
  {\partial\Basic_5^{D=6-2\e}\over\partial\al{i}}\ =\ (-1+2\eps)
   \LB \sum_{j=1}^5 {1\over2N_5}
     \L \eta_{ij} - {\ga{i}\ga{j}\over\Det_5} \R\,
    \Basic_4^{D=6-2\e\ (j)}
 \ +\ {\ga{i}\over\Det_5}\,\Basic_5^{D=6-2\e}\RB\ .
\eqn\PentagonSixDerivatives
$$
Since each term in this equation is nonsingular as $\e\to0$,
and since we need $\Basic_5^{D=6-2\e}$ only to $\Ord(\e^0)$,
we can set $\e=0$ in $\Basic_5^{D=6-2\e}$ and
$\Basic_4^{D=6-2\e\ (j)}$, and use in place
of~(\use\PentagonSixDerivatives) the slightly simpler equation
$$
  {\partial\Basic_5^{D=6}\over\partial\al{i}}\ =\ -
  \sum_j {1\over2N_5} \L \eta_{ij} - {\ga{i}\ga{j}\over\Det_5} \R\,
    \Basic_4^{D=6\ (j)}
   \ -\ {\ga{i}\over\Det_5}\,\Basic_5^{D=6}\ .
\eqn\NewPentagonSixDerivatives
$$
 
The $D=6$ scalar box integrals can be worked out directly,
or they can be determined from the $D=4-2\e$ box integrals and
triangle integrals, using equation~(\use\reducedInamixedeqn) with $n=4$.
For the box with one external mass, needed for the massless pentagon,
the explicit result is
$$
  \Basic_4^{D=6\ (j)}\ =\ -{4\,L_j \over \Det_5-\ga{j}^2}\ ,
\anoneqn$$
where $L_j$ is defined in equation~(\use\LDef).
 
Having eliminated its derivatives, we still have to deal with
the appearance of $\Basic_5^{D=6}$
itself in the integrals $\Basic_5[P(a_i)]$, for $m\geq2$.
The way to proceed is suggested by an argument due to Ellis, Giele and
Yehudai~[\ref\Private{R. K.\ Ellis, W. T. Giele and E. Yehudai,
private communication}].
They work in terms of loop-momentum integrals
directly, and use the Brown-Feynman or Passarino-Veltman procedure to
solve for the tensor pentagon integrals in terms of lower-order tensor
integrals (pentagons and boxes), all evaluated in $D=4-2\e$.
The quantity $\Basic_5^{D=6}$ does not appear at $\Ord(\e^0)$
in any momentum-space tensor integral.
This fact suggests that in our approach,
$\Basic_5^{D=6}$ will cancel out of the integral of any
Feynman parameter polynomial that
is the Feynman parametrization of some tensor integral in
momentum-space.
In appendix~\use\DFiveDrop,
we show explicitly that this is indeed true for integrals with up to
three loop-momenta inserted.  It is straightforward to
extend the argument to five loop-momenta, the maximum number
encountered in any gauge theory amplitude.  (Beyond five insertions of
the loop momentum, ultraviolet divergences of the integrals
complicate matters.)
 
While $\Basic_5^{D=6}$ disappears from the final answer in any
gauge theory calculation, it is still useful to know
with what coefficient it appears in any particular term.
One may use the vanishing of its coefficient in
the final expression as a check on the complete calculation.
Also, it is simple to write recursive formul\ae\ for the
integrals of monomials in the Feynman parameters
using this information.
Let us work out the coefficient of
$\Basic_5^{D=6}$ in $\Basic_5[a_{i_1}\ldots a_{i_m}]$.
Define $d_{i_1\ldots i_m}$ by
$$
 \Basic_5[a_{i_1}\ldots a_{i_m}]\ \equiv\
 { d_{i_1\ldots i_m}\over N_5 \, \Det_5^{m-1} }\ \Basic_5^{D=6}
  \ +\ \ldots,
\anoneqn$$
where `$\ldots$' denotes scalar box integrals (in $D=4-2\e$ and
in $D=6$) and their derivatives.
Notice from equations~(\use\NewPentagonSixDerivatives),
(\use\etadefn), and~(\use\gammadefn)
 that  $\sqrt{\Det_5}\Basic_5^{D=6}$ satisfies
a simple equation,
$$
{\partial\L\sqrt{\Det_5}\Basic_5^{D=6}\R\over\partial\al{i}}
\ =\ -{1\over2N_5} \sum_{j=1}^5 \L \eta_{ij}-{\ga{i}\ga{j}\over\Det_5}\R\,
       \sqrt{\Det_5}\,\Basic_4^{D=6\ (j)}\ .
\anoneqn
$$
Now write the term $(\eps/N_5)\,\Det_5\,\Basic_5^{D=6}$ in
equation~(\use\PreGeneratingFunction) as
$$
  {\eps\over N_5} \, \sqrt{\Det_5} \times
  \Bigl( \sqrt{\Det_5}\,\Basic_5^{D=6} \Bigr)\ ,
\anoneqn
$$
and apply the differentiation formula~(\use\BasicPentagon) to get
$$
 d_{i_1 \ldots i_m}\ =\
  {(-1)^m \, \Det_5^{m-1/2} \over 2(m-2)!}
   \ {\del^m \Det_5^{1/2} \over \del\al{i_1} \ldots \del\al{i_m} }
  \ .\qquad  m\geq2.
\eqn\dcoeffGeneral
$$
We have taken the limit $\e\to 0$ in the $\Gamma$-function
prefactor in~(\use\BasicPentagon), since we are working only
to $O(\e^0)$.
Carrying out the differentiations explicitly for the cases of
interest, $m=2,3,4,5$, we get
$$
\eqalign{
  d_{ij}\ &=\ \hf\left[ \eta_{ij}\,\Det_5 - \gamma_i\gamma_j \right]\ ,\cr
  d_{ijk}\ &=\ \hf\left[
  (\eta_{ij}\,\gamma_k + \eta_{jk}\,\gamma_i + \eta_{ki}\,\gamma_j) \Det_5
  - 3\,\gamma_i\gamma_j\gamma_k \right]\ ,\cr
  d_{ijkl}\ &=\ -\quarter\Bigl[
  (\eta_{ij} \eta_{kl} + \eta_{ik} \eta_{jl} + \eta_{il} \eta_{jk}) \Det_5^2\cr
   &\hskip 10mm
  - 3 (\eta_{ij}\gamma_k\gamma_l + \eta_{ik}\gamma_j\gamma_l
     + \eta_{il}\gamma_j\gamma_k + \eta_{jk}\gamma_i\gamma_l
     + \eta_{jl}\gamma_i\gamma_k + \eta_{kl}\gamma_i\gamma_j) \Det_5
        + 15\,\gamma_i\gamma_j\gamma_k\gamma_l \Bigr]\ ,\cr
  d_{ijklm}\ &=\ -\quarter\Bigl[
   \bigl( \eta_{ij} \eta_{kl}\gamma_m\ +\ {\rm perms\ of}\ ijklm\
   (15\ {\rm terms}) \bigr) \Det_5^2\cr
  &\hskip 10mm
  - 5 \bigl( \eta_{ij}\gamma_k\gamma_l\gamma_m\ +\ {\rm perms\ of}\ ijklm\
   (10\ {\rm terms}) \bigr) \Det_5
  + 35\,\gamma_i\gamma_j\gamma_k\gamma_l\gamma_m \Bigr]\ .\cr
}\eqn\dcoeffSpecial
$$
 
In some calculational schemes for gauge theory amplitudes,
the $D=6-2\eps$ pentagon integral will itself appear with
Feynman parameter polynomials of degree $m\leq3$ inserted.
It is then useful to know the coefficient
$d^{D=6}_{i_1\ldots i_m}$ defined by
$$
 \Basic_5^{D=6}[a_{i_1}\ldots a_{i_m}]\ \equiv\
 { d^{D=6}_{i_1\ldots i_m}\over \Det_5^{m} }\ \Basic_5^{D=6}
  \ +\ \ldots,
\anoneqn$$
where again `$\ldots$' denotes scalar box integrals
(in $D=6$) and their derivatives.
Writing
$\Basic_5^{D=6} = \Det_5^{-1/2}\times (\Det_5^{1/2}\Basic_5^{D=6})$
and repeating the above steps we find
$$
 d^{D=6}_{i_1 \ldots i_m}\ =\
  {(-1)^m \, \Det_5^{m+1/2} \over m!}
   \ {\del^m \Det_5^{-1/2} \over \del\al{i_1} \ldots \del\al{i_m} }
  \ .\qquad m\geq2.
\eqn\dsixcoeffGeneral
$$
The explicit values for the cases of interest are
$$
\eqalign{
  d_i\ &=\ \gamma_i\ ,\cr
  d_{ij}\ &=\ -\hf\left[ \eta_{ij}\,\Det_5 - 3\gamma_i\gamma_j \right]\ ,\cr
  d_{ijk}\ &=\ -\hf\left[
  (\eta_{ij}\,\gamma_k + \eta_{jk}\,\gamma_i + \eta_{ki}\,\gamma_j) \Det_5
  - 5\,\gamma_i\gamma_j\gamma_k \right]\ .\cr
}\eqn\dsixcoeffSpecial
$$
 
Now we shall give explicit formul\ae\ for the massless
pentagon integrals with
up to two Feynman parameters inserted, along with a simple recursion
relation for generating the remainder of the integrals.
 
For a single parameter insertion, equation~(\use\BasicPentagon) gives
$$
\Basic_5\Poly{a_i}\ =\ {1\over 1+2\eps} {\del\Basic_5\over\del\al{i}}
\ .
\anoneqn
$$
Thus we may differentiate the $\Ord(\e^0)$
expression~(\use\PentagonSolution) for $\Basic_5$, using
also equation~(\use\dilogdiff), to get
$$
\Basic_5\Poly{a_i}\ =\ \rg \, \LB {\al{i}^{2\e}\over \eps^2}
  \ +\ 2\, \Li_2\L 1-{\al{i+1}\over \al{i}} \R
  \ +\ 2\, \Li_2\L 1-{\al{i-1}\over \al{i}} \R
           -{\pi^2 \over 6} \RB\ +\ \Ord(\e).
\eqn\OneParamAnswer
$$

In the case of two Feynman parameter insertions, we have
$$
\Basic_5\Poly{a_i a_j}\ =\ {1\over 2\eps(1+ 2\eps)}\,
{\partial^2 \Basic_5\over\partial\al{i}\partial\al{j}}\;,
\eqn\TwoParameters
$$
which must now be applied to the
expression~(\use\PreGeneratingFunction) for $\Basic_5$.
The $\Ord(1)$ terms in $\Basic_5\Poly{a_i a_j}$
receive contributions both from $\Basic_5^{D=6}$
and from the $\Ord(\e)$ terms in the box integrals
$\Basic_4^{(j)}$.
Since $\ga{i}$ is linear in the $\al{i}$, only {\it derivatives}
of the $\Basic_4^{(j)}$ appear on the right-hand side
of~(\use\TwoParameters).
The derivatives $\del\Basic_4^{(j)}/\del\al{i}$ at $\Ord(\e)$
are nothing but the single insertions $\Basic_4^{(j)}[a_i]$
at $\Ord(1)$, thanks to the box differentiation formula
$$
\Basic_4^{(j)}\Poly{a_i}\ =\ {1\over 2\eps}\,
{\del\Basic_4^{(j)}\over\del\al{i}}\;.
\eqn\BoxOneParameter
$$
These integrals are tabulated in equation~(\use\MassBoxOneA).
Carrying out the differentiations in~(\use\TwoParameters), we find
$$
\eqalign{
  \Basic_5[a_ia_j]\ &=\ \rg\Biggl\{ {\delta_{i,j}\over\e^2}
    \biggl( { \alpha_{i+1}^\e \alpha_{i-1}^\e \over \alpha_i }
  + { \alpha_{i-2}^\e (\alpha_i^\e - \alpha_{i+1}^\e)
            \over \alpha_i - \alpha_{i+1} }
  + { \alpha_{i+2}^\e (\alpha_i^\e - \alpha_{i-1}^\e)
            \over \alpha_i - \alpha_{i-1} } \biggr) \cr
 &\qquad\qquad  -\ {\delta_{i+1,j}\,\alpha_{i-2}^\e
      +\delta_{j+1,i}\,\alpha_{j-2}^\e \over\e^2}
     { \alpha_i^\e - \alpha_j^\e \over \alpha_i - \alpha_j } \cr
 &\qquad\qquad +\ \sum_{k=1}^5 \biggl[
   \eta_{ik} \gamma_j\ +\ \eta_{jk} \gamma_i\ -\ \eta_{ik} \eta_{jk} \gamma_k
   \ -\ {\gamma_i\gamma_j\gamma_k \over \Det_5} \biggr]
   {L_k \over \Det_5-\gamma_k^2}
 \ +\ {d_{ij} \over \Det_5}\ \Basic_5^{D=6} \Biggr\}\ +\ \Ord(\eps).\cr}
\eqn\TwoParamAnswer
$$
For a generalization of this formula to arbitrary pentagon kinematics,
and also to hexagon ($n=6$) integrals, see
equation~(\use\twoparam) in appendix~\use\FPinHigherPointIntegrals.
 
\def\phys{{\rm phys}}
For more than two parameters inserted, we can proceed recursively.
Define some new quantities
$\Basic_5^{\phys}[a_{i_1}\ldots a_{i_m}]$ to be
the ``non-$\Basic_5^{D=6}$'' terms in
$\Basic_5[a_{i_1}\ldots a_{i_m}]$, i.e.
$$
   \Basic_5[a_{i_1}\ldots a_{i_m}]\ \equiv\
   \Basic_5^{\phys}[a_{i_1}\ldots a_{i_m}]\ +\
   {d_{i_1 \ldots i_m} \over \Det_5^{m-1}}\ \Basic_5^{D=6}\ .
\eqn\NonSixDefinition
$$
Then the differentiation formula~(\use\BasicPentagon)
along with~(\use\NewPentagonSixDerivatives) generates the following
recursion relation for $\Basic_5^{\phys}[a_{i_1}\ldots a_{i_m}]$:
$$
  \Basic_5^{\phys}[a_{i_1}\ldots a_{i_m}]
  \ =\ \left({-1\over m-2-2\e}\right)
   \left[ {\del\Basic_5^{\phys}[a_{i_1}\ldots a_{i_{m-1}}]
                 \over \del\alpha_{i_m}}
   \ +\ 4\, {d_{i_1\cdots i_{m-1}} \over \Det_5^{m-1}}
      \sum_{j=1}^5 { d_{i_m j} \, L_j \over \Det_5-\ga{j}^2} \right]\ .
\anoneqn
$$
In applying this formula, it is convenient to have a
differentiation formula for the $L_j$,
in terms of logarithms:
$$\eqalign{
  {\del L_j\over\del\alpha_i}\ &=\
 {1\over\al{i}}\left( \delta_{i,j+1}- \delta_{i,j+2} \right)
      \left[ -{ \alpha_{j+1}\ln(\alpha_{j+1}/\alpha_{j+2}) \over
         \alpha_{j+1}-\alpha_{j+2} }
           + \ln(\alpha_{j-1}/\alpha_{j-2}) \right] \cr
   &+\ {1\over\al{i}}\left( \delta_{i,j-1}- \delta_{i,j-2} \right)
      \left[ -{ \alpha_{j-1}\ln(\alpha_{j-1}/\alpha_{j-2}) \over
         \alpha_{j-1}-\alpha_{j-2} }
           + \ln(\alpha_{j+1}/\alpha_{j+2}) \right]\ . \cr
}\anoneqn
$$
 
This completes our prescription for evaluating massless pentagon
integrals with Feynman parameters inserted, in terms of dilogarithms
and logarithms.  The same basic procedure also works when external
and/or internal masses are present, provided that the relevant box
and triangle integrals are known through $\Ord(\eps^0)$.
(The triangles appear through equation~(\use\BoxOneParameter)
in combination with~(\reducedInaeqn).)
If all internal lines are massless, then all the requisite boxes and
triangles can be found in section~\use\PDESection,
except for the three-mass triangle.  This triangle may be computed in
$D=4$; see for example refs.~[\HVintegrals,\HungJung].
In appendix~\use\AllMassTriangle\ it is computed in $D=4-2\eps$ for
arbitrary $\eps$, as a further illustration of the partial differential
equation approach to scalar integrals.
 
 
\vskip .3 cm
\line{{\bf Acknowledgements}\hfill}
\vskip .2 cm
 
We thank R. K. Ellis and W. T. Giele for discussions, especially
regarding the cancellation of the six-dimensional pentagon from
all physical expressions.  We thank J. A. M. Vermaseren
and Z. Kunszt for comments on the manuscript, and J. A. M. Vermaseren
for other useful comments.
 
 
\appendix{Collection of Massless Pentagon and Scalar Box Results}
\tagappendix\CollectionAppendix
 
In this appendix we collect those results that are useful in an
explicit calculation.
The massless pentagon integral of interest is
$$
\Int_5 [P_m(\{ a_i\})] \ =\ \Gamma(3+\e)\int_0^1 d^5a_i\;
     {\delta\L 1 - \sum_i a_i \R P_m(\{ a_i\}) \over
         \LB - s_{12} a_1 a_3 - s_{23} a_2 a_4 - s_{34} a_3 a_5
- s_{45} a_4 a_1 - s_{51} a_5 a_2  - i\varepsilon\RB^{3+\eps}}\ ,
\anoneqn
$$
where $P_m(\{a_i\})$ is a polynomial in the $a_i$ of degree $m$.
For use in differentiation formulae we define the reduced integrals
$$
\Basic_n\Poly{\hat\generalPoly}\ =\ \biggl(\prod_{j=1}^n \al{j}\biggr)^{-1}
   \Int_n\Poly{P(\{a_i/\al{i}\})}\ .
\eqn\ReducedIntegralApp
$$
where $s_{i, i+1} = -1/(\alpha_i \alpha_{i+1})$ mod 5.
The basic differentiation formula for the pentagon is given by
$$
\Basic_5\Poly{\hat P_m(\{a_i\})}\ =\ {\Gamma(2-m+2\eps)\over\Gamma(2+2\eps)}
         \, P_m\L\left\{ \p{i} \right\}\R \,\Basic_5\Poly{1}\ .
\eqn\BasicPentagonApp
$$
 
Through $\Ord(1)$ the scalar pentagon is given by
$$
  \Basic_5[1]\ =\ \rg
  \sum_{j=1}^5 \alpha_j^{1+2\e} \biggl[ {1\over \eps^2}
  + 2\Li_2\Bigl(1 - {\alpha_{j+1} \over\alpha_j}\Bigr)
  + 2\Li_2\Bigl(1- {\alpha_{j-1}\over\alpha_j} \Bigr)
  -{\pi^2\over 6} \biggr]\ +\ \Ord(\eps) \; .
\eqn\PentagonSolutionApp
$$
where $\rg\ \equiv\ \Gamma(1+\eps)\Gamma^2(1-\eps)/\Gamma(1-2\eps)$.
The case of one Feynman parameter in the numerator may be obtained by
directly applying the differentiation formula (\use\BasicPentagonApp).
 
Beyond one Feynman parameter it is best to use the explicit value of the
two Feynman parameter integral as a generating function for integrals
with three or more Feynman parameters in the numerator.  The two
parameter integral is given by
$$
\eqalign{
  \Basic_5[a_ia_j]\ &=\ \rg\Biggl\{ {\delta_{i,j}\over\e^2}
    \biggl( { \alpha_{i+1}^\e \alpha_{i-1}^\e \over \alpha_i }
  + { \alpha_{i-2}^\e (\alpha_i^\e - \alpha_{i+1}^\e)
            \over \alpha_i - \alpha_{i+1} }
  + { \alpha_{i+2}^\e (\alpha_i^\e - \alpha_{i-1}^\e)
            \over \alpha_i - \alpha_{i-1} } \biggr) \cr
 &\qquad  -\ {\delta_{i+1,j}\,\alpha_{i-2}^\e
      +\delta_{j+1,i}\,\alpha_{j-2}^\e \over\e^2}
     { \alpha_i^\e - \alpha_j^\e \over \alpha_i - \alpha_j } \cr
 &\qquad +\ \sum_{k=1}^5 \biggl[
   \eta_{ik} \gamma_j\ +\ \eta_{jk} \gamma_i\ -\ \eta_{ik} \eta_{jk} \gamma_k
   \ -\ {\gamma_i\gamma_j\gamma_k \over \Det_5} \biggr]
   {L_k \over \Det_5-\gamma_k^2}
 \ +\ {d_{ij} \over \Det_5}\ \Basic_5^{D=6} \Biggr\}\ +\ \Ord(\eps),\cr}
\eqn\TwoParamAnswerApp
$$
where
$$
\eqalign{
&\Det_5\ \equiv\ \sum_{j=1}^5
   \bigl( \al{j}^2 - 2\al{j}\al{j+1} + 2\al{j}\al{j+2} \bigr) \cr
& \gamma_{i}\ \equiv\ {1\over2}{\partial\Det_5\over\partial\al{i}}
\ =\ \al{i-2} - \al{i-1} + \al{i} - \al{i+1} + \al{i+2}\;, \cr
&\eta_{ij}\ \equiv\ {\partial\gamma_{i}\over\partial\al{j}}
  \ =\ {1\over2}{\partial^2\Det_5\over\partial\al{i}\partial\al{j}}
  \ =\ \cases{     -1,  &$i=j\pm1$, \cr
                   +1,  &otherwise, \cr}  \cr
& d_{ij}\ =\ \hf\left[ \eta_{ij}\,\Det_5 - \gamma_i\gamma_j \right] \cr}
\eqn\DefinitionsApp
$$
and
$$
 L_i\ \equiv\ \Li_2\L 1-{\al{i+1}\over\al{i+2}} \R
              + \Li_2\L 1-{\al{i-1}\over\al{i-2}} \R
  + \ln\L{\al{i+1}\over\al{i+2}}\R \ln\L{\al{i-1}\over\al{i-2}}\R
                  -{\pi^2 \over 6}\ .
\eqn\LdefApp
$$
 
In calculations there is no need to know the explicit value of the
six-dimensional pentagon $\Basic_5^{D=6}$ since it cancels from all
quantities arising from loop momentum integrals. However, when applying
the differentiation formula (\use\BasicPentagonApp) the
terms containing $\Basic_5^{D=6}$
cannot be dropped since they generate $D=6$ box integrals via the
equation
$$
  {\partial\Basic_5^{D=6}\over\partial\al{i}}\ =\ -
  \sum_{j=1}^5 {1\over2} \L \eta_{ij} - {\gamma_{i}\gamma_{j}\over\Det_5} \R\,
    \Basic_4^{D=6\ (j)}
   \ -\ {\gamma_{i}\over\Det_5}\,\Basic_5^{D=6}\ .
\eqn\NewPentagonSixDerivativesApp
$$
Useful formul\ae\ when applying the differentiation
formula~(\BasicPentagonApp) are
$$
\eqalign{
  {\del L_j\over\del\alpha_i}\ &=\
 {1\over\alpha_i} \left( \delta_{i,j+1} - \delta_{i,j+2} \right)
      \left[ -{ \alpha_{j+1}\ln(\alpha_{j+1}/\alpha_{j+2}) \over
         \alpha_{j+1}-\alpha_{j+2} }
           + \ln(\alpha_{j-1}/\alpha_{j-2}) \right] \cr
   &+\ {1\over\alpha_i} \left( \delta_{i,j-1} - \delta_{i,j-2} \right)
      \left[ -{ \alpha_{j-1}\ln(\alpha_{j-1}/\alpha_{j-2}) \over
         \alpha_{j-1}-\alpha_{j-2} }
           + \ln(\alpha_{j+1}/\alpha_{j+2}) \right]\ . \cr}
\anoneqn
$$
and
$$
  \Basic_4^{D=6\ (j)}\ =\ -{4\,L_j \over \Det_5-\ga{j}^2}\ ,
\anoneqn$$
where $L_j$ is defined in equation~(\LdefApp).
 
We collect here the dimensionally-regulated scalar box integrals with
massless internal lines, but 0, 1, 2 or 3 nonzero external masses,
which appear in the process of evaluating ($n\geq5$)-point integrals,
and in subdiagrams in QCD loop calculations.  The integrals are defined
through equations~(\GeneralForm) and (\FourDenominator).
$$
  I_4^{0{\rm m}}(s,t)\ =\ {\rg \over st} \biggl\{
  {2\over\e^2} \Bigl[ (-s)^{-\e} + (-t)^{-\e} \Bigr]
  \ -\ \ln^2\left({s\over t}\right)\ -\ \pi^2 \biggr\}\ +\ \Ord(\e),
\eqn\newzeromass
$$
$$
\eqalign{
  I_4^{1{\rm m}}(s,t,m_4^2)\ &=\ {\rg \over st} \biggl\{
  {2\over\e^2} \Bigl[ (-s)^{-\e} + (-t)^{-\e} - (-m_4^2)^{-\e} \Bigr] \cr
 &\ -\ 2\ \Li_2\left(1-{m_4^2\over s}\right)
  \ -\ 2\ \Li_2\left(1-{m_4^2\over t}\right)
  \ -\ \ln^2\left({s\over t}\right)\ -\ {\pi^2\over3} \biggr\}
  \ +\ \Ord(\e), \cr}
\eqn\newonemass
$$
$$
\eqalign{
  I_4^{2{\rm m}e}(s,t,m_2^2,m_4^2)\ &=\ {\rg \over st-m_2^2m_4^2} \biggl\{
  {2\over\e^2} \Bigl[ (-s)^{-\e} + (-t)^{-\e}
              - (-m_2^2)^{-\e} - (-m_4^2)^{-\e} \Bigr] \cr
  &\ -\ 2\ \Li_2\left(1-{m_2^2\over s}\right)
   \ -\ 2\ \Li_2\left(1-{m_2^2\over t}\right)
   \ -\ 2\ \Li_2\left(1-{m_4^2\over s}\right)
   \ -\ 2\ \Li_2\left(1-{m_4^2\over t}\right) \cr
  &\ +\ 2\ \Li_2\left(1-{m_2^2m_4^2\over st}\right)
   \ - \ \ln^2\left({s\over t}\right) \biggr\}\ +\ \Ord(\e), \cr }
\eqn\neweasytwomass
$$
$$
\eqalign{
  I_4^{2{\rm m}h}(s,t,m_3^2,m_4^2)\ &=\ {\rg \over st} \biggl\{
  {2\over\e^2} \Bigl[ (-s)^{-\e} + (-t)^{-\e}
              - (-m_3^2)^{-\e} - (-m_4^2)^{-\e} \Bigr]
  \ +\ {1\over\e^2}
    { (-m_3^2)^{-\e}(-m_4^2)^{-\e} \over (-s)^{-\e} } \cr
 &\ -\ 2\ \Li_2\left(1-{m_3^2\over t}\right)
  \ -\ 2\ \Li_2\left(1-{m_4^2\over t}\right)
  \ -\ \ln^2\left({s\over t}\right)\biggr\}\ +\ \Ord(\e), \cr}
\eqn\newhardtwomass
$$
$$
\eqalign{
  I_4^{3{\rm m}}(s,t,m_i^2)\ &=\ {\rg \over st-m_2^2m_4^2} \biggl\{
  {2\over\e^2} \Bigl[ (-s)^{-\e} + (-t)^{-\e}
     - (-m_2^2)^{-\e}- (-m_3^2)^{-\e} - (-m_4^2)^{-\e} \Bigr] \cr
  &  \ +\ {1\over\e^2}
    { (-m_2^2)^{-\e}(-m_3^2)^{-\e} \over (-t)^{-\e} }
  \ +\ {1\over\e^2}
     { (-m_3^2)^{-\e}(-m_4^2)^{-\e} \over (-s)^{-\e} } \cr
  &\ -\ 2\ \Li_2\left(1-{m_2^2\over s}\right)
   \ -\ 2\ \Li_2\left(1-{m_4^2\over t}\right) \cr
  &\ +\ 2\ \Li_2\left(1-{m_2^2m_4^2\over st}\right)
      \ -\ \ln^2\left({s\over t}\right) \biggr\}\ +\ \Ord(\e). \cr }
\eqn\newthreemass
$$
 
 
\appendix{Connection with the Work of van Neerven and Vermaseren}
\tagappendix\VNVAppendix
 
\def\m#1{\hat m^2_{#1}}

Melrose~[\use\Melrose] and Van Neerven and Vermaseren~[\use\vNV]
were able to represent the general scalar pentagon integral in $D=4$
as a sum of five $D=4$ box integrals.
On the other hand, equation~(\use\reducedallNeqn)
expresses the pentagon integral in $D=4-2\e$
as a linear combination of five box integrals (also in $D=4-2\e$),
plus the pentagon in $D=6-2\e$ dimensions, so it can be thought
of as the dimensionally-regulated version of the equations in
refs.~[\use\Melrose,\use\vNV].
Indeed, the $D=4$ equation in ref.~[\use\vNV] was our motivation to find
an algebraic $D=4-2\e$ equation.
(Similar relations have recently been found using
momentum-space, rather than Feynman parameter, techniques by
Ellis, Giele and Yehudai~[\use\EGY].)
We would like to verify that the $D=4-2\e$ and $D=4$ equations
are consistent with each other, or in other words that the
($D$-independent) coefficients of the box integrals in
equation~(\use\reducedNeqfiveeqn) are equal to the corresponding
coefficients in refs.~[\use\Melrose,\use\vNV]
(up to normalization conventions for the integrals).
To do this, it is simplest to rewrite equation~(\use\reducedNeqfiveeqn)
in terms of unreduced integrals as
$$
  I_5[1]\ =\ {1\over2} \sum_{i=1}^5 c_i
       \ I_4^{(i)}[1]\ +\ \Ord(\e)
\eqn\unredpent
$$
where
$$
  c_i\ =\ {\alpha_i\gamma_i\over N_5}
     \ =\ \sum_{j=1}^5 S^{-1}_{ij}\ .
\eqn\ourci
$$
The second form of the $c_i$, in terms of more conventional kinematic
variables (the matrix $S$ is defined in~(\use\SDefinition)),
is the form in which the $c_i$ were obtained in ref.~[\OurAllN].
In this form the $c_i$ are manifestly the same as those found by
Melrose.
 
The coefficients found by van Neerven and
Vermaseren involve the $D=4$ Levi-Civita tensor, and are not manifestly
equal to~(\use\ourci).  Expressed in our notation, they are given by
$$
  c_1\ =\ -{ 4\Delta_5 - 2 \sum_{i=1}^4 v_i\cdot w
                  \over w^2-4\Delta_5 M_1^2 }\ ,
  \qquad\quad c_{i+1}\ =\ -{2 v_i\cdot w \over w^2-4\Delta_5 M_1^2 }\ ,
  \qquad i=1,2,3,4.
\eqn\cidef
$$
Here the ``axial vectors'' $v_i$ are the $D=4$ duals
of the vectors $p_i$ appearing in the momentum-space version of the
pentagon integral 
$$
\eqalign{
  v_1^\mu\ \equiv\ \pol^{\mu p_2p_3p_4},\quad
  v_2^\mu\ &\equiv\ \pol^{p_1\mu p_3p_4},\quad
  v_3^\mu\ \equiv\ \pol^{p_1p_2\mu p_4},\quad
  v_4^\mu\ \equiv\ \pol^{p_1p_2p_3\mu },\quad
  \cr
  p_i^\mu\ &\equiv\ \sum_{j=1}^i k_j^\mu,\qquad p_5^\mu\ =\ 0, \cr}
\eqn\vpdef
$$
where $\pol_{\mu p_2p_3p_4}$ is short for
$\pol_{\mu\mu_2\mu_3\mu_4} p_2^{\mu_2}p_3^{\mu_3}p_4^{\mu_4}$, etc.
The Gram determinant of the vectors $p_i$ is
$\Delta_5\ \equiv\ \pol^{p_1p_2p_3p_4}\pol_{p_1p_2p_3p_4}$,
and $w^\mu$ is defined by
$$
  w^\mu\ \equiv\ \sum_{i=1}^4 r_i v_i^\mu, \qquad\qquad
  r_i\ \equiv\ p_i^2 + M_1^2 - M_{i+1}^2, \quad i=1,2,3,4,
\eqn\wdef
$$
where $M_i$ are the masses on the internal lines.
The definition of the $c_i$ in~(\use\cidef) may seem to be tied
to $D=4$, because of the presence of the axial vectors.
However, the inner products $v_i\cdot v_j$ can be
eliminated in favor of the inverse of the matrix
$t_{ij} \equiv 2p_i\cdot p_j$, according to
$$
  v_i\cdot v_j\ =\ 2\Delta_5 \, (t^{-1})_{ij}\ , \qquad i,j=1,2,3,4.
\eqn\vvreplace
$$
Thus the $c_i$ can be written in a $D$-independent form:
$$
\eqalign{
  c_1\ &=\ { -2\ +\ 2\sum_{i,j=1}^4 (t^{-1})_{ij} \, r_j \over
     \sum_{k,l=1}^4 r_k \, (t^{-1})_{kl} \, r_l\ -\ 2M_1^2 }\ , \cr
  c_{i+1}\ \ &=\ { -2\sum_{j=1}^4 (t^{-1})_{ij} \, r_j \over
     \sum_{k,l=1}^4 r_k \, (t^{-1})_{kl} \, r_l\ -\ 2M_1^2 },
   \qquad i=1,2,3,4. \cr}
\eqn\newcidef
$$
 
To show that the $c_i$ in~(\use\newcidef) agree with those
in~(\use\ourci), it suffices to show that they obey
$$
  \sum_{j=1}^5 S_{ij} c_j\ =\ 1, \qquad i=1,\ldots,5,
\eqn\sceqone
$$
since $S$ is generically invertible for $n=5$.
When internal masses are also present, $S$ is given by
$$
  S_{ij}\ =\ \hf(M_i^2+M_j^2-(p_{i-1}-p_{j-1})^2)
        \ =\ \hf ( 2 M_1^2 - r_{i-1} - r_{j-1} + t_{i-1,j-1})\ ,
\eqn\srteqn
$$
and $r_{j-1} = t_{i-1,j-1} = 0$ for $j=1$,
so that
$$
  \sum_{j=1}^5 S_{ij} c_j\ =\
  {1\over2} ( 2 M_1^2 - r_{i-1}) \biggl(\sum_{j=1}^5 c_j \biggr)
  + {1\over2} \sum_{j=1}^4 (-r_j+t_{i-1,j}) c_{j+1})\ .
\eqn\sceqn
$$
Plugging in the values of $c_i$ from equation~(\use\newcidef), we
get
$$
  \sum_{j=1}^5 S_{ij} c_j\ =\
  { \bigl( -2 M_1^2\ +\ r_{i-1} \bigr)\ +\
 \bigl( \sum_{j,k=1}^4 r_j \, (t^{-1})_{jk} \, r_k\ -\ r_{i-1} \bigr)
   \over \sum_{k,l=1}^4 r_k \, (t^{-1})_{kl} \, r_l\ -\ 2M_1^2 }
   \ =\ 1,
\eqn\newsceqn
$$
as required.
 
In the same fashion, an equation obtained by van Neerven and
Vermaseren, relating hexagon integrals to pentagon integrals,
can be shown to be equivalent to equation~(\use\reducedallNeqn)
for $n=6$ (and $D=4$ external kinematics).

 
\appendix{Proof That  $I_5^{D=6}$ Drops Out}
\tagappendix\DFiveDrop
 
An explanation of why an explicit computation of
$I_5^{D=6}$ is not needed for the evaluation of pentagon integrals
near $D=4$ comes from
the momentum-space representation of tensor integrals;
when performing a Passarino-Veltman decomposition $I_5^{D=6}$ never
appears [\use\Private] and therefore it can be expected to cancel from
amplitudes evaluated using the Feynman parameter techniques discussed
in this paper.   In this appendix, we will demonstrate that
$I_5^{D=6}$ cancels
when summing over contributions which reconstruct the loop momentum integrals
appearing in dimensionally-regulated four-dimensional field theory
amplitudes.   Thus, there is no need to explicitly evaluate $I_5^{D=6}$.
(In this appendix we treat $I_5^{D=6}$ as equivalent to $I_5^{D=6-2\eps}$;
since $I_5^{D=6}$ is completely finite the difference between the two
is of $\Ord(\eps)$.)  The argument holds for general kinematics
(arbitrary external or internal masses), though here we suppress
internal masses.
 
Define the general pentagon integral by
$$
\hskip -.5 cm
\eqalign{
 I_5[P(p^\mu)] \equiv &  i\,(4\pi)^{2-\e} 4!
\int {d^{4-2\e}p \over (2\pi)^{4-2\e} }  \int d^5 a_i
   { \delta\bigl( 1-\sum a_i\bigr) P(p^\mu)
     \over \bigl(a_1 p^2 + a_2 (p-p_1)^2 + a_3 (p-p_2)^2
            + a_4 (p-p_3)^2 + a_5 (p-p_4)^2 \bigr)^5 } \cr }
\eqn\MomentumDef
$$
where $ p_i\ \equiv\ \sum_{j=1}^i k_j $
and $P(p^\mu)$ is some polynomial in the loop momentum $p^\mu$.
The normalization factor in front ensures that the integral, when
Feynman-parameterized, is normalized in the same way as the integrals
$I_5[P(a_i)]$ defined in section~\use\Properties.
 
In order to relate the integral (\use\MomentumDef)
to Feynman-parametrized integrals of the form~(\use\GeneralForm),
we complete the square and integrate out the loop
momentum in the usual fashion.
To complete the square in the denominator, we shift the loop-momentum
variables to
$$
p = q + \sum_{i=1}^4 a_{i+1}p_i \; .
\anoneqn
$$
Integrating out the loop momentum, for up to three powers of
loop momentum in the numerator, then gives
$$
\eqalign{
I_5[p^\mu] & = \sum_{i=1}^4 I_5[a_{i+1}] \, p_i^\mu, \cr
 I_5[p^\mu p^\nu] & =
-{1\over2}I_5^{D=6}[1]\ \delta^{\mu\nu}_{[4-2\e]}
 + \sum_{i,j=1}^4 p_i^\mu p_j^\nu I_5[a_{i+1}a_{j+1}]\ , \cr
I_5[p^\mu p^\nu p^\rho] & =
 -{1\over2}\Bigl(\delta^{\mu\nu}_{[4-2\e]}
\sum_i p_i^\rho I_5^{D=6}[a_{i+1}]
+\delta^{\mu\rho}_{[4-2\e]} \sum_i p_i^\nu I_5^{D=6}[a_{i+1}]
+\delta^{\nu\rho}_{[4-2\e]} \sum_i p_i^\mu I_5^{D=6}[a_{i+1}] \Bigr) \cr
\null & \hskip 1 cm
+ \sum_{ijk} p_i^\mu p_j^\nu p_k^\rho I_5[a_{i+1} a_{j+1} a_{k+1}]\ . \cr}
\eqn\sometensors
$$
Here we will explicitly consider only up to three
loop momenta; the other cases follow similarly.
 
For the case with one loop momentum inserted, since the explicit
value for $I_5[a_{i+1}]$ given in
equation~(\use\OneParamAnswer) does not contain
$I_5^{D=6}$, there is nothing to check.
Beyond this, we have from section~\use\GeneratingFunctionSection\
that the coefficient of
$I_5^{D=6}$ in the explicit value for
$I_5[a_1 \cdots a_k]$ is given by
$$
c_{i_1 i_2 \cdots i_m}\ =\
 {(-1)^{m} \Det_5^{1/2} \over 2 N_5 \, (m-2)! }
\al{i_1} \al{i_2} \cdots \al{i_m}
{\partial^m \Det_5^{1/2}\over
\partial\al{i_1} \partial\al{i_2} \cdots \partial\al{i_m}}
\hskip 1.5 cm (m \ge 2),
\anoneqn
$$
so that for $m=2,3$,
$$
\eqalign{
c_{ij}\ & =\ {\al{i}\al{j} \over 2 N_5 \, \Det_5}
\left[ \eta_{ij}\,\Det_5 - \ga{i}\ga{j} \right]\ ,\cr
c_{ijk}\ &=\ {\al{i}\al{j}\al{k}\over 2 N_5 \, \Det_5^2} \left[
  (\eta_{ij}\,\ga{k} + \eta_{jk}\,\ga{i} + \eta_{ki}\,\ga{j}) \Det_5
  - 3\,\ga{i}\ga{j}\ga{k} \right]\ ,\cr}
\anoneqn
$$
where $\ga{i}$ and $\eta_{ij}$ are defined in
equations~(\use\etadefn) and~(\use\gammadefn).
 
The identity that we will use to show that $I_5^{D=6}$ cancels is
$$
c_{ij}\ =\ {\alpha_i\alpha_j\over 2N_5\, \Det_5}
\left( \eta_{ij}\,\Det_5 - \ga{i}\ga{j} \right)
  \ =\ (t^{-1})_{i-1,j-1}\ , \qquad i,j=2,3,4,5,
\eqn\KeyIdentity
$$
where $t_{ij} = 2 p_i \cdot p_j$.
To verify the identity, we multiply it on the right by $t_{j-1,k-1}$,
which can be written~[\use\OurAllN] in terms of the matrix
$\rho = N_n\eta^{-1}$ using
$$
 p_{i-1} \cdot p_{j-1}\ =\
       {\rho_{ij}\over\alpha_i\alpha_j}
     - {\rho_{i1}\over\alpha_i\alpha_1}
     - {\rho_{1j}\over\alpha_1\alpha_j}
     + {\rho_{11}\over\alpha_1^2}\ ,\qquad\qquad i,j=2,3,\ldots,n.
\eqn\pidotpj
$$
Thus we have (using the equations~(\Nnetarhoeqn)--(\Rident) that relate
$\rho$, $\eta$, $\gamma_i$ and $\alpha_i$)
$$
\eqalign{
  \sum_{j=2}^5 c_{ij} t_{j-1,k-1}
  \ &=\ \sum_{j=1}^5 {\alpha_i \alpha_j\over N_5\,\Det_5}
  \left(\eta_{ij}\Det_5-\gamma_i\gamma_j\right)
   \left( {\rho_{jk}\over\alpha_j\alpha_k}
  - {\rho_{j1}\over\alpha_j\alpha_1}
  - {\rho_{1k}\over\alpha_1\alpha_k}
  + {\rho_{11}\over\alpha_1^2}  \right) \cr
  \ &=\ \delta_{ik}, \qquad i,k=2,3,4,5. \cr}
\eqn\derivekey
$$
Equation~(\use\KeyIdentity) implies that
$$
\sum_{i,j=1}^4 p_i^\mu p_j^\nu c_{i+1,j+1}
\ =\ \sum_{i,j=1}^4  p_i^\mu p_j^\nu (t^{-1})_{i, j}
\ =\ {1\over 2} \delta_{[4]}^{\mu\nu}\ ,
\eqn\newKeyIdentity
$$
since the four vectors $p_i^\mu$ span $D=4$ Minkowski space.
 
Using this identity and keeping only the $I_5^{D=6}$ content
we then have
$$
\eqalign{
  I_5[p_\mu p_\nu]\ &=\
    - {1\over2}I_5^{D=6}[1] \ \delta^{\mu\nu}_{[4-2\e]}
    + \sum_{i,j=1}^4 p_i^\mu p_j^\nu c_{i+1, j+1}
     I_5^{D=6}[1]+ \hbox{boxes} \cr
 \ &=\ -{1\over 2} \delta_{[-2\eps]} I_5^{D=6}[1]
         + \hbox{boxes} + \Ord(\eps)\cr
 \ &=\  \hbox{boxes}  + \Ord(\eps),\cr}
\anoneqn
$$
so that $I_5^{D=6}$ drops out as claimed.  To arrive at the last
line, we used the finiteness
of $I_5^{D=6}$ and that $\delta^{\mu\nu}_{[-2\eps]}$ can yield only
$\Ord(\eps$) contributions. This shows that there are no
`left-over' pieces of $I_5^{D=6}$ remaining when all the pieces
are combined to form an amplitude derived from a loop momentum integral
with up to two powers of momenta in the numerator.
 
The three Feynman parameter case is similar. Again applying the identity
(\use\KeyIdentity) we have
$$
\eqalign{
I_5(p^\mu p^\nu p^\rho)\ &=\
   -{1\over2}\Bigl(\delta^{\mu\nu}_{[4-2\e]} \sum_i p_i^\rho
    I_5^{D=6}[a_{i+1}]+\hbox{cyclic} \Bigr)
   + \sum_{ijk} p_i^\mu p_j^\nu p_k^\rho c_{ijk} I_5^{D=6} [1]
   +\hbox{boxes} \cr
  \ &=\ - {1\over 2} \Bigl(\delta^{\mu\nu}_{[-2\e]} \sum_i p_i^\rho
      {\al{i+1} \ga{i+1} \over \Det_5 }I_5^{D=6}[1]
      +\hbox{cyclic} \Bigr)  + \hbox{boxes} + \Ord(\eps) \cr
  \ &=\ \hbox{boxes} + \Ord(\eps), \cr}
\anoneqn
$$
where we used
$$
c_{ijk}\ =\ {\al{k} \ga{k}\over \Det_5}\ c_{ij} + \hbox{cyclic}
\anoneqn
$$
and
$$
I_5^{D=6}[a_{i}]\ =\
  {\al{i}\ga{i}\over\Det_5} I_5^{D=6}[1] + \hbox{boxes},
\anoneqn
$$
from equation~(\use\dsixcoeffSpecial).
 
It is straightforward to continue in this way, demonstrating that
$I_5^{D=6}$
drops out from the loop momentum integrals encountered in relativistic
field theories.  For gauge theories,
up to five factors of the loop momentum in the numerator can appear.
 
 
\appendix{Constants of Integration for Box Integrals}
\tagappendix\TwoMassBoxConstantIntegral
 
In this appendix we evaluate the constant of integration for the
box with two adjacent massive legs, or with three massive legs,
by performing the integral at the point where all the $\al{i}$
are equal.  The constant of integration for the adjacent two-mass
box is a special case of that for the three-mass box, with $\lambda=0$.
We have
$$
  \Basic_0\ \equiv\
  \Basic_4^\threemass(\alpha_i=1,\lambda)\ =\ \Gamma(2+\e)
  \int d^4u_i\ {\delta\bigl( 1-\sum u_i \bigr)\over
   \bigl[ (u_1+\lambda u_2)u_3 + u_4(1-u_4) \bigr]^{2+\e}}\ .
\eqn\alphaone
$$
We let
$$
u_1\ =\ z(1-y),\quad u_2\ =\ (1-z)(1-y), \quad u_3\ =\ y(1-x), \quad
u_4\ =\ xy.
\eqn\xyzdef
$$
The $z$ integral is elementary and leads to
$$
 \Basic_0\ =\ -{\Gamma(1+\e)\over 1-\lambda} \int_0^1 dx
 \int_0^1 dy {y^{-1-\e} \over 1-x} \Bigl\{
  \bigl[ (1-x)(1-y)+x(1-xy) \bigr]^{-1-\e}
  -   \bigl[ \lambda (1-x)(1-y)+x(1-xy) \bigr]^{-1-\e} \Bigr\}\ .
\eqn\xyintegral
$$
The $y$ integral can be done in terms of hypergeometric functions,
$$
\eqalign{
 \Basic_0\ &=\ { \Gamma(1+\e) \over \e \, (1-\l) }
  \int_0^1 { dx \over 1-x} \biggl\{
  \F21(1+\e,-\e;1-\e;1-x+x^2)  \cr
  &\ -\ \bigl( x + \l (1-x) \bigr)^{-1-\e}
         \F21\left(1+\e,-\e;1-\e;{ \l(1-x)+x^2 \over \l(1-x)+x } \right)
        \biggr\}\ . \cr}
\eqn\xintegral
$$
 
For $\lambda\neq 0$,
the integrand has no singularities as $\e\rightarrow0$,
so we may expand it in $\e$; the hypergeometric functions have the
following expansion for small $\e$,
$$
  \F21(1+\e,-\e;1-\e;v)\ =\ 1\ +\ \e\,\ln(1-v)
  \ +\ \e^2\left[ -2\ \Li_2(v) - \hf\ln^2(1-v) \right]\ +\ {\cal O}(\e^3)
\eqn\hyperexpand
$$
(we only need the first two terms here),
which leads to
$$
\eqalign{
 \Basic_0\ &=\ \rg \int_0^1 dx \left[
    { -{1\over\e} - \ln x - \ln(1-x)
     + 2\ln(\l + (1-\l)x) \over \l + (1-\l)x }
     + { 2\ln(\l + (1-\l)x) \over (1-\l)(1-x) } \right]  \cr
  &=\ {\rg\over 1-\l} \left\{ {\ln\l\over\e}
  \ +\ \int_\l^1 {du\over u}
      \bigl( - \ln(u-\l) - \ln(1-u) + 2\ln u + 2\ln(1-\l) \bigr)
      \ +\ 2\int_0^{1-\l} {dv \over v} \ln(1-v) \right\} \cr
  &=\ {\rg\over 1-\l}
    \biggl[ {\ln\l\over\e}\ -\ {1\over2} \ln^2\l \biggr]\ . \cr}
\eqn\alphaoneanswer
$$
 
In the case $\lambda=0$, we add and subtract terms in~(\use\xintegral)
to obtain
$$
\eqalign{
 \Basic_0(\lambda=0)\ &=\ { \Gamma(1+\e) \over \e}
  \int_0^1 { dx \over 1-x} \biggl\{
   \F21(1+\e,-\e;1-\e;1-x+x^2) - x^{-\e} \F21\L 1+\e,-\e;1-\e; x \R
        \biggr\}\  \cr
&\quad -{\Gamma(1+\e)\over\e}
  \int_0^1 dx\ x^{-1-\e} \L \F21(1+\e,-\e;1-\e;x)-1\R
- { \Gamma(1+\e) \over \e} \int_0^1 dx\ x^{-1-\e}\ .\cr
}\anoneqn
$$
In the first and second integrals,
the integrand is again nonsingular everywhere,
and we can expand in $\e$; the third is elementary:
$$
\eqalign{
 \Basic_0(\lambda=0)\ &=\ 2\,\Gamma(1+\e) \int_0^1 dx {\ln x \over 1-x}
- \Gamma(1+\e) \int_0^1  dx {\ln(1-x)\over x}
+ { \Gamma(1+\e) \over \e^2} \cr
&=\ { \Gamma(1+\e) \over \e^2} \L 1 -{\pi^2\over6} \e^2\R \cr
&=\ { r_\Gamma \over \e^2}\ ,\cr
}\anoneqn
$$
so that $c_0(0)=0$.
 
 
\appendix{The Triangle with Three External Masses}
\tagappendix\AllMassTriangle
 
The differential equations approach also provides an easy way to
derive a compact expression for the three-mass triangle integral
to all orders in $\e$.
(The integral is in fact finite, so only the leading order is
needed in practical calculations; but in order to examine explicitly
the limit in which one of the external masses vanishes, it is convenient
to have the forms derived here, or ones equivalent to
them~[\ref\OtherThreeMassTriangles{D. Kreimer, Mainz preprint
MZ--TH--92--20\semi
A. Davydychev, J. Phys. A25:5587 (1992)}].)
 
The three-mass triangle with massless internal lines
satisfies the following system of equations
(using $N_3=1$):
$$
{\partial\L \Det_3^{1/2-\eps}\Basic_3\R\over\partial\al{i}}
  \ =\ -\hf(1-2\e) \Det_3^{1/2-\eps}
\sum_{j=1}^3 \L \eta_{ij} - {\ga{i}\ga{j}\over\Det_3}\R\, \Basic_2^{(j)}\ ,
\eqn\tripde
$$
where
$$
\Det_3\ =\ -\al1^2-\al2^2-\al3^2 + 2\al1\al2+2\al2\al3+2\al3\al1,
\eqn\tridet
$$
so that
$$\eqalign{
\ga{i}\ &=\ \sum_{j=1}^3 \al{j} - 2\al{i},\cr
\Det_3\ &=\ \ga1\ga2+\ga2\ga3+\ga3\ga1.\cr
}\eqn\trigamdet
$$
Also, the two-point integrals $\hat I_2^{(i)}$ are very simple,
$$
\eqalign{
  \hat I_2^{(i)}\ &=\ \Gamma(\e)\ (\alpha_{i+1}\alpha_{i-1})^{\e-1}
   \int_0^1 dx\ x^{-\e}(1-x)^{-\e} \cr
     &=\ {\rg\over \e(1-2\e)} (\alpha_{i+1}\alpha_{i-1})^{\e-1}\ .
 \cr}
\eqn\bubble
$$
 
Notice that a function of
$$
  \delta_j\ \equiv\ {\gamma_j\over\sqrt{\Det_3}}
$$
obeys
$$
  {\del F(\delta_j)\over\del\alpha_i}
  \ =\ \Det_3^{-1/2}\L \eta_{ij}-{\gamma_i\gamma_j\over\Det_3} \R
           F^\prime(\delta_j),
\eqn\fdelta
$$
and that
$$
  \alpha_{i+1}\alpha_{i-1}
   \ =\ \coeff14 (\gamma_i+\gamma_{i-1}) (\gamma_i+\gamma_{i+1})
   \ =\ \coeff14\,\Det_3\,(1+\delta_i^2).
\eqn\alalsimple
$$
Therefore we may solve the differential equations~(\use\tripde) by
$$
  \Basic_3\ =\ \Det_3^{-1/2+\eps}\LB F(\delta_1)+F(\delta_2)+F(\delta_3)
    + C \RB\ ,
\eqn\triansatz
$$
where $F(\delta)$ satisfies
$$
F^\prime(\delta)\ =\ -\hf(1-2\e)\,\Det_3^{1-\eps}\
  \left[ {\rg\over \e(1-2\e)}
    \bigl(\coeff14\,\Det_3\,(1+\delta^2)\bigr)^{\e-1} \right]
 \ =\ -{2^{1-2\e}\rg\over\e}\ (1+\delta^2)^{\e-1},
\eqn\fdiffeqn
$$
and $C$ is a constant of integration.
 
We need the integral
$$\eqalign{
\int_0^\delta dz\; (1+z^2)^{\e-1}\ &=\
  \int_0^\delta dz\; (1+iz)^{\e-1}(1-i z)^{\e-1}\cr
 &=\ -i \int_1^{1+i\delta} dw\; w^{\e-1} 2^{\e-1} (1-w/2)^{\e-1}\cr
 &=\ -{2^{\e-1}i\over\e} \left[ (1+i\delta)^{\e}
   \F21\L1-\e,\e;1+\e; {1+i\delta\over2}\R\ -\
   \F21\L1-\e,\e;1+\e; {1\over2}\R \right]  \cr
 &=\ -{2^{2\e-1}i\over\e}\Biggl[ \L {1+i\delta\over 1-i\delta}\R^{\e}
    \F21\L2\e,\e;1+\e; - {1+i\delta\over 1-i\delta}\R
   \ -\ \F21\L2\e,\e;1+\e; -1\R \Biggr] \cr
 &=\ {4^{\e-1}\over\e}{1\over i}
   \Biggl[ \L {1+i\delta\over 1-i\delta} \R^\e
    \F21\L 2\e,\e;1+\e; - {1+i\delta\over 1-i\delta}\R \cr
 &\qquad\quad \ -\ \L {1-i\delta\over 1+i\delta} \R^\e
    \F21\L 2\e,\e;1+\e; - {1-i\delta\over 1+i\delta}\R
   \Biggr]\ , \cr
 }
\eqn\onetriint
$$
where we have symmetrized the result in the last line.
Alternatively, we may do the integral as
$$
\eqalign{
  \int_0^\delta dz\ (1+z^2)^{\e-1}
  \ &=\ \int_0^\delta dz\ \sum_{m=0}^\infty (\e-1)(\e-2)\cdots(\e-m)
    {z^{2m}\over m!} \cr
  \ &=\ \sum_{m=0}^\infty (\e-1)(\e-2)\cdots(\e-m)
    {\delta^{2m+1}\over m!(2m+1)} \cr
  \ &=\ \delta \sum_{m=0}^\infty {(1-\e)(2-\e)\cdots(m-\e)
   (\hf)(\coeff32)\cdots (m-\hf)\ (-\delta^2)^m \over
     (\coeff32)\cdots (m-\hf)(m+\hf)\ m!} \cr
  \ &=\ \delta\ \F21(1-\e,\hf;\coeff32;-\delta^2). \cr}
\eqn\twotriint
$$
The two expressions for the integral can be related using
a variety of hypergeometric identities.
 
Thus we have
$$
  \hat I_3(\alpha_i)
  \ =\ -{1\over2}{\rg\over\e^2} \Det_3^{-1/2+\e}
  \bigl[ f(\delta_1) + f(\delta_2) + f(\delta_3) + c \bigr]\ ,
\eqn\finaltrians
$$
where
$$
\eqalign{
  f(\delta)\ &=\ \e\,4^{1-\e}\,\delta\ \F21(1-\e,\hf;\coeff32;-\delta^2)
  \cr
  &=\ {1\over i}
   \left[ \L {1+i\delta\over 1-i\delta}\R^{\e}
    \F21\L2\e,\e;1+\e; - {1+i\delta\over 1-i\delta}\R
  \ -\ \L {1-i\delta\over 1+i\delta}\R^{\e}
    \F21\L2\e,\e;1+\e; - {1-i\delta\over 1+i\delta}\R
   \right]\ . \cr}
\eqn\finalfform
$$
 
To fix the constant $c$, it is easiest to consider the integral at
the following, somewhat asymmetric, kinematic point:
$$
  s_{12}\ =\ {-1\over\alpha_3\alpha_1}\ = -{1\over2}\ , \quad
  s_{23}\ =\ {-1\over\alpha_1\alpha_2}\ = -{1\over2}\ , \quad
  s_{31}\ =\ {-1\over\alpha_2\alpha_3}\ = -1,
\eqn\schoice
$$
or
$$
  \alpha_1\ =\ 2,\quad \alpha_2\ =\ \alpha_3\ = 1;
  \qquad \Det_3\ =\ 4;
  \qquad \delta_1\ =\ 0, \quad \delta_2\ =\ \delta_3\ =\ 1.
\eqn\alphachoice
$$
At this kinematic point, we make the change of variables
$$
 a_1\ =\ 1-y, \qquad a_2\ =\ xy, \qquad a_3\ =\ (1-x)y,
\eqn\trichangeofvar
$$
with Jacobian equal to $y$, and obtain
$$
\eqalign{
  \hat I_3(2,1,1)\ &=\ {\Gamma(1+\e)\over2}
    \int_0^1 d^3a_i\
    { \delta(1-{\textstyle \sum} a_i) \over
     (\coeff12 a_3a_1 + \coeff12 a_1a_2 + a_2a_3)^{1+\e} } \cr
  &=\ 2^\e\Gamma(1+\e) \int_0^1 dx \int_0^1 dy\ y^{-\e}
    \bigl[ 1 - (1- 2x(1-x)) y \bigr]^{-1-\e}  \cr
  &=\ 2^\e {\Gamma(1+\e)\over 1-\e} \int_0^1 dx
  \ \F21(1+\e,1-\e;2-\e; 1-2x(1-x))\ . \cr }
\eqn\starteval
$$
 
Next we use the change-of-variables,
$$
  x\ =\ {1-\sqrt{1-z} \over 2}\ , \qquad z\ =\ 4x(1-x),
\eqn\xtoz
$$
and a hypergeometric identity, to get
$$
\eqalign{
  \hat I_3(2,1,1)\ &=\ 2^{\e-1} {\Gamma(1+\e)\over 1-\e}
 \int_0^1 dz\ (1-z)^{-1/2} \Biggl[
 {\Gamma(2-\e)\Gamma(-\e)\over\Gamma(1-2\e)}\F10(1-\e;z/2) \cr
 &\qquad +\ {\Gamma(2-\e)\Gamma(\e)\over\Gamma(1+\e)\Gamma(1-\e)}
  2^\e \, z^{-\e}\, \F21(1,1-2\e;1-\e;z/2) \Biggr] \cr
  &=\ 2^{\e-1} {\Gamma(1+\e)\over 1-\e} \LB
  2 {\Gamma(2-\e)\Gamma(-\e)\over\Gamma(1-2\e)}
   \F21(1-\e,1;\coeff32;\coeff12)\RP\cr
  &\qquad \LP
  \ +\ 2^\e {\Gamma(2-\e)\Gamma(\e)\Gamma(\hf)
       \over\Gamma(1+\e)\Gamma(\coeff32-\e)}
    \F21(1,1-2\e;\coeff32-\e;\hf) \right]\ , \cr }
\eqn\nexteval
$$
We then use the following hypergeometric identities,
$$\eqalign{
  \F21(1-\e,1;\coeff32;\coeff12)
  \ &=\ 2^{1-\e} \F21(1-\e,\hf;\coeff32;-1)\ ,\cr
  \F21(1,1-2\e;\coeff32-\e;\hf)\ &=\ \F21(\hf,\hf-\e;\coeff32-\e;1)
  \ =\ {\Gamma(\coeff32-\e)\Gamma(\hf)\over\Gamma(1-\e)}\ ,\cr
}\eqn\gridents
$$
to get
$$
  \hat I_3(2,1,1)\ =\ -{2\rg\over\e} \F21(1-\e,\hf;\coeff32;-1)
    \ +\ {4^\e \pi\Gamma(1+\e)\over 2\e}\ .
\eqn\directtrisp
$$
 
On the other hand, plugging the values of $\Det_3$ and $\delta_i$
from~(\use\alphachoice) into equation~(\use\finaltrians), we have
$$
  \hat I_3(2,1,1)\ =\ -{4^{-1/2+\e}\rg\over2\e^2}
  \Bigl[ 2 \, \e\ 4^{1-\e}\, \F21(1-\e,\hf;\coeff32;-1) + c \bigr]\ .
\eqn\plugtrisp
$$
Comparing equations~(\use\directtrisp) and~(\use\plugtrisp), we find
$$
  c\ =\ -{2\pi\e\Gamma(1+\e)\over\rg}
  \ =\ -2\pi\,\e\,{\Gamma(1-2\e)\over\Gamma^2(1-\e)}.
\eqn\fixc
$$
 
Despite its appearance, equation~(\use\finaltrians) does have a finite
limit as $\e\to0$,
$$
  \hat I_3(\alpha_i)\ =\ {i\over \sqrt{\Det_3}}  \sum_{j=1}^3
  \left[ \Li_2\left(-\left({1+i\delta_j \over 1-i\delta_j}\right)\right)
       - \Li_2\left(-\left({1-i\delta_j \over 1+i\delta_j}\right)\right)
  \right]\ + \ \Ord(\e),
\eqn\deqfourtrians
$$
which is the form given in ref.~[\HungJung].
 
 
\appendix{Higher-Point Scalar Integrals}
\tagappendix\HigherPointScalarIntegrals
 
In this appendix, we discuss formul\ae\ allowing the evaluation
of higher-point scalar integrals ($n>5$), in part to
correct some statements we made in a previous paper~[\use\OurAllN].
The corrected results will be similar to results obtained
previously by Melrose, and by van Neerven and
Vermaseren~[\use\Melrose,\use\vNV].  The main difference is that the
present results allow for external kinematics in the full
$4-2\e$ dimensions, which is useful for obtaining tensor integrals by
the differentiation method discussed in sections~\use\Properties,
\use\GeneratingFunctionSection, and
appendix~\use\FPinHigherPointIntegrals.
 
We begin by recalling equation~(\use\reducedallNeqn), which we rewrite
here in a slightly different form,
$$
  \Basic_n\ =\ \sum_{i=1}^n {\gamma_i\over 2N_n} \,
   \Basic_{n-1}^{(i)}\ +\ (n-5+2\e) \, {\Det_n\over 2N_n} \,
   \Basic_n^{D=6-2\e}\ .
\eqn\rallNapp
$$
For $n\geq6$, in order to use equation~(\rallNapp) to evaluate scalar
integrals, it is desirable to
take the external momenta $k_1,k_2,\ldots,k_n$ to be restricted to
$D=4$.  The loop momenta have to remain in $D=4-2\e$ in order to
regulate infrared divergences.
In the 't~Hooft-Veltman variant of
dimensional regularization, the external momenta appearing in the
one-loop integral in a next-to-leading-order calculation are
indeed taken to be four-dimensional.
In the conventional dimensional regularization
scheme, the external momenta are taken to be
$4-2\eps$-dimensional, but this will generally lead to
only $\Ord(\eps)$ corrections, since the integrals $I_n^{D=6-2\eps}$
are finite as $\eps \rightarrow 0$ for $n\geq4$.
 
In reference~[\use\OurAllN], we argued that the
term containing $\Basic_n^{D=6-2\e}$ in equation~(\rallNapp)
could be dropped for $n\geq6$, when the external momenta are
restricted to $D=4$.
The argument was based on the fact that for $n\geq6$ the Gram
determinant $\Det_n$ appearing in equation~(\use\rallNapp)
vanishes for $D=4$ kinematics,
due to the linear dependence of the $(n-1)$ vectors
$k_1,k_2,\ldots,k_{n-1}$~[\use\ByK,\ref\Gram{%
V. E.\ Asribekov, Sov.\ Phys.\ -- JETP 15:394 (1962)\semi
N. Byers and C. N. Yang, Rev.\ Mod.\ Phys.\ 36:595 (1964)}].
If the $\Basic_n^{D=6-2\e}$ term could be dropped, then
equation~(\rallNapp) would reduce to a simple
recursion relation expressing the scalar integrals $\Basic_n$ as
a linear combination of the $n$ $(n-1)$-point integrals $\Basic_{n-1}$.
For $n=6$, the argument does indeed hold, and the scalar hexagon
integral is given by
$$
  \Basic_6\ =\ \sum_{i=1}^6 {\gamma_i\over 2N_6} \, \Basic_{5}^{(i)}
    \qquad\qquad\hbox{($D=4$ kinematics)}.
\eqn\rneqsixNapp
$$
 
Unfortunately, for $n\geq7$ the situation is more complicated.
It is true that for $n\geq6$ the Gram determinant $\Det_n$
vanishes for $D=4$ external kinematics.
However, for $n\geq7$, the factor of $N_n$ in the
denominator also vanishes.
Indeed, $N_n$ is given by $N_n = 2^{n-1}\det\rho =
2^{n-1}\left(\prod_{i=1}^n\alpha_i\right)^2 \det S$,
and the dimension of the null space of the $n\times n$ matrix
$S_{ij}$ is $n-6$ for $D=4$ kinematics~[\use\Melrose].
Therefore, for $n\geq7$, the coefficients appearing in
equation~(\use\rallNapp) are not well-defined for $D=4$ external
kinematics (which is where we would like to use the equation).
 
Notice that both numerator and denominator of the coefficient
ratios $\gamma_i/2N_n$ and $\Det_n/2N_n$ vanish for $D=4$ kinematics:
The matrices that give rise to $\Det_n$ and to
$\gamma_i= \half(\del\Det_n/\del\alpha_i)$ have null spaces of
dimension $n-5$ and $n-6$ respectively.
Based on the dimensions of the corresponding null spaces, we can argue
that $\Det_n$ vanishes ``faster'' than $N_n$, and $\gamma_i$
vanishes ``equally fast'', as $D=4$ kinematics are approached.
Thus we might expect that a modification of equation~(\rallNapp) should
exist, which is well-defined for $D=4$ kinematics, and for which
the coefficient of $\Basic_n^{D=6-2\e}$ vanishes in this limit.
In fact, van Neerven and Vermaseren~[\use\vNV] have shown how to obtain
such an equation, which expresses an $n$-point scalar integral in terms
of six $(n-1)$-point integrals.  (Their derivation was carried out for
$D=4$ loop momenta; however it is easy to see that it is equally valid for
$D=4-2\e$ loop momenta as well, as long as the external momenta are
restricted to $D=4$.)
 
Here we will obtain an equation similar to~(\rallNapp), but where the
coefficients have $N_{n-1}^{(k)}$ in the denominator instead of $N_n$.
Since $N_6$ is nonzero for generic $D=4$ kinematics,
this equation will be well-defined for the heptagon integral
($n=7$) in $D=4$.  It reduces to the above-mentioned equation of van
Neerven and Vermaseren in $D=4$, but it is also well-defined away from
$D=4$, which makes it a useful starting point if one wishes to apply
the differentiation approach of sections~\use\Properties\ and
\use\GeneratingFunctionSection\ to compute
tensor integrals.
The reason why restricting external kinematics to $D=4$ complicates
the differentiation approach is that the $\alpha_i$ variables are then
subject to various Gram-determinental constraints~[\use\ByK,\use\Gram],
which would have to be respected in performing the differentiations.
After carrying out the differentiations it is permissible, and usually
desirable, to restrict the external kinematics to $D=4$, in order
to take advantages of certain simplifications.
An example of this procedure, for the
one-parameter heptagon integrals, is provided in the next appendix.
 
To derive the new scalar equation, we first need some general relations
between the quantities $\Det_n$, $\gamma_i$ and $N_n$,
which are associated with the
integral $\Basic_n$, and the corresponding quantities
$\Det_{n-1}^{(k)}$, $\gamma_i^{(k)}$ and $N_{n-1}^{(k)}$ associated
with the $(n-1)$-point ``daughter'' integral $\Basic_{n-1}^{(k)}$.
As in section~\use\AllNSection, we choose the $\alpha_i$ variables for the
daughter and parent integrals to be the same.
We also take the kinematics to be general for now, i.e. not restricted to
$D=4$, so that all quantities are well-defined.
The necessary relations follow from the observation:
   If $A$ is a symmetric $n\times n$ matrix, and $B_{(k)}$
is the $(n-1)\times(n-1)$ matrix formed by crossing out the
$k^{\rm th}$ row and $k^{\rm th}$ column of $A$, then the inverse of
$B_{(k)}$ can be computed as
$$
  \left( B^{-1}_{(k)} \right)_{ij}\ =\
  A^{-1}_{ij} - { A^{-1}_{ik} A^{-1}_{kj}\over A^{-1}_{kk} }\ ,
\qquad\qquad i\neq k,\ j\neq k.
\eqn\Binv
$$
The proof is simply to multiply equation~(\Binv) on the left by
$(B_{(k)})_{\ell i} = A_{\ell i}$, and simplify.
Note also that $\left( B^{-1}_{(k)} \right)_{ij}$
vanishes for either $i=k$ or $j=k$.
 
Starting with the expression~(\use\etadefn) for $\Det_{n-1}^{(k)}$ in
terms of $\alpha_i$ and the matrix $\eta^{(k)}$, and using
equations~(\use\Nnetarhoeqn) and (\use\gammadefn), we can rewrite
$\Det_{n-1}^{(k)}$ as
$$
\eqalign{
  \Det_{n-1}^{(k)}\ &=\
  \sum_{i,j\neq k} \alpha_i \eta_{ij}^{(k)} \alpha_j
  \ =\ \alpha^{\rm T} \eta^{(k)} \alpha \cr
   &=\ \gamma^{\rm T} \eta^{-1} \eta^{(k)} \eta^{-1} \gamma
  \ =\ {N_{n-1}^{(k)}\over N_n^2} \gamma^{\rm T} \rho
  \Bigl( \rho_{n-1}^{(k)} \Bigr)^{-1} \rho \gamma \cr
 &=\ {N_{n-1}^{(k)}\over N_n^2} \sum_{i,j=1}^n
  (\gamma^{\rm T} \rho)_i
   \left( \rho^{-1}_{ij}
     - {\rho^{-1}_{ik} \rho^{-1}_{kj}\over \rho^{-1}_{kk} } \right)
  (\rho \gamma)_j \cr
 &=\ {N_{n-1}^{(k)}\over N_n^2} \left[ \gamma^{\rm T} \rho \gamma
    - {\gamma_k^2\over\rho^{-1}_{kk}} \right]\ . \cr}
\eqn\firstdeteval
$$
Using the definitions~(\use\Nnetarhoeqn) $N_n = 2^{n-1}\det\rho$,
$N_{n-1}^{(k)} = 2^{n-2}\det\rho^{(k)}$,
and the fact that
$\det\rho^{(k)}\ =\ \rho^{-1}_{kk}\,\det\rho$ is the cofactor of
the $kk$ element of $\rho$, we have
$$
  {\eta_{kk}\over N_n}\ =\ \rho^{-1}_{kk}\ =\ {2 N_{n-1}^{(k)}\over N_n}
  \ .
\eqn\Nnminusone
$$
Using equations~(\use\firstdeteval), (\use\Nnminusone) and
the relation $\gamma^{\rm T} \rho \gamma = N_n \Det_n$ which follows
from equation~(\use\gammaconj),
we obtain expressions for $\Det_{n-1}^{(k)}$ and its derivatives
with respect to $\alpha_i$:
$$
\eqalign{
  \Det_{n-1}^{(k)}\ &=\ {\eta_{kk}\Det_n-\gamma_k^2\over2N_n}\ , \cr
  \gamma_i^{(k)}\ &=\ {\eta_{kk}\gamma_i-\eta_{ik}\gamma_k\over2N_n}\ , \cr
  \eta_{ij}^{(k)}\ &=\ {\eta_{kk}\eta_{ij}-\eta_{ik}\eta_{kj}\over2N_n}
  \ . \cr}
\eqn\detgammaetaminusone
$$
One can iterate this procedure to get expressions for
$\Det_{n-2}^{(k,p)}$, etc., if necessary.
 
Now we proceed to derive the new scalar equation which is of use
for $n=7$.
To do this, we consider equation~(\rallNapp), and also
the one-parameter equation
$$
 \Basic_n[a_k]
  \ =\  \sum_{i=1}^n {\eta_{ki}\over 2 N_n}\ \Basic_{n-1}^{(i)}
   \ +\ (n-5+2\e)\,{\gamma_k\over 2 N_n}\ \Basic_n^{D=6-2\e}\ .
\eqn\rInamixedapp
$$
Multiply equation~(\rInamixedapp) by $\gamma_k/\eta_{kk}$
and subtract it from equation~(\rallNapp), to get
$$
  \Basic_n\ =\ \sum_{i=1}^n
  \biggl[ {\gamma_i\over 2 N_n} -
  {\eta_{ik}\gamma_k\over \eta_{kk}\cdot 2N_n} \biggr]\ \Basic_{n-1}^{(i)}
   \ +\ {\gamma_k\over\eta_{kk}} \, \Basic_n[a_k]
   \ +\ (n-5+2\e)\, \biggl[ {\Det_n\over 2 N_n} -
  {\gamma_k^2\over\eta_{kk}\cdot 2N_n} \biggr]\ \Basic_n^{D=6-2\e}\ .
\eqn\intermeqn
$$
 
The coefficients in brackets in equation~(\intermeqn)
can now be rewritten in terms of $(n-1)$-point quantities
using equations~(\detgammaetaminusone).
We get
$$
  \Basic_n\ =\
  \sum_{i=1}^n {\gamma_i^{(k)}\over 2 N_{n-1}^{(k)}}\ \Basic_{n-1}^{(i)}
   \ +\ {\gamma_k\over 2N_{n-1}^{(k)}} \, \Basic_n[a_k]
   \ +\ (n-5+2\e)\, {\Det_{n-1}^{(k)}\over 2 N_{n-1}^{(k)}}
   \ \Basic_n^{D=6-2\e}\ .
\eqn\scalarneqn
$$
Any value of $k=1,2,\ldots,n$ may be used in this formula.
Note that $\gamma_k^{(k)}=0$, so there are only $n-1$ terms in
equation~(\scalarneqn).
 
For $n=7$ and (generic) $D=4$ kinematics, we have $N_6^{(k)} \neq 0$,
while $\Det_6^{(k)}=0$ and $\gamma_k=0$.  So equation~(\scalarneqn)
reduces to
$$
  \Basic_7\ =\
  \sum_{i=1}^7 {\gamma_i^{(k)}\over 2 N_{6}^{(k)}}\ \Basic_{6}^{(i)}
  \qquad\qquad\hbox{($D=4$ kinematics)},
\eqn\scalarhepteqn
$$
which contains only six hexagons due to the vanishing of
$\gamma_k^{(k)}$.
Indeed, the formula can be shown to be equivalent to the
scalar integral formula of Melrose, and van Neerven and
Vermaseren~[\use\Melrose,\use\vNV].
For $n>7$, equation~(\scalarneqn) is still ill-defined.
Presumably one could go on to construct equations in terms of
$\gamma^{(k,l)}$, $\Det_{n-2}^{(k,l)}$, etc. that will be well-defined
for $n=8$, and so on.  This would be useful for evaluating the
corresponding tensor integrals via differentiation.
 
 
\appendix{Higher-Point Tensor Integrals}
\tagappendix\FPinHigherPointIntegrals
 
In this appendix, we derive formul\ae\ allowing the evaluation
of tensor integrals for the pentagon ($n=5$) and hexagon ($n=6$)
integrals, for arbitrary internal and external masses.
We also briefly discuss tensor heptagon ($n=7$) integrals.
The situation regarding tensor integrals is similar to the case of
the pentagon discussed in section~\use\GeneratingFunctionSection\ and
appendix~\use\DFiveDrop.  In order
to effectively use the differentiation approach, one must
show two things:  First, that the $1/\e$ pole encountered in the basic
formula~(\use\BasicNPoint), at the level of $n-3$ Feynman parameter
insertions in the $n$-point integral, does not present any problems;
and second, that the ``hard'' six-dimensional integrals
$\Basic_n^{D=6-2\e}$ (for $n\geq5$)
always drop out of any ``physical'' tensor integral, i.e. any integral
which is the Feynman-parametrization of some loop-momentum integral.
We discuss these issues here to some extent for $n=6,7$; presumably
both points can be shown to hold for arbitrary $n$.
 
For $n=5$ and $n=6$, the insertion of a single Feynman parameter
can be treated using equation~(\use\rInamixedapp).
For $n=5$, the term containing the $D=6-2\e$ scalar integral
$\Basic_n^{D=6-2\eps}$ is $\Ord(\e)$ and can be ignored.
For $n=6$, this term is $\Ord(1)$, and so
we would like to show that for ``physical''
one-parameter integrals (linear combinations of the parameters
corresponding to Feynman-parametrization of some loop-momentum
integral), and for $D=4$ external kinematics,
the integral $\Basic_6^{D=6-2\e}$ drops out.
Feynman parametrization of the loop-momentum integral $I_n[p^\mu]$
leads to a linear combination of one-parameter integrals, similar to
the first equation in~(\use\sometensors),
$$
  \sum_{i=2}^n p_{i-1}^\mu I_n[a_i]
  \ \propto\ \sum_{i=2}^n p_{i-1}^\mu \alpha_i  \Basic_n[a_i]\ .
\eqn\oneparamlc
$$
So we can show that $\Basic_6^{D=6-2\eps}$ drops out by showing that
$$
  \sum_{i=2}^6 p_{i-1}^\mu {\alpha_i\gamma_i\over N_6}\ =\ 0
\eqn\wanttoshow
$$
for $D=4$ external kinematics.
To show that equation~(\use\wanttoshow)
holds, it suffices to contract the equation with a set of vectors
$p_{j-1}^\mu$ that span $D=4$ (we can pick
any four of $j=2,\ldots,6$ for nonexceptional momentum
configurations).
We then use equation~(\use\pidotpj) to write
$p_{i-1}\cdot p_{j-1}$ in terms of the matrix $\rho = N_6\eta^{-1}$,
and equation~(\use\gammaconj) to simplify the sum:
$$
\eqalign{
  \sum_{i=2}^6 {\alpha_i\gamma_i\over N_6} \ p_{i-1} \cdot p_{j-1}
  \ &=\ \sum_{i=1}^6 {\alpha_i\gamma_i\over N_6} \left(
    {\rho_{ij}\over\alpha_i\alpha_j}
  - {\rho_{i1}\over\alpha_i\alpha_1}
  - {\rho_{1j}\over\alpha_1\alpha_j}
  + {\rho_{11}\over\alpha_1^2}  \right) \cr
  \ &=\ \sum_{i=1}^6 {\gamma_i\over N_6} \left(
    {\rho_{ij}\over\alpha_j} - {\rho_{i1}\over\alpha_1} \right)
  \ +\ {\Det_6\over N_6}{1\over\alpha_1}
    \left( - {\rho_{1j}\over\alpha_j}
    + {\rho_{11}\over\alpha_1} \right) \cr
  \ &=\ {\Det_6\over N_6}{1\over\alpha_1}
    \left( - {\rho_{1j}\over\alpha_j}
    + {\rho_{11}\over\alpha_1} \right)\ . \cr}
\eqn\oneparamdrop
$$
But $\Det_6 = 0$ while $N_6\neq0$ for $D=4$ external kinematics, so
$\Basic_6^{D=6-2\eps}$ does drop out as desired.
 
We turn next to the insertion of two Feynman parameters.
The first part of the derivation parallels the derivation of
the one-parameter equations~(\use\allnmost) and (\use\reducedInaeqn)
in section~\use\AllNSection.  We consider the integrals
$$\eqalign{
J_{n;i}[a_k]\ &\equiv\ \Gamma(n-3+\e)
\int_0^1 da_{n-1} \int_0^{1-a_{n-1}} da_{n-2} \,\cdots\,
\int_0^{1-a_1-a_2-\cdots-\widehat{a_i}-\cdots- a_{n-1}} da_i \cr
  &\qquad\qquad \times {d\over d a_i}
{a_k \over \LB \sum_{i,j=1}^n S_{ij} a_i a_j\RB^{n-3+\eps} }
    \Biggr\vert_{ a_n = 1 - a_1 - a_2 - \cdots - a_{n-1} }
  \ , \cr}
\eqn\Jnia
$$
evaluated two different ways, to obtain the set of equations
$$
  \sum_{j=1}^n \left( {\rho_{ij}\over\alpha_i}
                    - {\rho_{nj}\over\alpha_n} \right)
  \Basic_n[a_ja_k]\ =\
  {1\over2}\left[ {\Basic_{n-1}^{(i)}[a_k]\over\alpha_i}
                - {\Basic_{n-1}^{(n)}[a_k]\over\alpha_n} \right]
 +   {1\over2}\left( {\delta_{ik}\over\alpha_i}
                - {\delta_{nk}\over\alpha_n} \right)
          \Basic_n^{D=6-2\eps}[1]\ .
\anoneqn
$$
Solving for $\Basic_n[a_ia_j]$, we get
$$
  \Basic_n[a_ia_j]\ =\ {1\over2N_n} \sum_{\ell=1}^n
  \left( \eta_{j\ell} - {\gamma_j\gamma_\ell\over \Det_n} \right)
    \Basic_{n-1}^{(\ell)}[a_i]
  + {\gamma_j\over\Det_n} \Basic_n[a_i]
  + {1\over2N_n}
    \left( \eta_{ij} - {\gamma_i\gamma_j\over \Det_n} \right)
    \Basic_n^{D=6-2\e}\ .
\eqn\firsttwoparam
$$
We can rewrite the right-hand-side of equation~(\use\firsttwoparam) in
terms of scalar integrals only, with the help of
equation~(\use\reducedInamixedeqn):
$$
\eqalign{
  \Basic_n[a_ia_j]\ &=\ {1\over2N_n} \sum_{\ell=1}^n
  \left( \eta_{j\ell} - {\gamma_j\gamma_\ell\over \Det_n} \right)
  {1\over2N_{n-1}^{(\ell)}}
   \Biggl[ \sum_{{p=1\atop p\neq\ell}}^{n} \eta^{(\ell)}_{ip}\
     \Basic_{n-2}^{(\ell,p)}
    \ +\ (n-6+2\e)\, \gamma_i^{(\ell)}\
    \Basic_{n-1}^{D=6-2\eps\ (\ell)}\Biggr] \cr
 &\quad + {\gamma_j\over\Det_n} {1\over2N_n}
     \Biggl[ \sum_{\ell=1}^n \eta_{i\ell}\ \Basic_{n-1}^{(\ell)}
    \ +\ (n-5+2\e)\, \gamma_i\ \Basic_n^{D=6-2\eps} \Biggr] \cr
 &\quad + {1\over2N_n}
  \left( \eta_{ij} - {\gamma_i\gamma_j\over \Det_n} \right)
    \Basic_n^{D=6-2\e}\ .\cr}
\eqn\secondtwoparam
$$
 
In this equation, $\Basic_{n-2}^{(\ell,p)}$ is the $(n-2)$-point
scalar integral obtained from $\Basic_{n-1}^{(\ell)}$ by eliminating
the $p$-th propagator.  We keep the original kinematic
$\alpha_i$-variables defined for $\Basic_n$;
$\Basic_{n-2}^{(\ell,p)}$ will be independent of $\al{\ell}$
and $\al{p}$.  The other quantities ---
$N_{n-1}^{(\ell)}$, $\Det_{n-1}^{(\ell)}$ and its
derivatives --- refer to the normalization, rescaled Gram determinant,
and so on, associated with $\Basic_{n-1}^{(\ell)}$.
We can eliminate $\Basic_{n-1}^{(\ell)}$ from
equation~(\secondtwoparam) in favor of
$\Basic_{n-2}^{(\ell,p)}$ and $\Basic_{n-1}^{(\ell)\ D=6-2\e}$,
and use equations~(\use\detgammaetaminusone) to simplify things.
We get finally
$$
\eqalign{
  \Basic_n[a_ia_j]\ &=\ {\eta_{ij}\Det_n + (n-6+2\e) \gamma_i\gamma_j
    \over 2N_n\Det_n}\ \Basic_n^{D=6-2\e} \cr
&\quad + {n-6+2\e\over 4N_n^2} \sum_{\ell=1}^n \left[
   \eta_{i\ell}\gamma_j + \eta_{j\ell}\gamma_i
 - {\eta_{i\ell}\eta_{j\ell}\gamma_\ell\over\eta_{\ell\ell}}
 - {\gamma_i\gamma_j\gamma_\ell\over\Det_n} \right]
      \Basic_{n-1}^{D=6-2\e\ (\ell)}\cr
&\quad + {1\over 4N_n^2} \sum_{\ell,p=1}^n \left[
   {\eta_{ip}\eta_{j\ell}\eta_{\ell\ell}
    - \eta_{i\ell}\eta_{j\ell}\eta_{\ell p}
   \over\eta_{\ell\ell}} \right] \Basic_{n-2}^{(\ell,p)}\ . \cr }
\eqn\twoparam
$$
This formula merits several comments:\hfil\break
\item 1) The expression $\Basic_{n-2}^{(\ell,p)}$ has no meaning for
$\ell=p$; however, $\ell\neq p$ is enforced automatically by the prefactor.
 
\item 2) For $n=5$, and all-massless kinematics,
this equation reduces to equation~(\use\TwoParamAnswer);
notice that $\eta_{\ell\ell}=1$ in this case,
and that we wrote out the $\Basic_{n-2}$ terms
--- in this case triangles --- more explicitly there.
 
\item 3) For $n=5$ and general kinematics, we now have $\Basic_5[a_ia_j]$ to
${\cal O}(1)$, which means that we have surmounted the
``$1/\e$ barrier'' for the pentagon.  That is, insertions of
more than two Feynman parameters can be obtained using just the
differentiation formula~(\use\BasicNPoint), and equation~(\use\twoparam)
evaluated to $\Ord(1)$.
The argument in appendix~(\use\DFiveDrop) for the cancellation
of $\Basic_5^{D=6-2\e}$ from physical quantities
works for general kinematics too.
 
\item 4) For $n=6$, we will have surmounted the ``$1/\e$ barrier'' if
we can produce $\Basic_6[a_ia_ja_k]$ to ${\cal O}(1)$,
or alternatively the derivative $\del/\del\alpha_k$
of equation~(\use\twoparam)
to ${\cal O}(\e)$.  The $\Basic_{n-2}^{(\ell,p)}$ term presents no problem,
because we can easily compute the first derivatives of box integrals to
${\cal O}(\e)$, using equation~(\use\NPointPDEs) with $n=4$.
The $\Basic_{n-1}^{D=6-2\e\ (\ell)}$ term also presents no problem,
due to the manifest $\e$ prefactor
for $n=6$.  Finally, the $\Basic_n^{D=6-2\e}$ term works out as well: the
$\gamma_i\gamma_j$ term has a manifest $\e$, and the $\eta_{ij}$ term
requires us to know the first derivatives of $\Basic_6^{D=6-2\e}$
to ${\cal O}(\e)$; which we can again compute using
equation~(\use\NPointPDEs), this time with $n=6$ and $\e\to\e-1$,
in terms of the integrals
$\Basic_5^{D=6-2\e\ (\ell)}$ and $\Basic_6^{D=6-2\e}$ through $\Ord(1)$.
 
There is one last step to surmounting the ``$1/\e$ barrier'' for the
hexagon, which is showing that $\Basic_6^{D=6-2\e}$ and
$\Basic_5^{D=6-2\e\ (\ell)}$ drop out of ``physical'' quantities.
Before looking at the three-parameter expression, let's look at the
two-parameter expression~(\use\twoparam) again and see how
how $\Basic_6^{D=6-2\e}$ drops out of Feynman-parametrized loop
integrals.
Feynman parametrization of the integral $I_n[p^\mu p^\nu]$ leads to
$$
  -{1\over2} \delta^{\mu\nu}_{[4-2\e]} I_n^{D=6-2\e}[1]
   \ +\ \sum_{i,j=2}^n p_{i-1}^\mu p_{j-1}^\nu I_n[a_ia_j]\ ,
\eqn\munubasis
$$
which means that we want to show that
$$
  \sum_{i,j=2}^6 p_{i-1}^\mu p_{j-1}^\nu \left[
  {\eta_{ij}\over 2N_6} + {\e \gamma_i\gamma_j \over N_6\Det_6}
  \right] \alpha_i\alpha_j\ =\ {1\over2} \delta^{\mu\nu}_{[4]} + \ordereps.
\eqn\newwanttoshow
$$
Because of the factor of $\Det_6$ in the denominator,
we should really be slightly more careful about how we go to
``$D=4$ kinematics'', than in the one-parameter discussion above.
We choose four of the vectors $p_{i-1}^\mu$ to lie in $D=4$ and therefore
to span $D=4$; we will permit the remaining two vectors to have
components in the $[-2\e]$ directions, and we will only
take $\e\to0$ at the end.
In order to prove that equation~(\use\newwanttoshow) holds,
it suffices to contract it with
$p_{i'-1}^\mu p_{j'-1}^\nu$, where $i'$ and $j'$ each run over the set of
four vectors spanning $D=4$.  (In the expression~(\use\munubasis) we can
consider $\mu$ and $\nu$ to belong to $D=4$, not $[-2\e]$, since we
intend to contract the result with $D=4$ vectors.)
The derivation of equation~(\use\oneparamdrop) continues to be
valid, since we are taking $p_{j-1}$ to be one of the momenta
that lie in $D=4$.  Thus each factor of $\gamma_i$ in~(\newwanttoshow)
will end up with a factor of $\Det_6$,
and the $\gamma_i\gamma_j$ term in the equation drops out
in the limit $\e\rightarrow 0$.
The $\eta_{ij}$ term has a smooth limit; using equation~(\use\pidotpj)
it is easy to show that it gives the desired result,
$\hf p_{i^\prime-1}\cdot p_{j^\prime-1}$.
 
We now sketch how  $\Basic_6^{D=6-2\e}$ drops out of the integral
$I_6[p^\mu p^\nu p^\lambda]$, which after Feynman-parametrization
becomes
$$
  \sum_{i,j,k=2}^6 p_{i-1}^\mu p_{j-1}^\nu p_{k-1}^\lambda
     I_6[a_ia_ja_k]
     - {1\over2} \Bigl( \delta^{\mu\nu}_{[4-2\e]}
       \sum_{k=2}^6  p_{k-1}^\lambda
        I_6^{D=6-2\e}[a_k]\ +\ {\rm cyclic} \Bigr)\ .
\eqn\munulambdabasis
$$
In order that the coefficient of $\Basic_6^{D=6-2\e}$ vanishes from the
combination~(\use\munulambdabasis) we find, after
differentiating~(\twoparam), that we must have
$$
  \sum_{i,j,k=2}^6 p_{i-1}^\mu p_{j-1}^\nu p_{k-1}^\lambda
   \left(
   {\eta_{ij}\gamma_k + \eta_{ik}\gamma_j + \eta_{jk}\gamma_i
      \over 2N_6\Det_6} \right) \alpha_i\alpha_j\alpha_k
   \ =\ {1\over2} \delta^{\mu\nu}_{[4]}
   \sum_{k=2}^6 p_{k-1}^\lambda
   \left( {\gamma_k\over\Det_6} \right) \alpha_k\ +\ {\rm cyclic}.
\eqn\lastwanttoshow
$$
We have already dropped terms with more $\gamma_i$'s in their
numerators than $\Det_6$'s in their denominators, which will vanish
in the $D=4$ limit, following the same logic used earlier.  If we now
use equation~(\use\newwanttoshow), and again drop terms
vanishing in the $D=4$ limit, then we can see that
equation~(\use\lastwanttoshow) is indeed true, and so
$\Basic_6^{D=6-2\e}$ drops out of a momentum-space integral with three
loop-momentum insertions.  Similar considerations
apply to $\Basic_5^{D=6-2\e\ (\ell)}$.
 
For the case of heptagon ($n=7$) tensor integrals, here we will
be content to obtain a well-defined one-parameter equation.
By similar manipulations to those giving equation~(\scalarneqn),
we can get the one-parameter equation,
$$
  \Basic_n[a_i] - {\eta_{ik}\over\eta_{kk}}\Basic_n[a_k]\ =\
  \sum_{j=1}^n {\eta_{ij}^{(k)}\over 2 N_{n-1}^{(k)}}\ \Basic_{n-1}^{(j)}
   \ +\ (n-5+2\e)\, {\gamma_i^{(k)}\over 2 N_{n-1}^{(k)}}
   \ \Basic_n^{D=6-2\e}\ .
\eqn\firstoneparameqn
$$
This equation is not adequate as it stands, since $\Basic_n[a_i]$
appears twice; however, by differentiating equation~(\scalarneqn) with
respect to $\alpha_i$, we get a second one-parameter equation,
$$
\eqalign{
  \Basic_n[a_i]\ &=\ {1\over n-4+2\e} {\del\Basic_n\over\del\alpha_i}
  \cr
 \ &=\ {1\over n-4+2\e} \Biggl[
  \sum_{j=1}^n \left(
  {\eta_{ij}^{(k)}\over 2 N_{n-1}^{(k)}}\ \Basic_{n-1}^{(j)}
 + {\gamma_{j}^{(k)}\over 2 N_{n-1}^{(k)}}\
 {\del\Basic_{n-1}^{(j)}\over\del\alpha_i} \right)
  + {\eta_{ik}\over 2N_{n-1}^{(k)}} \, \Basic_n[a_k] \cr
 &\qquad
  \ +\ (n-5+2\e)\, \left( {2\gamma_i^{(k)}\over 2 N_{n-1}^{(k)}}
   \ \Basic_n^{D=6-2\e} +  {\Det_{n-1}^{(k)}\over 2 N_{n-1}^{(k)}}
   \ {\del\Basic_n^{D=6-2\e}\over\del\alpha_i} \right) \Biggr]\ .\cr}
\eqn\secondoneparameqn
$$
Solving the two equations~(\firstoneparameqn) and~(\secondoneparameqn)
for $\Basic_n[a_i]$, we get
$$
  \Basic_n[a_i]\ =\
  \sum_{j=1}^n {\gamma_{j}^{(k)}\over 2 N_{n-1}^{(k)}}\
  \Basic_{n-1}^{(j)}[a_i]\ +\
  {\gamma_i^{(k)}\over 2 N_{n-1}^{(k)}}\ \Basic_n^{D=6-2\e}
 \ +\ {\Det_{n-1}^{(k)}\over 2 N_{n-1}^{(k)}}
 \ {\del\Basic_n^{D=6-2\e}\over\del\alpha_i}\ .
\eqn\finaloneparameqn
$$
 
For $n=7$ and $D=4$ kinematics, we have $\Det_6^{(k)} = 0$ while
$N_6^{(k)} \neq 0$, so we can drop the last term, to get:
$$
  \Basic_7[a_i]\ =\
  \sum_{j=1}^7 {\gamma_{j}^{(k)}\over 2 N_6^{(k)}}\
  \Basic_6^{(j)}[a_i]\ +\
  {\gamma_i^{(k)}\over 2 N_6^{(k)}}\ \Basic_7^{D=6-2\e}
  \qquad\qquad\hbox{($D=4$ kinematics)}.
\eqn\finaloneparamhepteqn
$$
As with the scalar heptagon equation~(\use\scalarhepteqn), in this
equation any value of $k$, $k=1,\ldots,7$, may be chosen.
In using equation~(\finaloneparamhepteqn),
we would like to know that $\Basic_7^{D=6-2\e}$
drops out of ``physical'' quantities.  This amounts to showing that
$$
  \sum_{i=2}^7 p_{i-1}^\mu {\alpha_i\gamma_i^{(k)}\over 2 N_6^{(k)}}
  \ =\ 0,
\eqn\havetoshow
$$
for $D=4$ external kinematics.  To show that equation~(\havetoshow)
holds, we contract it with four independent vectors spanning $D=4$
Minkowski space, namely $p_{j-1}^\mu$ for $j\in\{2,\ldots,n\}$, $j\neq
k$.  We thus have to show
$$
  \sum_{{i=2\atop i\neq k}}^7 {\alpha_i\gamma_i^{(k)}\over 2 N_6^{(k)}}
  p_{i-1}\cdot p_{j-1}\ =\ 0.
\eqn\newhavetoshow
$$
But this is the same sum encountered in showing that
$\Basic_6^{(k)\ D=6-2\e}$ drops out of ``physical'' linear combinations
of $\Basic_6^{(k)}[a_i]$, which we have already shown above.
 
Finally, the linear combinations of $\Basic_6^{(j)}[a_i]$ that
appear explicitly in equation~(\finaloneparamhepteqn), namely
$\sum_{i=2}^7 p_{i-1}^\mu \, \alpha_i \, a_i$, are also the same as
those occurring in ``physical'' one parameter hexagon integrals
(using $\Basic_6^{(j)}[a_j] = 0$).
So $\Basic_6^{(j)\ D=6-2\e}$ drops out there too.
Therefore ``physical'' combinations of $\Basic_7[a_i]$ in
equation~(\finaloneparamhepteqn) are given
in terms of well-defined, $D=4$ quantities, as desired.
 
To get heptagon integrals with two or more Feynman parameters inserted,
one can differentiate equation~(\finaloneparameqn) with
respect to the $\alpha_i$, and then take the limit of $D=4$ kinematics;
it remains to show that the unwanted six-dimensional integrals drop out
for ``physical'' quantities.
 
\vfill\eject\immediate\closeout\rfile
\centerline{{\bf References}}\bigskip\frenchspacing%
\input refs.tmp\vfill\eject\nonfrenchspacing
 
 
\centerline{\bf Figure Captions}
 
\vskip .2 cm
\noindent
{\bf Fig.~1:}
A schematic depiction of equation (\use\reducedNeqfiveeqn), with the
coefficients suppressed.
 
\noindent
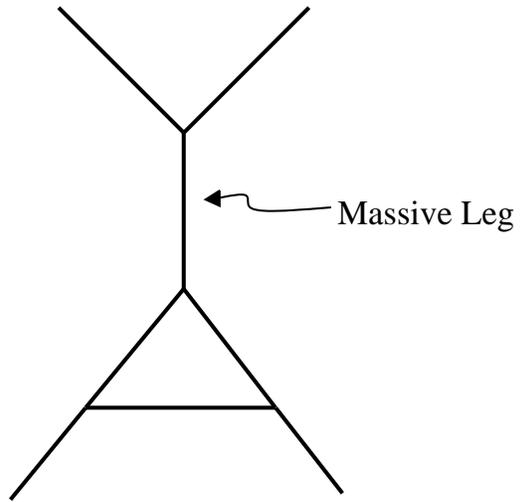
{\bf Fig.~2:} A diagram containing a triangle loop with one
massive (or off-shell) leg.
 
\noindent
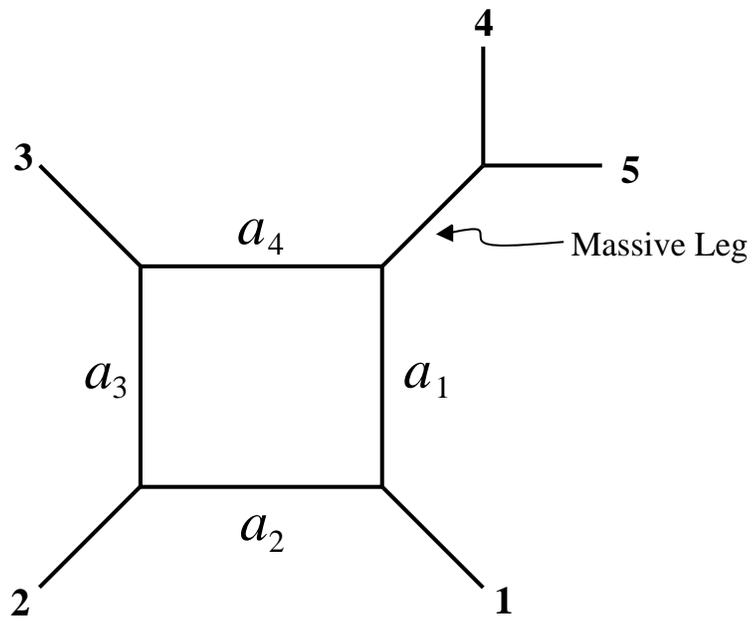
{\bf Fig.~3:}
A diagram containing a box loop with one massive leg.

\bye